\documentclass[aps,prd,reprint,preprintnumbers,showpacs,floatfix,nofootinbib,superscript address]{revtex4-2}
\usepackage{parskip}
\usepackage{printlen}
\usepackage{mathrsfs}
\usepackage{amssymb}
\usepackage{stix}
\usepackage{accents}
\usepackage{listings}
\usepackage{hhline}
\usepackage{xspace}
\usepackage{amsmath}
\usepackage{mathtools}
\usepackage{lipsum}
\usepackage[dvipsnames]{xcolor}
\usepackage{xspace}
\usepackage{multirow,tabularx}
\usepackage[separate-uncertainty=true,multi-part-units=repeat,binary-units]{siunitx}
\usepackage{upgreek}
\usepackage{multirow}
\usepackage{graphicx}
\usepackage{xstring}
\usepackage{etoolbox}
\usepackage{tocbasic}
\usepackage{natbib}
\usepackage{lineno}
\usepackage{tensor}
\usepackage[thinlines,thicklines]{easybmat}
\usepackage{tikz}
\usetikzlibrary{positioning}
\usetikzlibrary{shapes.geometric,arrows}
\usepackage{bm}

\DeclareTOCStyleEntry[linefill=\bfseries\TOCLineLeaderFill]{tocline}{section}
\DeclareTOCStyleEntry[entryformat=\textit,numwidth=10pt,linefill=\TOCLineLeaderFill]{tocline}{subsubsection}

\DeclareSIUnit\parsec{pc} 
\DeclareSIUnit\littleh{\mathsf{h}} 
\DeclareSIUnit\ccs{{m_{\text{p}}}^2} 
\DeclareSIUnit\nothing{\relax} 

\lstdefinelanguage[HiGGS]{Mathematica}[]{Mathematica}{
  morekeywords=[2]{green,mx,Association,KernelID,OptionsPattern,OptionValue,Private,Head,DistributeDefinitions},
morekeywords=[3]{blue,xAct,xTensor,xCore,xPerm,xTras,SymManipulator,AllowUpperDerivatives,Antisymmetric,ChangeCovD,Christoffel,ConstantSymbolQ,ContractMetric,ContractMetrics,DefConstantSymbol,DefScalarFunction,DefTensor,delta,DependenciesOfTensor,IndicesOfVBundle,Labels,LI,LieDToCovD,MakeRule,NoScalar,OverDerivatives,ParamD,PD,PrintAs,Projected,ScalarFunctionQ,ScreenDollarIndices,SeparateMetric,SlotsOfTensor,Symmetric,SymmetryGroupOfTensor,ToCanonical,CommuteCovDs,xTensorQ,Zero,Tensors,ConstantSymbols,DefManifold,IndexRange,DefMetric,FlatMetric,SymCovDQ,RiemannCD,RicciCD,RicciScalarCD,DefCovD},
morekeywords=[4]{red,§WorkingDirectory,HiGGS,BuildHiGGS,Import,DefTheory,StudyTheory,ViewTheory,PoissonBracket,ToNesterForm,ToBasicForm,Velocity,thr,svy,bin,build,HiGGS_sources,HiGGS_variations,
R, T, W, RLambda, TLambda, R1, R2, R3, R4, R5, R6, T1, T2, T3, 
RLambda1, RLambda2, RLambda3, RLambda4, RLambda5, RLambda6, TLambda1, 
TLambda2, TLambda3, Spin1, Spin2, Spin3, STensor, PR1, PR2, PR3, PR4, 
PR5, PR6, PW, PT1, PT2, PT3, Rc, Rs, Tc, V, Lapse, Ji, J, APi, APiP, 
BPi, BPiP, H, B, A, G3, Eps, FoliG, HComp, PPerp, PPara, DVDB, DHDB, 
DJDB, DJiDB, DLapseDB, X, TP, RP, RLambdaP, TLambdaP, TPerp, RPerp, 
TLambdaPerp, RLambdaPerp, DV, DJ, DpJ, DpV, Q, CDAInert, PThreePara, 
PThreePerp, PAPerp, PAPara, PBPerp, PBPara, PA0p, PA1p, PA2p, PA0m, 
PA1m, PA2m, PB0p, PB1p, PB2p, PB1m, PB0pT, PB1pT, PB2pT, PB1mT, 
PA0pT, PA1pT, PA2pT, PA0mT, PA1mT, PA2mT, PT0m, PT1p, PT1m, PT2m, 
PR0p, PR0m, PR1p, PR1m, PR2p, PR2m, PPerpTPerp, PPerpTPara, 
PPerpRPerp, PPerpRPara, PPerpT0p, PPerpT1p, PPerpT1m, PPerpT2p, 
PPerpR0p, PPerpR0m, PPerpR1p, PPerpR1m, PPerpR2p, PPerpR2m, PhiB0p, 
PhiB1p, PhiB1m, PhiB2p, PhiA0p, PhiA0m, PhiA1p, PhiA1m, PhiA2p, 
PhiA2m, BPhi, APhi, ChiB0p, ChiB1p, ChiB1m, ChiB2p, ChiA0p, ChiA0m, 
ChiA1p, ChiA1m, ChiA2p, ChiA2m, ChiPerpB0p, ChiPerpB1p, ChiPerpB1m, 
ChiPerpB2p, ChiPerpA0p, ChiPerpA0m, ChiPerpA1p, ChiPerpA1m, 
ChiPerpA2p, ChiPerpA2m, BChiPerp, AChiPerp, ChiSingB0p, ChiSingB1p, 
ChiSingB1m, ChiSingB2p, ChiSingA0p, ChiSingA0m, ChiSingA1p, 
ChiSingA1m, ChiSingA2p, ChiSingA2m, BChiSingExtra, AChiSingExtra, 
UB0p, UB1p, UB1m, UB2p, UA0p, UA0m, UA1p, UA1m, UA2p, UA2m, PiPB0p, 
PiPB1p, PiPB1m, PiPB2p, PiPA0p, PiPA0m, PiPA1p, PiPA1m, PiPA2p, 
PiPA2m, TP0m, TP1p, TP1m, TP2m, RP0p, RP0m, RP1p, RP1m, RP2p, RP2m, 
PTPerp, PTPara, PRPerp, PRPara, TLambdaP0m, TLambdaP1p, TLambdaP1m, 
TLambdaP2m, RLambdaP0p, RLambdaP0m, RLambdaP1p, RLambdaP1m, 
RLambdaP2p, RLambdaP2m, TPerp0p, TPerp1p, TPerp1m, TPerp2p, RPerp0p, 
RPerp0m, RPerp1p, RPerp1m, RPerp2p, RPerp2m, TLambdaPerp0p, 
TLambdaPerp1p, TLambdaPerp1m, TLambdaPerp2p, RLambdaPerp0p, 
RLambdaPerp0m, RLambdaPerp1p, RLambdaPerp1m, RLambdaPerp2p, 
RLambdaPerp2m, RPPara, RPPerp, RLambdaPPara, RLambdaPPerp, RPerpPerp, 
RPerpPara, RLambdaPerpPerp, RLambdaPerpPara, ChiParaB0m, ChiParaB1p, 
ChiParaB1m, ChiParaB2m, ChiParaA0p, ChiParaA0m, ChiParaA1p, 
ChiParaA1m, ChiParaA2p, ChiParaA2m, DPiPB0p, DPiPB1p, DPiPB1m, 
DPiPB2p, DPiPA0p, DPiPA0m, DPiPA1p, DPiPA1m, DPiPA2p, DPiPA2m, 
DpPiPB0p, DpPiPB1p, DpPiPB1m, DpPiPB2p, DpPiPA0p, DpPiPA0m, DpPiPA1p, 
DpPiPA1m, DpPiPA2p, DpPiPA2m, DTP0m, DTP1p, DTP1m, DTP2m, DRP0p, 
DRP0m, DRP1p, DRP1m, DRP2p, DRP2m, DTLambdaP0m, DTLambdaP1p, 
DTLambdaP1m, DTLambdaP2m, DRLambdaP0p, DRLambdaP0m, DRLambdaP1p, 
DRLambdaP1m, DRLambdaP2p, DRLambdaP2m, DTLambdaPerp0p, 
DTLambdaPerp1p, DTLambdaPerp1m, DTLambdaPerp2p, DRLambdaPerp0p, 
DRLambdaPerp0m, DRLambdaPerp1p, DRLambdaPerp1m, DRLambdaPerp2p, 
DRLambdaPerp2m, DHComp, DpTP0m, DpTP1p, DpTP1m, DpTP2m, DpRP0p, 
DpRP0m, DpRP1p, DpRP1m, DpRP2p, DpRP2m, DpTLambdaP0m, DpTLambdaP1p, 
DpTLambdaP1m, DpTLambdaP2m, DpRLambdaP0p, DpRLambdaP0m, DpRLambdaP1p, 
DpRLambdaP1m, DpRLambdaP2p, DpRLambdaP2m, DpTLambdaPerp0p, 
DpTLambdaPerp1p, DpTLambdaPerp1m, DpTLambdaPerp2p, DpRLambdaPerp0p, 
DpRLambdaPerp0m, DpRLambdaPerp1p, DpRLambdaPerp1m, DpRLambdaPerp2p, 
DpRLambdaPerp2m, DpHComp, SuperHamiltonian0p, LinearSuperMomentum1m, 
RotationalSuperMomentum1m, RotationalSuperMomentum1p, RPShellPara, 
RPShellPerp, PerpBComplement, OrigBComplement, SingBComplement, 
PerpAComplement, OrigAComplement, SingAComplement, KX, KKX, KXP, 
KKXP, DummyGradient, DummyHessian, DummyGradientGreek, 
DummyHessianGreek, RD, RDS1, RDS2, RDS3, TD, TDS1, TDS2, TDS3, 
PhiDB0p, PhiDS1B0p, PhiDS2B0p, PhiDS3B0p, PhiDB1p, PhiDS1B1p, 
PhiDS2B1p, PhiDS3B1p, PhiDB1m, PhiDS1B1m, PhiDS2B1m, PhiDS3B1m, 
PhiDB2p, PhiDS1B2p, PhiDS2B2p, PhiDS3B2p, PhiDA0p, PhiDS1A0p, 
PhiDS2A0p, PhiDS3A0p, PhiDA0m, PhiDS1A0m, PhiDS2A0m, PhiDS3A0m, 
PhiDA1p, PhiDS1A1p, PhiDS2A1p, PhiDS3A1p, PhiDA1m, PhiDS1A1m, 
PhiDS2A1m, PhiDS3A1m, PhiDA2p, PhiDS1A2p, PhiDS2A2p, PhiDS3A2p, 
PhiDA2m, PhiDS1A2m, PhiDS2A2m, PhiDS3A2m, QD, QDS1, QDS2, QDS3, JD, 
JDS1, JDS2, JDS3, LapseD, LapseDS1, LapseDS2, LapseDS3, S1, S2, S3,  
Prt, cAlp1, cAlp2, cAlp3, cAlp4, cAlp5, cAlp6, gAlp1, gAlp2, gAlp3, 
gAlp4, gAlp5, gAlp6, cAlpParaPara0p, cAlpParaPara0m, cAlpParaPara1p, 
cAlpParaPara1m, cAlpParaPara2p, cAlpParaPara2m, cAlpPerpPerp0p, 
cAlpPerpPerp0m, cAlpPerpPerp1p, cAlpPerpPerp1m, cAlpPerpPerp2p, 
cAlpPerpPerp2m, cAlpPerpPara0p, cAlpPerpPara0m, cAlpPerpPara1p, 
cAlpPerpPara1m, cAlpPerpPara2p, cAlpPerpPara2m, cAlpParaPerp0p, 
cAlpParaPerp0m, cAlpParaPerp1p, cAlpParaPerp1m, cAlpParaPerp2p, 
cAlpParaPerp2m, cBet1, cBet2, cBet3, cBet4, cBet5, cBet6, gBet1, 
gBet2, gBet3, gBet4, gBet5, gBet6, cBetParaPara0p, cBetParaPara0m, 
cBetParaPara1p, cBetParaPara1m, cBetParaPara2p, cBetParaPara2m, 
cBetPerpPerp0p, cBetPerpPerp0m, cBetPerpPerp1p, cBetPerpPerp1m, 
cBetPerpPerp2p, cBetPerpPerp2m, cBetPerpPara0p, cBetPerpPara0m, 
cBetPerpPara1p, cBetPerpPara1m, cBetPerpPara2p, cBetPerpPara2m, 
cBetParaPerp0p, cBetParaPerp0m, cBetParaPerp1p, cBetParaPerp1m, 
cBetParaPerp2p, cBetParaPerp2m, mAlp0, mAlp1, mAlp2, mAlp3, mAlp4, 
mAlp5, mAlp6, Alp0, Alp1, Alp2, Alp3, Alp4, Alp5, Alp6, mBet1, mBet2, 
mBet3, mBet4, mBet5, mBet6, Bet1, Bet2, Bet3, Bet4, Bet5, Bet6, 
cPerpA0p, cPerpA0m, cPerpA1p, cPerpA1m, cPerpA2p, cPerpA2m, cPerpB0p, 
cPerpB0m, cPerpB1p, cPerpB1m, cPerpB2p, cPerpB2m, BetPerpPerp0p, 
BetPerpPerp0m, BetPerpPerp1p, BetPerpPerp1m, BetPerpPerp2p, 
BetPerpPerp2m, AlpPerpPerp0p, AlpPerpPerp0m, AlpPerpPerp1p, 
AlpPerpPerp1m, AlpPerpPerp2p, AlpPerpPerp2m, cParaA0p, cParaA0m, 
cParaA1p, cParaA1m, cParaA2p, cParaA2m, cParaB0p, cParaB0m, cParaB1p, 
cParaB1m, cParaB2p, cParaB2m, AlpPerpPara0p, AlpPerpPara0m, 
AlpPerpPara1p, AlpPerpPara1m, AlpPerpPara2p, AlpPerpPara2m, 
BetPerpPara0p, BetPerpPara0m, BetPerpPara1p, BetPerpPara1m, 
BetPerpPara2p, BetPerpPara2m, AlpParaPerp0p, AlpParaPerp0m, 
AlpParaPerp1p, AlpParaPerp1m, AlpParaPerp2p, AlpParaPerp2m, 
BetParaPerp0p, BetParaPerp0m, BetParaPerp1p, BetParaPerp1m, 
BetParaPerp2p, BetParaPerp2m, AlpParaPara0p, AlpParaPara0m, 
AlpParaPara1p, AlpParaPara1m, AlpParaPara2p, AlpParaPara2m, 
BetParaPara0p, BetParaPara0m, BetParaPara1p, BetParaPara1m, 
BetParaPara2p, BetParaPara2m, ShellParaA0p, ShellParaA0m, 
ShellParaA1p, ShellParaA1m, ShellParaA2p, ShellParaA2m, ShellOrigA0p, 
ShellOrigA0m, ShellOrigA1p, ShellOrigA1m, ShellOrigA2p, ShellOrigA2m, 
ShellPerpA0p, ShellPerpA0m, ShellPerpA1p, ShellPerpA1m, ShellPerpA2p, 
ShellPerpA2m, ShellSingA0p, ShellSingA0m, ShellSingA1p, ShellSingA1m, 
ShellSingA2p, ShellSingA2m, ShellPrimA0p, ShellPrimA0m, ShellPrimA1p, 
ShellPrimA1m, ShellPrimA2p, ShellPrimA2m, ShellParaB0p, ShellParaB0m, 
ShellParaB1p, ShellParaB1m, ShellParaB2p, ShellParaB2m, ShellOrigB0p, 
ShellOrigB0m, ShellOrigB1p, ShellOrigB1m, ShellOrigB2p, ShellOrigB2m, 
ShellPerpB0p, ShellPerpB0m, ShellPerpB1p, ShellPerpB1m, ShellPerpB2p, 
ShellPerpB2m, ShellSingB0p, ShellSingB0m, ShellSingB1p, ShellSingB1m, 
ShellSingB2p, ShellSingB2m, ShellPrimB0p, ShellPrimB0m, ShellPrimB1p, 
ShellPrimB1m, ShellPrimB2p, ShellPrimB2m,a,b,c,d,e,f,g,h,i,j,k,l,m,n,o,p,q,r,s,t,u,v,w,x,y,z,a1,b1,c1,d1,e1,f1,g1,h1,i1,j1,k1,l1,m1,n1,o1,p1,q1,r1,s1,t1,u1,v1,w1,x1,y1,z1,CD,G,M4,dimension,TangentM4,G3,H,B,A,BPi,PiPToPi,PADMActivate,R,T,ExpandStrengths,StrengthLambdaSO13Activate,StrengthSO13Activate,PActivate,PiPToPiPO3,StrengthDecompose,StrengthLambdaDecompose,StrengthPToStrengthPO3,StrengthLambdaPToStrengthLambdaPO3,StrengthLambdaPerpToStrengthLambdaPerpO3,PADMPiActivate,PO3PiActivate,PO3TActivate,PO3RActivate,StrengthPerpToStrengthPerpO3,PR0p,DRPDeactivate,DPiPDeactivate,DPiPActivate,DRPActivate,DpRPActivate,DpPiPActivate,DpRPDeactivate,DpPiPDeactivate,ToShell,StrengthPShellToStrengthPO3,PiPShellToPiPPO3,TheoryCDPiPToCDPiPO3,TheoryPiPToPiPO3,epsilonG
}
}

\definecolor{backing}{rgb}{0.95,1,0.8}

\def\jcap{\rm{J. Cosmology Astropart. Phys.}}

\def\prd{\rm{Phys.~Rev.~D}}

\def\procrsa{\rm{Proc.~Math.~Phys.~Eng.~Sci.}}

\def\annphys{\rm{Ann.~Physics (N.~Y.)}}

\def\ptrsla{\rm{Philos.~Trans.~R.~Soc~A}}

\def\jmp{\rm{J.~Math.~Phys.}}

\def\physrev{\rm{Phys.~Rev.}}

\def\epjc{\rm{Eur.~Phys.~J.~C}}

\def\ijmpd{\rm{Int.~J.~Mod.~Phys.~D}}

\def\jhep{\rm{J.~High~Energy~Phys.}}

\def\rmphyo{\rm{Rev.~Mod.~Phys}} 
\def\progtp{\rm{Prog.~Theor.~Phys.}}
\def\appb{\rm{Acta~Phys.~Pol.~B}}

\def\aaca{\rm{Adv.~Appl.~Clifford~Algebras}}

\def\ijgmmp{\rm{Int.~J.~Geom.~Meth.~Mod.~Phys.}}

\newrobustcmd{\ppn}[1]{%
	\IfEqCase{#1}{%
		{1}{\mathcal{O}\left(\varepsilon\right)}%
	}[\mathcal{O}\left(\varepsilon^{#1}\right)]
}

\newcommand{\HiGGS}{\emph{HiGGS}}
\newcommand{\Mathematica}{\emph{Mathematica}}
\newcommand{\SLURM}{\emph{SLURM}}
\newcommand{\TORQUE}{\emph{TORQUE}}

\newcommand{\xAct}{\emph{xAct}}
\newcommand{\WSTP}{\emph{WSTP}}

\makeatletter
\newcommand{\dalembertian}{\mathop{\mathpalette\dalembertian@\relax}}
\newcommand{\dalembertian@}[2]{%
  \begingroup
  \sbox\z@{$\m@th#1\square$}%
  \dimen0=\fontdimen8
    \ifx#1\displaystyle\textfont\else
    \ifx#1\textstyle\textfont\else
    \ifx#1\scriptstyle\scriptfont\else
    \scriptscriptfont\fi\fi\fi3
  \makebox[\wd\z@]{%
    \hbox to \ht\z@{%
      \vrule width \dimen0
      \kern-\dimen0
      \vbox to \ht\z@{
        \hrule height \dimen0 width \ht\z@
        \vss
        \hrule height 2\dimen0
      }%
      \kern-2.5\dimen0
      \vrule width 2.5\dimen0
    }%
  }%
  \endgroup
}
\makeatother

\newrobustcmd{\Mgra}[1]{%
  {\tensor{M}{_{\text{#1}}}}%
}
\newrobustcmd{\Mpro}[1]{%
  {\tensor{\mathproper{M}}{_{\text{#1}}}}%
}
\newrobustcmd{\Malt}[1]{%
  {\tensor{\mathscr{M}}{_{\text{#1}}}}%
}
\newrobustcmd{\Mkom}[1]{%
  {\tensor{\mathfrak{M}}{_{\text{#1}}}}%
}

\newrobustcmd{\Mtotal}{%
  {\tensor{M}{_{\text{T}}}}%
}

\newrobustcmd{\Qtotal}{%
  {\tensor{Q}{_{\text{T}}}}%
}

\newrobustcmd{\Qtotalcal}{%
  {\tensor{\mathcal{  Q}}{_{\text{T}}}}%
}

\newrobustcmd{\action}[1]{%
  {\tensor{S}{_{\text{#1}}}}%
}

\newrobustcmd{\lagrangian}[1]{%
  {\tensor{L}{_{\text{#1}}}}%
}

\newrobustcmd{\lagrangianprop}[1]{%
  {\tensor{\mathproper{L}}{_{\text{#1}}}}%
}

\newrobustcmd{\epl}{%
  {\tensor{\mathsf{e}}{_+}}%
}

\newrobustcmd{\epe}{%
  {\tensor{\mathsf{e}}{_\perp}}%
}

\newrobustcmd{\qz}{%
  {\text{\color{orange}\cmark}}%
}
\newrobustcmd{\jz}{%
  {\text{\color{red}\xmark}}%
}

\newrobustcmd{\projmatrix}[2][placeholder]{%
  {\tensor*{M}{_{#1}^{#2}}}
}
\newrobustcmd{\projorthhum}[2][placeholder]{%
  {\tensor[^#2]{\smash{\check{\mathcal{  P}}}}{#1}}
}
\newrobustcmd{\projorthhumu}[2][placeholder]{%
  {\tensor[^#2]{\smash{\check{\mathcal{  P}}}}{#1}}
}
\newrobustcmd{\projorth}[2][placeholder]{%
  {\tensor[^#2]{\smash{\hat{\mathcal{  P}}}}{#1}}
}
\newrobustcmd{\projlore}[2][placeholder]{%
  {\tensor[^#2]{\hat{\mathcal{  P}}}{#1}}
}
\newrobustcmd{\qprojlore}[2][placeholder]{%
  {\tensor[^#2]{\hat{  P}}{#1}}
}

\newrobustcmd{\gensec}[3][placeholder]{%
  {\tensor*[^#1]{\chi}{^{#2}_{\acu{#3}}}}
}

\newrobustcmd{\glfourr}{%
  {\mathrm{GL}(4,\mathbb{R})}%
}

\newrobustcmd{\sltwoc}{%
  {\mathrm{SL}(2,\mathbb{C})}%
}

\newrobustcmd{\poincare}{%
  {\mathbb{R}^{1,3}\rtimes\mathrm{SO}^+(1,3)}%
}

\newrobustcmd{\poincaref}{%
  {\mathrm{P}(1,3)}%
}

\newrobustcmd{\weyl}{%
  {\mathrm{W}(1,3)}%
}

\newrobustcmd{\conformal}{%
  {\mathrm{C}(1,3)}%
}

\newrobustcmd{\diffeomorphism}{%
  {\mathbb{R}^{1,3}}%
}

\newrobustcmd{\soonethree}{%
  {\mathrm{SO}^+(1,3)}%
}

\newrobustcmd{\othree}{%
  {\mathrm{SO}(3)}%
}

\newrobustcmd{\sothree}{%
  {\mathrm{SO}(3)}%
}

\newrobustcmd{\sotwo}{%
  {\mathrm{SO}(2)}%
}

\newrobustcmd{\suthreec}{%
  {\mathrm{SU}(3)_{\text{c}}}%
}

\newrobustcmd{\sutwol}{%
  {\mathrm{SU}(2)_{\text{L}}}%
}

\newrobustcmd{\uoney}{%
  {\mathrm{U}(1)_{\text{Y}}}%
}

\newrobustcmd{\uone}{%
  {\mathrm{U}(1)}%
}

\newrobustcmd{\uoneem}{%
  {\mathrm{U}(1)_{\text{em}}}%
}

\newrobustcmd{\sutwo}{%
  {\mathrm{SU}(2)}%
}

\newrobustcmd{\eplus}{%
  {\tensor{\mathsf{e}}{_{+}}}%
}

\newrobustcmd{\esf}[1]{%
  {\tensor{\mathsf{e}}{_{#1}}}
}%
\newrobustcmd{\esfu}[1]{%
  {\tensor{\mathsf{e}}{^{#1}}}
}%
\newrobustcmd{\gam}[1]{%
  {\tensor{\gamma}{_{#1}}}
}%
\newrobustcmd{\gamu}[1]{%
  {\tensor{\gamma}{^{#1}}}
}%

\newrobustcmd{\planck}{%
  {m_{\text{p}}}%
}

\newrobustcmd{\caligR}{%
  {\mathcal{R}}%
}
\newrobustcmd{\caligT}{%
  {\mathcal{T}}%
}

\newrobustcmd{\pgt}{%
  PGT\textsuperscript{q,+}\ %
}

\newrobustcmd{\unl}[1]{%
  {\mathfrak{#1}}%
}
\newrobustcmd{\ovl}[1]{%
\overline{#1}%
}
\newrobustcmd{\acu}[1]{%
\acute{#1}%
}

\newrobustcmd{\indiq}[2][placeholder]{%
\IfEqCase{#1}{%
  {placeholder}{%
    \IfEqCase{#2}{%
      {1}{\ovl{k}}%
      {2}{\ovl{kl}}%
      {3}{\ovl{klm}}%
    }%
  }%
}[#1]%
}%

\newrobustcmd{\indaq}[2][placeholder]{%
\IfEqCase{#1}{%
  {placeholder}{%
    \IfEqCase{#2}{%
      {1}{\overline{k}}%
      {2}{\overline{kl}}%
      {3}{\overline{klm}}%
    }%
  }%
}[#1]%
}%

\newrobustcmd{\indeq}[2][placeholder]{%
\IfEqCase{#1}{%
  {placeholder}{%
    \IfEqCase{#2}{%
      {1}{k}%
      {2}{kl}%
      {3}{klm}%
    }%
  }%
}[#1]%
}%

\newrobustcmd{\indoq}[2][placeholder]{%
\IfEqCase{#1}{%
  {placeholder}{%
    \IfEqCase{#2}{%
      {1}{\alpha}%
      {2}{\alpha\beta}%
      {3}{\alpha\beta\gamma}%
    }%
  }%
}[#1]%
}%

\newrobustcmd{\foli}[1]{%
\tensor{n}{_{#1}}%
}

\newrobustcmd{\foliu}[1]{%
\tensor{n}{^{#1}}%
}

\newrobustcmd{\covderl}[1]{%
\tensor{\mathcal{D}}{^{\flat}_{\indiq[#1]{1}}}%
}

\newrobustcmd{\covder}[1]{%
\tensor{\mathcal{D}}{_{\indiq[#1]{1}}}%
}
\newrobustcmd{\coder}[1]{%
\tensor{D}{_{\indiq[#1]{1}}}%
}

\newrobustcmd{\deltal}[2]{%
  \tensor*{\delta}{_{\phantom{\flat}}^{\flat}_{#1}^{#2}}%
}

\newrobustcmd{\deltaud}[2]{%
  \tensor*{\delta}{^{#1}_{#2}}%
}
\newrobustcmd{\etau}[1]{%
\tensor{\eta}{^{\indiq[#1]{2}}}%
}
\newrobustcmd{\etaul}[1]{%
\tensor{\eta}{^{\flat}^{\indiq[#1]{2}}}%
}
\newrobustcmd{\etad}[1]{%
\tensor{\eta}{_{\indiq[#1]{2}}}%
}
\newrobustcmd{\etadl}[1]{%
\tensor{\eta}{^{\flat}_{\indiq[#1]{2}}}%
}

\newrobustcmd{\epsul}[1]{%
\tensor{\epsilon}{^{\flat}^{\indiq[#1]{3}}^{\perp}}
}
\newrobustcmd{\epsdl}[1]{%
\tensor{\epsilon}{^{\flat}_{\indiq[#1]{3}}_{\perp}}
}
\newrobustcmd{\epsd}[1]{%
\tensor{\epsilon}{_{\indiq[#1]{3}}_{\perp}}
}
\newrobustcmd{\epsu}[1]{%
\tensor{\epsilon}{^{\indiq[#1]{3}}^{\perp}}
}

\newrobustcmd{\hfl}[2]{%
  \tensor{h}{^{\flat}_{#1}^{#2}}
}

\newrobustcmd{\bet}[1]{%
  \tensor{\hat{\beta}}{_{#1}}
}
\newrobustcmd{\alp}[1]{%
  \tensor{\hat{\alpha}}{_{#1}}
}
\newrobustcmd{\cgalp}{\tensor{\alpha}{_{\text{CG}}}}

\newrobustcmd{\cbet}[1]{%
  \tensor{\bar{\beta}}{_{#1}}
}
\newrobustcmd{\calp}[1]{%
  \tensor{\bar{\alpha}}{_{#1}}
}

\newrobustcmd{\alpg}[1]{%
  \tensor{\check{\alpha}}{_{#1}}
}
\newrobustcmd{\betg}[1]{%
  \tensor{\check{\beta}}{_{#1}}
}
\newrobustcmd{\calpg}[1]{%
  \tensor{\acu{\alpha}}{_{#1}}
}
\newrobustcmd{\cbetg}[1]{%
  \tensor{\acu{\beta}}{_{#1}}
}

\newrobustcmd{\hub}{%
  {\underline{\mathsf{h}}}
}
\newrobustcmd{\hubm}{%
  {\underline{\mathsf{h}}^{-1}}
}
\newrobustcmd{\hob}{%
  {\bar{\mathsf{h}}}
}
\newrobustcmd{\hobm}{%
  {\bar{\mathsf{h}}^{-1}}
}
\newrobustcmd{\hdet}{%
  {\det \mathsf{h}}
}
\newrobustcmd{\hmdet}{%
  {\det \mathsf{h}^{-1}}
}
\newrobustcmd{\Rsf}{%
  {\mathsf{R}}
}

\newrobustcmd{\alpm}[2][placeholder]{%
  {\tensor*{\hat{\alpha}}{_{#1}^{#2}}}
}
\newrobustcmd{\calpm}[2][placeholder]{%
  \tensor*{\bar{\alpha}}{_{#1}^{#2}}
}

\newrobustcmd{\betm}[2][placeholder]{%
  {\tensor*{\hat{\beta}}{_{#1}^{#2}}}
}
\newrobustcmd{\cbetm}[2][placeholder]{%
  \tensor*{\bar{\beta}}{_{#1}^{#2}}
}

\newrobustcmd{\lamr}{%
  {\lambda_{\mathcal{  R}} }
}
\newrobustcmd{\barlamr}{%
  {\bar{\lambda}_{\mathcal{  R}} }
}
\newrobustcmd{\lamt}{%
  {\lambda_{\mathcal{  T}} }
}
\newrobustcmd{\barlamt}{%
  {\bar{\lambda}_{\mathcal{  T}} }
}

\newrobustcmd{\atmp}[1]{%
  \tensor{\hat{a}}{_{#1}}
}
\newrobustcmd{\btmp}[1]{%
  \tensor{b}{_{#1}}
}

\newrobustcmd{\ctmp}[2][placeholder]{%
  {\tensor*{c}{_{#1}^{#2}}}
}
\newrobustcmd{\dtmp}[2][placeholder]{%
  {\tensor*{d}{_{#1}^{#2}}}
}

\newrobustcmd{\etmp}[1]{%
  \tensor{e}{_{#1}}
}

\newrobustcmd{\batmp}[1]{%
  \tensor{\ovl{a}}{_{#1}}
}
\newrobustcmd{\bbtmp}[1]{%
  \tensor{\ovl{b}}{_{#1}}
}
\newrobustcmd{\bctmp}[1]{%
  \tensor{\ovl{c}}{_{#1}}
}
\newrobustcmd{\bdtmp}[1]{%
  \tensor{\ovl{d}}{_{#1}}
}
\newrobustcmd{\betmp}[1]{%
  \tensor{\ovl{e}}{_{#1}}
}

\newrobustcmd{\ptl}[1]{%
  \tensor{\partial}{#1}
}

\newrobustcmd{\etaf}[1]{%
  \tensor{\eta}{#1}
}

\newrobustcmd{\epsf}[1]{%
  \tensor{\epsilon}{#1}
}

\newrobustcmd{\RSO}[2][placeholder]{%
  {\tensor[^{#2}]{\mathcal{  R}}{#1}}
}
\newrobustcmd{\TSO}[2][placeholder]{%
  {\tensor[^{#2}]{\mathcal{  T}}{#1}}
}
\newrobustcmd{\FSO}[2][placeholder]{%
  {\tensor[^{#2}]{\mathcal{  F}}{#1}}
}
\newrobustcmd{\spinSO}[2][placeholder]{%
  {\tensor[^{#2}]{\sigma}{#1}}
}
\newrobustcmd{\RLambdaSO}[2][placeholder]{%
  {\tensor[^{#2}]{\lambda}{#1}}
}
\newrobustcmd{\TLambdaSO}[2][placeholder]{%
  {\tensor[^{#2}]{\lambda}{#1}}
}
\newrobustcmd{\KSO}[2][placeholder]{%
  {\tensor[^{#2}]{\mathcal{  K}}{#1}}
}

\newrobustcmd{\bper}[2][placeholder]{%
\IfEqCase{#2}{%
  {s}{\tensor{\mathfrak{s}}{#1}}%
  {a}{\tensor{\mathfrak{a}}{#1}}%
  {sbar}{\tensor{\bar{\mathfrak{s}}}{#1}}%
}[\packageError{cosmicclass}{Unidentified Critical Case: #1}{}]%
}

\newrobustcmd{\Jl}{%
  {J^{\flat}}%
}%

\newrobustcmd{\Nl}{%
  {N^{\flat}}%
}%

\newrobustcmd{\haml}[2][placeholder]{%
\IfEqCase{#2}{%
{mom0p}{\tensor{\mathcal{H}}{^{\flat}_{\perp}}}%
{mom1m}{\tensor{\mathcal{H}}{^{\flat}_{\indoq[#1]{1}}}}%
{rot1p}{\tensor{\mathcal{H}}{^{\flat}_{\indaq[#1]{2}}}}%
{rot1m}{\tensor{\mathcal{H}}{^{\flat}_{\perp}_{\indaq[#1]{1}}}}%
}[\packageError{cosmicclass}{Unidentified Critical Case: #1}{}]%
}

\newrobustcmd{\arc}[2][placeholder]{%
\IfEqCase{#2}{%
{B1p}{\tensor{\vartheta}{_{\perp\indiq[#1]{2}}}}%
{B2m}{\tensor[^{\text{T}}]{\vartheta}{_{\indiq[#1]{3}}}}%
{A0m}{\tensor[^{\text{P}}]{\vartheta}{}}%
{A1p}{\tensor{\overset{\wedge}{\vartheta}}{_{\perp\indiq[#1]{2}}}}%
{A1m}{\tensor{\overset{\rightharpoonup}{\vartheta}}{_{\indiq[#1]{1}}}}%
{A2p}{\tensor{\overset{\sim}{\vartheta}}{_{\perp\indiq[#1]{2}}}}%
{A2m}{\tensor[^{\text{T}}]{\vartheta}{_{\perp\indiq[#1]{3}}}}%
}[\packageError{cosmicclass}{Unidentified Critical Case: #1}{}]%
}

\newrobustcmd{\pic}[2][placeholder]{%
\IfEqCase{#2}{%
{B0p}{\varphi}%
{B1p}{\tensor{\overset{\wedge}{\varphi}}{_{\indiq[#1]{2}}}}%
{B1m}{\tensor{\varphi}{_{\perp\indiq[#1]{1}}}}%
{B2p}{\tensor{\overset{\sim}{\varphi}}{_{\indiq[#1]{2}}}}%
{A0p}{\tensor{\varphi}{_\perp}}%
{A0m}{\tensor[^{\text{P}}]{\varphi}{}}%
{A1p}{\tensor{\overset{\wedge}{\varphi}}{_{\perp\indiq[#1]{2}}}}%
{A1m}{\tensor{\overset{\rightharpoonup}{\varphi}}{_{\indiq[#1]{1}}}}%
{A2p}{\tensor{\overset{\sim}{\varphi}}{_{\perp\indiq[#1]{2}}}}%
{A2m}{\tensor[^{\text{T}}]{\varphi}{_{\indiq[#1]{3}}}}%
}[\packageError{cosmicclass}{Unidentified Critical Case: #1}{}]%
}

\newrobustcmd{\picu}[2][placeholder]{%
\IfEqCase{#2}{%
{B0p}{\varphi}%
{B1p}{\tensor{\smash{\overset{\wedge}{\varphi}}}{^{\indiq[#1]{2}}}}%
{B1m}{\tensor{\varphi}{^{\perp\indiq[#1]{1}}}}%
{B2p}{\tensor{\smash{\overset{\sim}{\varphi}}}{^{\indiq[#1]{2}}}}%
{A0p}{\tensor{\varphi}{_\perp}}%
{A0m}{\tensor[^{\text{P}}]{\varphi}{}}%
{A1p}{\tensor{\smash{\overset{\wedge}{\varphi}}}{^{\perp\indiq[#1]{2}}}}%
{A1m}{\tensor{\smash{\overset{\rightharpoonup}{\varphi}}}{^{\indiq[#1]{1}}}}%
{A2p}{\tensor{\smash{\overset{\sim}{\varphi}}}{^{\perp\indiq[#1]{2}}}}%
{A2m}{\tensor[^{\text{T}}]{\varphi}{^{\indiq[#1]{3}}}}%
}[\packageError{cosmicclass}{Unidentified Critical Case: #1}{}]%
}

\newrobustcmd{\picl}[2][placeholder]{%
\IfEqCase{#2}{%
{B0p}{\tensor{\varphi}{^{\flat}}}%
{B1p}{\tensor{\smash{\overset{\wedge}{\varphi}}}{^{\flat}_{\indiq[#1]{2}}}}%
{B1m}{\tensor{\varphi}{^{\flat}_{\perp}_{\indiq[#1]{1}}}}%
{B2p}{\tensor{\smash{\overset{\sim}{\varphi}}}{^{\flat}_{\indiq[#1]{2}}}}%
{A0p}{\tensor{\varphi}{_\perp}^{\flat}}%
{A0m}{\tensor[^{\text{P}}]{\varphi}{^{\flat}}}%
{A1p}{\tensor{\smash{\overset{\wedge}{\varphi}}}{^{\flat}_{\perp\indiq[#1]{2}}}}%
{A1m}{\tensor{\smash{\overset{\rightharpoonup}{\varphi}}}{^{\flat}_{\indiq[#1]{1}}}}%
{A2p}{\tensor{\smash{\overset{\sim}{\varphi}}}{^{\flat}_{\perp\indiq[#1]{2}}}}%
{A2m}{\tensor[^{\text{T}}]{\varphi}{^{\flat}_{\indiq[#1]{3}}}}%
}[\packageError{cosmicclass}{Unidentified Critical Case: #1}{}]%
}

\newrobustcmd{\mull}[2][placeholder]{%
\IfEqCase{#2}{%
{B0p}{\tensor{u}{^{\flat}}}%
{B1p}{\tensor{\smash{\overset{\wedge}{u}}}{^{\flat}_{\indiq[#1]{2}}}}%
{B1m}{\tensor{u}{^{\flat}_{\perp}_{\indiq[#1]{1}}}}%
{B2p}{\tensor{\smash{\overset{\sim}{u}}}{^{\flat}_{\indiq[#1]{2}}}}%
{A0p}{\tensor{u}{_\perp}^{\flat}}%
{A0m}{\tensor[^{\text{P}}]{u}{^{\flat}}}%
{A1p}{\tensor{\smash{\overset{\wedge}{u}}}{^{\flat}_{\perp\indiq[#1]{2}}}}%
{A1m}{\tensor{\smash{\overset{\rightharpoonup}{u}}}{^{\flat}_{\indiq[#1]{1}}}}%
{A2p}{\tensor{\smash{\overset{\sim}{u}}}{^{\flat}_{\perp\indiq[#1]{2}}}}%
{A2m}{\tensor[^{\text{T}}]{u}{^{\flat}_{\indiq[#1]{3}}}}%
}[\packageError{cosmicclass}{Unidentified Critical Case: #1}{}]%
}

\newrobustcmd{\PiP}[2][placeholder]{%
\IfEqCase{#2}{%
{B0p}{\hat{\pi}}%
{B1p}{\tensor{\overset{\wedge}{\hat{\pi}}}{_{\indiq[#1]{2}}}}%
{B1m}{\tensor{\hat{\pi}}{_{\perp\indiq[#1]{1}}}}%
{B2p}{\tensor{\overset{\sim}{\hat{\pi}}}{_{\indiq[#1]{2}}}}%
{A0p}{\tensor{\hat{\pi}}{_\perp}}%
{A0m}{\tensor[^{\text{P}}]{\hat{\pi}}{}}%
{A1p}{\tensor{\overset{\wedge}{\hat{\pi}}}{_{\perp\indiq[#1]{2}}}}%
{A1m}{\tensor{\overset{\rightharpoonup}{\hat{\pi}}}{_{\indiq[#1]{1}}}}%
{A2p}{\tensor{\overset{\sim}{\hat{\pi}}}{_{\perp\indiq[#1]{2}}}}%
{A2m}{\tensor[^{\text{T}}]{\hat{\pi}}{_{\indiq[#1]{3}}}}%
}[\packageError{cosmicclass}{Unidentified Critical Case: #1}{}]%
}

\newrobustcmd{\PiPu}[2][placeholder]{%
\IfEqCase{#2}{%
{B0p}{\hat{\pi}}%
{B1p}{\tensor{\smash{\overset{\wedge}{\hat{\pi}}}}{^{\indiq[#1]{2}}}}%
{B1m}{\tensor{\smash{\hat{\pi}}}{^{\perp\indiq[#1]{1}}}}%
{B2p}{\tensor{\smash{\overset{\sim}{\hat{\pi}}}}{^{\indiq[#1]{2}}}}%
{A0p}{\tensor{\smash{\hat{\pi}}}{^\perp}}%
{A0m}{\tensor[^{\text{P}}]{\smash{\hat{\pi}}}{}}%
{A1p}{\tensor{\smash{\overset{\wedge}{\hat{\pi}}}}{^{\perp\indiq[#1]{2}}}}%
{A1m}{\tensor{\smash{\overset{\rightharpoonup}{\hat{\pi}}}}{^{\indiq[#1]{1}}}}%
{A2p}{\tensor{\smash{\overset{\sim}{\hat{\pi}}}}{^{\perp\indiq[#1]{2}}}}%
{A2m}{\tensor[^{\text{T}}]{\smash{\hat{\pi}}}{^{\indiq[#1]{3}}}}%
}[\packageError{cosmicclass}{Unidentified Critical Case: #1}{}]%
}

\newrobustcmd{\sicl}[2][placeholder]{%
\IfEqCase{#2}{%
{B0p}{\tensor{\chi}{^{\flat}}}%
{B1p}{\tensor{\smash{\overset{\wedge}{\chi}}}{^{\flat}_{\indiq[#1]{2}}}}%
{B1m}{\tensor{\chi}{^{\flat}_{\perp}_{\indiq[#1]{1}}}}%
{B2p}{\tensor{\smash{\overset{\sim}{\chi}}}{^{\flat}_{\indiq[#1]{2}}}}%
{A0p}{\tensor{\chi}{^{\flat}_\perp}}%
{A0m}{\tensor[^{\text{P}}]{\chi}{^{\flat}}}%
{A1p}{\tensor{\smash{\overset{\wedge}{\chi}}}{^{\flat}_{\perp\indiq[#1]{2}}}}%
{A1m}{\tensor{\smash{\overset{\rightharpoonup}{\chi}}}{^{\flat}_{\indiq[#1]{1}}}}%
{A2p}{\tensor{\smash{\overset{\sim}{\chi}}}{^{\flat}_{\perp\indiq[#1]{2}}}}%
{A2m}{\tensor[^{\text{T}}]{\chi}{^{\flat}_{\indiq[#1]{3}}}}%
}[\packageError{cosmicclass}{Unidentified Critical Case: #1}{}]%
}

\newrobustcmd{\ticl}[2][placeholder]{%
\IfEqCase{#2}{%
{B0p}{\tensor{\zeta}{^{\flat}}}%
{B1p}{\tensor{\smash{\overset{\wedge}{\zeta}}}{^{\flat}_{\indiq[#1]{2}}}}%
{B1m}{\tensor{\zeta}{^{\flat}_{\perp}_{\indiq[#1]{1}}}}%
{B2p}{\tensor{\smash{\overset{\sim}{\zeta}}}{^{\flat}_{\indiq[#1]{2}}}}%
{A0p}{\tensor{\zeta}{^{\flat}_\perp}}%
{A0m}{\tensor[^{\text{P}}]{\zeta}{^{\flat}}}%
{A1p}{\tensor{\smash{\overset{\wedge}{\zeta}}}{^{\flat}_{\perp\indiq[#1]{2}}}}%
{A1m}{\tensor{\smash{\overset{\rightharpoonup}{\zeta}}}{^{\flat}_{\indiq[#1]{1}}}}%
{A2p}{\tensor{\smash{\overset{\sim}{\zeta}}}{^{\flat}_{\perp\indiq[#1]{2}}}}%
{A2m}{\tensor[^{\text{T}}]{\zeta}{^{\flat}_{\indiq[#1]{3}}}}%
}[\packageError{cosmicclass}{Unidentified Critical Case: #1}{}]%
}

\newrobustcmd{\PiPl}[2][placeholder]{%
\IfEqCase{#2}{%
{B0p}{\tensor{\hat{\pi}}{^{\flat}}}%
{B1p}{\tensor{\smash{\overset{\wedge}{\hat{\pi}}}}{^{\flat}_{\indiq[#1]{2}}}}%
{B1m}{\tensor{\hat{\pi}}{^{\flat}_{\perp}_{\indiq[#1]{1}}}}%
{B2p}{\tensor{\smash{\overset{\sim}{\hat{\pi}}}}{^{\flat}_{\indiq[#1]{2}}}}%
{A0p}{\tensor{\hat{\pi}}{_\perp}^{\flat}}%
{A0m}{\tensor[^{\text{P}}]{\hat{\pi}}{^{\flat}}}%
{A1p}{\tensor{\smash{\overset{\wedge}{\hat{\pi}}}}{^{\flat}_{\perp\indiq[#1]{2}}}}%
{A1m}{\tensor{\smash{\overset{\rightharpoonup}{\hat{\pi}}}}{^{\flat}_{\indiq[#1]{1}}}}%
{A2p}{\tensor{\smash{\overset{\sim}{\hat{\pi}}}}{^{\flat}_{\perp\indiq[#1]{2}}}}%
{A2m}{\tensor[^{\text{T}}]{\hat{\pi}}{^{\flat}_{\indiq[#1]{3}}}}%
}[\packageError{cosmicclass}{Unidentified Critical Case: #1}{}]%
}

\newrobustcmd{\sic}[2][placeholder]{%
\IfEqCase{#2}{%
{B0p}{\chi}%
{B1p}{\tensor{\overset{\wedge}{\chi}}{_{\indiq[#1]{2}}}}%
{B1m}{\tensor{\chi}{_{\perp\indiq[#1]{1}}}}%
{B2p}{\tensor{\overset{\sim}{\chi}}{_{\indiq[#1]{2}}}}%
{A0p}{\tensor{\chi}{_\perp}}%
{A0m}{\tensor[^{\text{P}}]{\chi}{}}%
{A1p}{\tensor{\overset{\wedge}{\chi}}{_{\perp\indiq[#1]{2}}}}%
{A1m}{\tensor{\overset{\rightharpoonup}{\chi}}{_{\indiq[#1]{1}}}}%
{A2p}{\tensor{\overset{\sim}{\chi}}{_{\perp\indiq[#1]{2}}}}%
{A2m}{\tensor[^{\text{T}}]{\chi}{_{\indiq[#1]{3}}}}%
}[\packageError{cosmicclass}{Unidentified Critical Case: #1}{}]%
}

\newrobustcmd{\lorsicpar}[2][placeholder]{%
\IfEqCase{#2}{%
{B0m}{\tensor*[^{\text{P}}]{\smash{\underline{\chi}}}{^{\parallel}}}%
{B1p}{\tensor*{\smash{\overset{\wedge}{\chi}}}{^{\parallel}_{\indiq[#1]{2}}}}%
{B1m}{\tensor*{\chi}{^{\parallel}_{\perp\indiq[#1]{1}}}}%
{B2m}{\tensor*[^{\text{T}}]{\smash{\underline{\chi}}}{^{\parallel}_{\indiq[#1]{3}}}}%
{A0p}{\tensor*{\chi}{^{\parallel}_\perp}}%
{A0m}{\tensor*[^{\text{P}}]{\chi}{^{\parallel}}}%
{A1p}{\tensor*{\smash{\overset{\wedge}{\chi}}}{^{\parallel}_{\perp\indiq[#1]{2}}}}%
{A1m}{\tensor*{\smash{\overset{\rightharpoonup}{\chi}}}{^{\parallel}_{\indiq[#1]{1}}}}%
{A2p}{\tensor*{\smash{\overset{\sim}{\chi}}}{^{\parallel}_{\perp\indiq[#1]{2}}}}%
{A2m}{\tensor*[^{\text{T}}]{\chi}{^{\parallel}_{\indiq[#1]{3}}}}%
}[\packageError{cosmicclass}{Unidentified Critical Case: #1}{}]%
}

\newrobustcmd{\lorsicpir}[2][placeholder]{%
\IfEqCase{#2}{%
{B0p}{\tensor*{\chi}{^{\vDash}}}%
{B1p}{\tensor*{\smash{\overset{\wedge}{\chi}}}{^{\vDash}_{\indiq[#1]{2}}}}%
{B1m}{\tensor*{\chi}{^{\vDash}_{\perp\indiq[#1]{1}}}}%
{B2p}{\tensor*{\smash{\overset{\sim}{\chi}}}{^{\vDash}_{\indiq[#1]{2}}}}%
{A0p}{\tensor*{\chi}{^{\vDash}_\perp}}%
{A0m}{\tensor*[^{\text{P}}]{\chi}{^{\vDash}}}%
{A1p}{\tensor*{\smash{\overset{\wedge}{\chi}}}{^{\vDash}_{\perp\indiq[#1]{2}}}}%
{A1m}{\tensor*{\smash{\overset{\rightharpoonup}{\chi}}}{^{\vDash}_{\indiq[#1]{1}}}}%
{A2p}{\tensor*{\smash{\overset{\sim}{\chi}}}{^{\vDash}_{\perp\indiq[#1]{2}}}}%
{A2m}{\tensor*[^{\text{T}}]{\chi}{^{\vDash}_{\indiq[#1]{3}}}}%
}[\packageError{cosmicclass}{Unidentified Critical Case: #1}{}]%
}

\newrobustcmd{\lorsicper}[2][placeholder]{%
\IfEqCase{#2}{%
{B0p}{\tensor*{\chi}{^{\perp}}}%
{B1p}{\tensor*{\smash{\overset{\wedge}{\chi}}}{^{\perp}_{\indiq[#1]{2}}}}%
{B1m}{\tensor*{\chi}{^{\perp}_{\perp\indiq[#1]{1}}}}%
{B2p}{\tensor*{\smash{\overset{\sim}{\chi}}}{^{\perp}_{\indiq[#1]{2}}}}%
{A0p}{\tensor*{\chi}{^{\perp}_\perp}}%
{A0m}{\tensor*[^{\text{P}}]{\chi}{^{\perp}}}%
{A1p}{\tensor*{\smash{\overset{\wedge}{\chi}}}{^{\perp}_{\perp\indiq[#1]{2}}}}%
{A1m}{\tensor*{\smash{\overset{\rightharpoonup}{\chi}}}{^{\perp}_{\indiq[#1]{1}}}}%
{A2p}{\tensor*{\smash{\overset{\sim}{\chi}}}{^{\perp}_{\perp\indiq[#1]{2}}}}%
{A2m}{\tensor*[^{\text{T}}]{\chi}{^{\perp}_{\indiq[#1]{3}}}}%
}[\packageError{cosmicclass}{Unidentified Critical Case: #1}{}]%
}

\newrobustcmd{\Tl}[2][placeholder]{%
\IfEqCase{#2}{%
{B0p}{\tensor{\chi}{^{\flat}}}%
{B1p}{\tensor{\smash{\overset{\wedge}{\chi}}}{^{\flat}_{\indiq[#1]{2}}}}%
{B1m}{\tensor{\chi}{^{\flat}_{\perp}_{\indiq[#1]{1}}}}%
{B2p}{\tensor{\smash{\overset{\sim}{\chi}}}{^{\flat}_{\indiq[#1]{2}}}}%
{A0p}{\tensor{\chi}{^{\flat}_\perp}}%
{A0m}{\tensor[^{\text{P}}]{\mathcal{T}}{^{\flat}}}%
{A1p}{\tensor{\smash{\overset{\wedge}{\chi}}}{^{\flat}_{\perp\indiq[#1]{2}}}}%
{A1m}{\tensor{\smash{\overset{\rightharpoonup}{\mathcal{T}}}}{^{\flat}_{\indiq[#1]{1}}}}%
{A2p}{\tensor{\smash{\overset{\sim}{\chi}}}{^{\flat}_{\perp\indiq[#1]{2}}}}%
{A2m}{\tensor[^{\text{T}}]{\mathcal{T}}{^{\flat}_{\indiq[#1]{3}}}}%
}[\tensor{\mathcal{T}}{^{\flat}_{\indiq[#1]{3}}}]%
}

\newrobustcmd{\cT}[2][placeholder]{%
\IfEqCase{#2}{%
{B1p}{\tensor{\mathcal{T}}{_{\perp\indiq[#1]{2}}}}%
{B1m}{\tensor{\overset{\rightharpoonup}{\mathcal{T}}}{_{\indiq[#1]{1}}}}%
{A0m}{\tensor[^{\text{P}}]{\mathcal{T}}{}}%
{A2m}{\tensor[^{\text{T}}]{\mathcal{T}}{_{\indiq[#1]{3}}}}%
}[\packageError{cosmicclass}{Unidentified Critical Case: #1}{}]%
}

\newrobustcmd{\cTLambda}[2][placeholder]{%
\IfEqCase{#2}{%
{B1p}{\tensor{\lambda}{_{\perp\indiq[#1]{2}}}}%
{B1m}{\tensor{\overset{\rightharpoonup}{\lambda}}{_{\indiq[#1]{1}}}}%
{A0m}{\tensor[^{\text{P}}]{\lambda}{}}%
{A2m}{\tensor[^{\text{T}}]{\lambda}{_{\indiq[#1]{3}}}}%
}[\packageError{cosmicclass}{Unidentified Critical Case: #1}{}]%
}

\newrobustcmd{\cTl}[2][placeholder]{%
\IfEqCase{#2}{%
{B1p}{\tensor{\mathcal{T}}{^{\flat}_{\perp\indiq[#1]{2}}}}%
{B1m}{\tensor{\smash{\overset{\rightharpoonup}{\mathcal{T}}}}{^{\flat}_{\indiq[#1]{1}}}}%
{A0m}{\tensor[^{\text{P}}]{\mathcal{T}}{^{\flat}}}%
{A2m}{\tensor[^{\text{T}}]{\mathcal{T}}{^{\flat}_{\indiq[#1]{3}}}}%
}[\packageError{cosmicclass}{Unidentified Critical Case: #1}{}]%
}

\newrobustcmd{\cTu}[2][placeholder]{%
\IfEqCase{#2}{%
{B1p}{\tensor{\mathcal{T}}{^{\perp\indiq[#1]{2}}}}%
{B1m}{\tensor{\smash{\overset{\rightharpoonup}{\mathcal{T}}}}{^{\indiq[#1]{1}}}}%
{A0m}{\tensor[^{\text{P}}]{\mathcal{T}}{}}%
{A2m}{\tensor[^{\text{T}}]{\mathcal{T}}{^{\indiq[#1]{3}}}}%
}[\packageError{cosmicclass}{Unidentified Critical Case: #1}{}]%
}

\newrobustcmd{\ncTLambda}[2][placeholder]{%
\IfEqCase{#2}{%
{B0p}{\tensor{\lambda}{^{\indiq[#1]{1}}_{\indiq[#1]{1}\perp}}}%
{B1p}{\tensor{\lambda}{_{[\indiq[#1]{2}]\perp}}}%
{B1m}{\tensor{\lambda}{_{\perp\indiq[#1]{1}\perp}}}%
{B2p}{\tensor{\lambda}{_{\langle\indiq[#1]{2}\rangle\perp}}}%
}[\packageError{cosmicclass}{Unidentified Critical Case: #1}{}]%
}

\newrobustcmd{\ncT}[2][placeholder]{%
\IfEqCase{#2}{%
{B0p}{\tensor{\mathcal{T}}{^{\indiq[#1]{1}}_{\indiq[#1]{1}\perp}}}%
{B1p}{\tensor{\mathcal{T}}{_{[\indiq[#1]{2}]\perp}}}%
{B1m}{\tensor{\mathcal{T}}{_{\perp\indiq[#1]{1}\perp}}}%
{B2p}{\tensor{\mathcal{T}}{_{\langle\indiq[#1]{2}\rangle\perp}}}%
}[\packageError{cosmicclass}{Unidentified Critical Case: #1}{}]%
}

\newrobustcmd{\mul}[2][placeholder]{%
\IfEqCase{#2}{%
{B0p}{\tensor{u}{}}%
{B1p}{\tensor{\smash{\overset{\wedge}{u}}}{_{\indiq[#1]{2}}}}%
{B1m}{\tensor{u}{_{\perp}_{\indiq[#1]{1}}}}%
{B2p}{\tensor{\smash{\overset{\sim}{u}}}{_{\indiq[#1]{2}}}}%
{A0p}{\tensor{u}{_\perp}}%
{A0m}{\tensor[^{\text{P}}]{u}{}}%
{A1p}{\tensor{\smash{\overset{\wedge}{u}}}{_{\perp\indiq[#1]{2}}}}%
{A1m}{\tensor{\smash{\overset{\rightharpoonup}{u}}}{_{\indiq[#1]{1}}}}%
{A2p}{\tensor{\smash{\overset{\sim}{u}}}{_{\perp\indiq[#1]{2}}}}%
{A2m}{\tensor[^{\text{T}}]{u}{_{\indiq[#1]{3}}}}%
}[\packageError{cosmicclass}{Unidentified Critical Case: #1}{}]%
}

\newrobustcmd{\ncTmul}[2][placeholder]{%
\IfEqCase{#2}{%
{B0p}{\tensor{\upsilon}{^{\indiq[#1]{1}}_{\indiq[#1]{1}\perp}}}%
{B1p}{\tensor{\upsilon}{_{[\indiq[#1]{2}]\perp}}}%
{B1m}{\tensor{\upsilon}{_{\perp\indiq[#1]{1}\perp}}}%
{B2p}{\tensor{\upsilon}{_{\langle\indiq[#1]{2}\rangle\perp}}}%
}[\packageError{cosmicclass}{Unidentified Critical Case: #1}{}]%
}

\newrobustcmd{\cTmul}[2][placeholder]{%
\IfEqCase{#2}{%
{B1p}{\tensor{\upsilon}{_{\perp\indiq[#1]{2}}}}%
{B1m}{\tensor{\overset{\rightharpoonup}{\upsilon}}{_{\indiq[#1]{1}}}}%
{A0m}{\tensor[^{\text{P}}]{\upsilon}{}}%
{A2m}{\tensor[^{\text{T}}]{\upsilon}{_{\indiq[#1]{3}}}}%
}[\packageError{cosmicclass}{Unidentified Critical Case: #1}{}]%
}

\newrobustcmd{\cTpic}[2][placeholder]{%
\IfEqCase{#2}{%
{B1p}{\tensor{\phi}{_{\perp\indiq[#1]{2}}}}%
{B1m}{\tensor{\overset{\rightharpoonup}{\phi}}{_{\indiq[#1]{1}}}}%
{A0m}{\tensor[^{\text{P}}]{\phi}{}}%
{A2m}{\tensor[^{\text{T}}]{\phi}{_{\indiq[#1]{3}}}}%
}[\packageError{cosmicclass}{Unidentified Critical Case: #1}{}]%
}

\newrobustcmd{\ncTpic}[2][placeholder]{%
\IfEqCase{#2}{%
{B0p}{\tensor{\phi}{^{\indiq[#1]{1}}_{\indiq[#1]{1}\perp}}}%
{B1p}{\tensor{\phi}{_{[\indiq[#1]{2}]\perp}}}%
{B1m}{\tensor{\phi}{_{\perp\indiq[#1]{1}\perp}}}%
{B2p}{\tensor{\phi}{_{\langle\indiq[#1]{2}\rangle\perp}}}%
}[\packageError{cosmicclass}{Unidentified Critical Case: #1}{}]%
}

\newrobustcmd{\cR}[2][placeholder]{%
\IfEqCase{#2}{%
{A0p}{\tensor{\underline{\mathcal{R}}}{}}%
{A0m}{\tensor[^{\text{P}}]{\mathcal{R}}{_{\perp\circ}}}%
{A1p}{\tensor{\underline{\mathcal{R}}}{_{[\indiq[#1]{2}]}}}%
{A1m}{\tensor{\mathcal{R}}{_{\perp\indiq[#1]{1}}}}%
{A2p}{\tensor{\underline{\mathcal{R}}}{_{\langle\indiq[#1]{2}\rangle}}}%
{A2m}{\tensor[^{\text{T}}]{\mathcal{R}}{_{\perp\indiq[#1]{3}}}}%
}[\packageError{cosmicclass}{Unidentified Critical Case: #1}{}]%
}

\newrobustcmd{\cRLambda}[2][placeholder]{%
\IfEqCase{#2}{%
{A0p}{\tensor{\underline{\lambda}}{}}%
{A0m}{\tensor[^{\text{P}}]{\lambda}{_{\perp\circ}}}%
{A1p}{\tensor{\underline{\lambda}}{_{[\indiq[#1]{2}]}}}%
{A1m}{\tensor{\lambda}{_{\perp\indiq[#1]{1}}}}%
{A2p}{\tensor{\underline{\lambda}}{_{\langle\indiq[#1]{2}\rangle}}}%
{A2m}{\tensor[^{\text{T}}]{\lambda}{_{\perp\indiq[#1]{3}}}}%
}[\packageError{cosmicclass}{Unidentified Critical Case: #1}{}]%
}

\newrobustcmd{\cRl}[2][placeholder]{%
\IfEqCase{#2}{%
  {A0p}{\tensor{\underline{\mathcal{R}}}{^{\flat}}}%
{A0m}{\tensor[^{\text{P}}]{\mathcal{R}}{^{\flat}_{\perp\circ}}}%
{A1p}{\tensor{\underline{\mathcal{R}}}{^{\flat}_{[\indiq[#1]{2}]}}}%
{A1m}{\tensor{\mathcal{R}}{^{\flat}_{\perp\indiq[#1]{1}}}}%
{A2p}{\tensor{\underline{\mathcal{R}}}{^{\flat}_{\langle\indiq[#1]{2}\rangle}}}%
{A2m}{\tensor[^{\text{T}}]{\mathcal{R}}{^{\flat}_{\perp\indiq[#1]{3}}}}%
}[\packageError{cosmicclass}{Unidentified Critical Case: #1}{}]%
}

\newrobustcmd{\cRu}[2][placeholder]{%
\IfEqCase{#2}{%
{A0p}{\tensor{\underline{\mathcal{R}}}{}}%
{A0m}{\tensor[^{\text{P}}]{\mathcal{R}}{_{\perp\circ}}}%
{A1p}{\tensor{\underline{\mathcal{R}}}{^{[\indiq[#1]{2}]}}}%
{A1m}{\tensor{\mathcal{R}}{^{\perp\indiq[#1]{1}}}}%
{A2p}{\tensor{\underline{\mathcal{R}}}{^{\langle\indiq[#1]{2}\rangle}}}%
{A2m}{\tensor[^{\text{T}}]{\mathcal{R}}{^{\perp\indiq[#1]{3}}}}%
}[\packageError{cosmicclass}{Unidentified Critical Case: #1}{}]%
}

\newrobustcmd{\ncR}[2][placeholder]{%
\IfEqCase{#2}{%
  {A0p}{\tensor{\mathcal{R}}{_{\perp\perp}}}%
{A0m}{\tensor[^{\text{P}}]{\mathcal{R}}{_{\circ\perp}}}%
{A1p}{\tensor{\mathcal{R}}{_{\perp[\indiq[#1]{2}]\perp}}}%
{A1m}{\tensor{\mathcal{R}}{_{\indiq[#1]{1}\perp}}}%
{A2p}{\tensor{\mathcal{R}}{_{\perp\langle\indiq[#1]{2}\rangle\perp}}}%
{A2m}{\tensor[^{\text{T}}]{\mathcal{R}}{_{\indiq[#1]{3}\perp}}}%
}[\packageError{cosmicclass}{Unidentified Critical Case: #1}{}]%
}

\newrobustcmd{\ncRLambda}[2][placeholder]{%
\IfEqCase{#2}{%
  {A0p}{\tensor{\lambda}{_{\perp\perp}}}%
{A0m}{\tensor[^{\text{P}}]{\lambda}{_{\circ\perp}}}%
{A1p}{\tensor{\lambda}{_{\perp[\indiq[#1]{2}]\perp}}}%
{A1m}{\tensor{\lambda}{_{\indiq[#1]{1}\perp}}}%
{A2p}{\tensor{\lambda}{_{\perp\langle\indiq[#1]{2}\rangle\perp}}}%
{A2m}{\tensor[^{\text{T}}]{\lambda}{_{\indiq[#1]{3}\perp}}}%
}[\packageError{cosmicclass}{Unidentified Critical Case: #1}{}]%
}

\newrobustcmd{\Proj}[2][placeholder]{%
\IfEqCase{#2}{%
  {A2m}{\tensor[^{\text{T}}]{\check{\mathcal{P}}}{#1}}%
}[\packageError{cosmicclass}{Unidentified Critical Case: #1}{}]%
}

\newrobustcmd{\Projl}[2][placeholder]{%
\IfEqCase{#2}{%
  {A2m}{\tensor[^{\text{T}}]{\check{\mathcal{P}}}{^{\flat}#1}}%
}[\packageError{cosmicclass}{Unidentified Critical Case: #1}{}]%
}

\newrobustcmd{\fA}{%
  {\tensor{\mathcal{  A}}{_{\acu{u}}}}%
}
\newrobustcmd{\fB}{%
  {\tensor{\mathcal{  B}}{_{\acu{v}}}}%
}
\newrobustcmd{\fC}{%
  {\tensor{\mathcal{  C}}{^{\acu{v}}}}%
}
\newrobustcmd{\fphi}{%
  {\tensor{\phi}{^{\acu{w}}}}%
}
\newrobustcmd{\fpi}{%
  {\tensor{\pi}{_{\acu{w}}}}%
}

\newrobustcmd{\covard}[2]{%
  {\frac{\bar{\delta}#1}{\bar{\delta}#2}}
}
\newrobustcmd{\copard}[2]{%
  {\frac{\bar{\partial}#1}{\bar{\partial}#2}}
}
\newrobustcmd{\pard}[2]{%
  {\frac{\partial #1}{\partial #2}}
}

\def\jcap{\rm{J. Cosmology Astropart. Phys.}}

\def\prd{\rm{Phys.~Rev.~D}}

\def\procrsa{\rm{Proc.~Math.~Phys.~Eng.~Sci.}}

\def\annphys{\rm{Ann.~Physics (N.~Y.)}}

\def\ptrsla{\rm{Philos.~Trans.~R.~Soc~A}}

\def\jmp{\rm{J.~Math.~Phys.}}

\def\physrev{\rm{Phys.~Rev.}}

\def\epjc{\rm{Eur.~Phys.~J.~C}}

\def\ijmpd{\rm{Int.~J.~Mod.~Phys.~D}}

\def\jhep{\rm{J.~High~Energy~Phys.}}

\def\rmphyo{\rm{Rev.~Mod.~Phys}}

\def\progtp{\rm{Prog.~Theor.~Phys.}}
\def\appb{\rm{Acta~Phys.~Pol.~B}}

\def\aaca{\rm{Adv.~Appl.~Clifford~Algebras}}

\def\ijgmmp{\rm{Int.~J.~Geom.~Meth.~Mod.~Phys.}}

\DeclareRobustCommand{\ppmfish}{%
  \begingroup\normalfont
  \includegraphics[height=0.7\fontcharht\font`\B]{fish_2}%
  \endgroup
}

\newrobustcmd{\PPM}[1]{%
  {\left[\tensor*{\mathsf{M}}{_{\ \ppmfish}^{\left(\text{#1}\right)}}\right]}%
}

\newrobustcmd{\listingsout}[1]{%
	{\vspace{-0pt}\input{figures/#1}}
}

\usepackage{hyperref}
\hypersetup{%
     colorlinks = true,%
     linkcolor = Blue,%
     citecolor = Blue,%
     filecolor = Blue,%
     urlcolor = Blue%
     }%
\usepackage[capitalize]{cleveref} 
\allowdisplaybreaks

\begin{document}


\title{Supercomputers against strong coupling in gravity with curvature and torsion}
\author{W.E.V. Barker}
\email{wb263@cam.ac.uk}
\affiliation{Astrophysics Group, Cavendish Laboratory, JJ Thomson Avenue, Cambridge CB3 0HE, UK}
\affiliation{Kavli Institute for Cosmology, Madingley Road, Cambridge CB3 0HA, UK}

\begin{abstract}
  Many theories of gravity are spoiled by strongly coupled modes: the high computational cost of Hamiltonian analysis can obstruct the identification of these modes.
  A computer algebra implementation of the Hamiltonian constraint algorithm for curvature and torsion theories is presented. 
  These non-Riemannian or Poincar\'e gauge theories suffer notoriously from strong coupling.
  The implementation forms a package (the `Hamiltonian Gauge Gravity Surveyor' -- \emph{HiGGS}) for the \emph{xAct} tensor manipulation suite in \emph{Mathematica}.
Poisson brackets can be evaluated in parallel, meaning that Hamiltonian analysis can be done on silicon, and at scale. Accordingly \emph{HiGGS} is designed to survey the whole Lagrangian space with high-performance computing resources (clusters and supercomputers).
 To demonstrate this, the space of `outlawed' Poincar\'e gauge theories is surveyed, in which a massive parity-even/odd vector or parity-odd tensor torsion particle accompanies the usual graviton.
 The survey spans possible configurations of teleparallel-style multiplier fields which might be used to kill-off the strongly coupled modes, with the results to be analysed in subsequent work. 
   All brackets between the known primary and secondary constraints of all theories are made available for future study. 
   Demonstrations are also given for using \emph{HiGGS} -- on a desktop computer -- to run the Dirac--Bergmann algorithm on specific theories, such as Einstein--Cartan theory and its minimal extensions.
 \end{abstract}

\pacs{04.50.Kd, 04.60.-m, 04.20.Fy, 02.70.-c, 07.05.Bx}

\maketitle

\tableofcontents

\section{Introduction}\label{introduction}

\begin{figure*}[t!]
  \center
  \includegraphics[width=\textwidth]{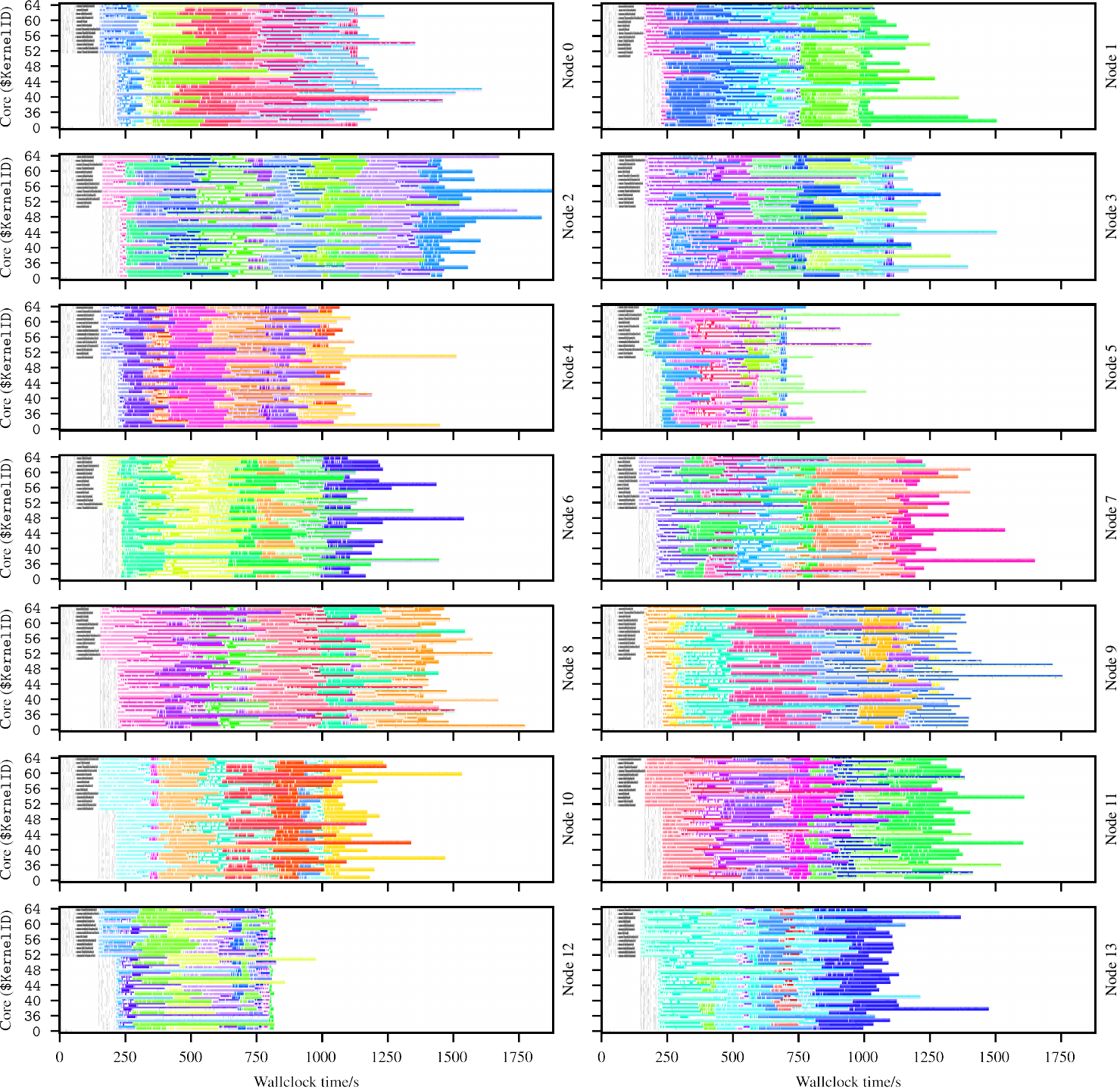}
	\caption{\label{hpcsurvey} 
	  On a supercomputer, 448 processors obtain simplified, covariant expressions for all nonlinear Poisson brackets, among all discovered primary and secondary constraints, for 192 modified gravity theories. These theories are based on Einstein--Cartan gravity with an extra, massive spin-parity $1^+$, $1^-$ or $2^+$ torsion particle, in which various parts ($2^3\times 2^3$ choices) of the curvature and torsion are disabled by multipliers. Each colour is a different theory, black/gray is initialisation. 
	  The objective is eventually to discover whether multipliers can be used to suppress the strongly-coupled $1^-$, $1^+$ or $2^-$ modes, which respectively ruin these theories. Ready access to the Poisson bracket structure is vital for this analysis.
	}
\end{figure*}

The modern frontier in the search for alternatives to general relativity (GR) is characterised by a very large number of competing models~\cite{Golovnev:2022bfm,Nojiri:2017ncd,Clifton:2011jh}.
These models are problematic by dint of their heterogeneity: differences in the mathematical formulation hinder comparison between different classes (e.g. mimetic~\cite{Sebastiani:2016ras,Sebastiani:2016ras} and MOND~\cite{Famaey:2011kh,Skordis:2020eui} theories). 
This seems intractable, but we expect the number of classes to grow only in proportion to the community.
A more serious problem presents when a class contains a large number of parameters, e.g. the couplings in a Lagrangian.
Critical theoretical properties, such as the number and health of propagating degrees of freedom (d.o.f) may be sensitive to these parameters in ways which are hard not only to characterise in general, but even to calculate in detail for a given parameter choice. 

This scenario commonly arises in the Hamiltonian analysis, used routinely to determine the d.o.fs of a proposed theory~\cite{Henneaux:1992ig,blagojevic2002gravitation,1951PhRv...83.1018A,1955PhRv...98..531B,1982AnPhy.143..357C}.
If the coupling parameters elimniate some velocity $\dot\psi$ from the motivated Lagrangian $\mathcal{L}$, then the total Hamiltonian $\mathcal{H}_\text{T}$ must be modified to express the constraint $\pi_\psi\sim\partial_{\dot\psi}\mathcal{L}\approx 0$. That constraint is only preserved if $\dot{\pi}_\psi\sim \{\pi_\psi,\mathcal{H}_\text{T}\}\approx 0$, which either vanishes identically or constitutes a \emph{new} constraint, ad infinitum. Each constraint subtracts d.o.fs from the countable fields $\psi$ na\"ively present in $\mathcal{L}$.
The full chain of constraints is systematically elucidated by means of the \emph{Dirac--Bergmann} algorithm~\cite{1958RSPSA.246..326D,1955PhRv...98..531B,Henneaux:1992ig}, the fundamental computational unit of which is the Poisson bracket. In theories constructed from the higher-spin representations of the Lorentz group, including most tensor theories of gravity, a single covariant bracket can be surprisingly cumbersome for manual evaluation~\cite{chapter4,mythesis}; less so, as we will find, for computer assistance (see~\cref{hpcsurvey}). Through the algorithm, the d.o.f and symmetry structure depend not only on eliminated velocities, but also on all brackets between all constraints, which themselves may be contingent on the couplings in ways that are, ab initio, unknowable.

The \emph{strong coupling problem} is commonly diagnosed in the Hamiltonian picture. It is usually preferred that two gravitational d.o.f propagate~\cite{WEINBERG198059,York:1971hw}. Additional d.o.f may be tolerated unless ghostly or tachyonic\footnote{Perhaps with some interesting exceptions~\cite{Arkani-Hamed:2003pdi,Bagla:2002yn}.}, but they must kept under close theoretical and phenomenological control (e.g. by large~\cite{Carney:2022gse} -- or small~\cite{deRham:2016nuf,Klaer:2017ond} -- masses, screening~\cite{Burrage:2017qrf} or other measures). Frequently we start with many fields $\psi$ (ten in the case of GR), so that reduction to two d.o.f is an achivement in the linearised Hamiltonian picture: strongly coupled modes may \emph{increase} this number the nonlinear analysis~\cite{2002IJMPD..11..747Y}.
Strong coupling can be imagined as a finely-tuned suppression of the kinetic coefficients\footnote{Where permitted, a canonical normalisation recasts this as a divergent mass.} in a mode's linear wave equation~\cite{Baumann:2011dt}. The nonlinear operators are generally still present, however, and describe a non-perturbative dynamics which may be unacceptable, having for example an elliptic or parabolic character. The linearisation in this case refers to some motivated exact vacuum solution. This solution need not be Minkowskian: for GR in higher (odd) dimensions, a fine-tuned admixture of the Gauss--Bonnet invariant strongly couples the whole graviton on a maximally symmetric but curved background~\cite{Charmousis:2008ce}.
In many versions of Ho\v{r}ava gravity~\cite{Wang:2017brl}, the `detailed balance' which defines the Lorentz-asymmetric theory has a similar effect~\cite{Charmousis:2009tc,Papazoglou:2009fj}.

There are some prominent scenarios where strong coupling seems desirable. The linearised, massive graviton of Fierz and Pauli~\cite{Fierz:1939ix} appears troubled even in the massless limit by its fifth d.o.f --- the helicity-0 mode or van Dam--Veltman--Zakharov (vDVZ) scalar~\cite{vanDam:1970vg,Zakharov:1970cc} --- which couples to the matter trace. However, nonlinear completions~\cite{Boulware:1972yco,Arkani-Hamed:2002bjr} of Fierz--Pauli theory revealed that rather than persisting as a light d.o.f, the vDVZ scalar instead becomes strongly coupled, i.e. Vainshtein-screened~\cite{Vainshtein:1972sx,Deffayet:2001uk}. Unfortunately, these same theories were also plagued by a sixth d.o.f --- the nonlinear Boulware--Deser ghost~\cite{Boulware:1972yco} --- whose origin was apparently connected back to the same strong coupling effect~\cite{Deffayet:2005ys}. These matters remain strongly contested~\cite{deRham:2014zqa}, and massive gravity continues to evolve~\cite{deRham:2010ik,deRham:2010kj}.

Strong coupling in massive gravity had another (equally contested~\cite{deRham:2014zqa}) association with superluminality~\cite{Motloch:2015gta,Hinterbichler:2009kq}. Both properties are sometimes conflated with acausality, and ill-posedness of the Cauchy--Kovalevskaya problem. While systems might exist in which these are all mutually implicated pathologies, it is by no means general. In each case one should study the characteristic surfaces and sift for coordinate artifacts~\cite{2009PhRvD..79d3525M,2007PhRvD..75h3513A}. Strongly coupled or \emph{singular} surfaces in the Hamiltonian phase space, where the rank of the matrix of constraint brackets changes, have also been accused of causality violation in the \emph{Poincar\'e gauge theory} (PGT) of gravity~\cite{1998AcPPB..29..961C,1999IJMPD...8..459Y,2002IJMPD..11..747Y}. PGT is the general theory which may be constructed from spacetime curvature and torsion. Whether or not the same caveats resolve the causal question here~\cite{chapter4}, the proclivity for strong coupling in this class remains, and has yet to be seriously addressed. To this end, we target the Poincar\'e gauge theory class in this paper.

\vspace{10pt}

The Poincar\'e gauge theory encompasses a large sector of the general category of \emph{non-Riemannian} theories, whose current popularity warrants a quick introduction. Geometrically interpreted, we retain for non-Riemannian gravity the system of clocks and rulers in the gravitational metric potential $\tensor{g}{_{\mu\nu}}$. In GR, however, the connection $\tensor{\Gamma}{^\mu_{\nu\sigma}}$ is fixed to the Levi--Civita form $\tensor*{C}{^\mu_{\nu\sigma}}\equiv \tensor*{C}{^\mu_{(\nu\sigma)}}\equiv \frac{1}{2}\tensor{g}{^{\mu\rho}}(\tensor{\partial}{_\nu}\tensor{g}{_{\rho\sigma}}+\tensor{\partial}{_\sigma}\tensor{g}{_{\rho\nu}}-\tensor{\partial}{_\rho}\tensor{g}{_{\nu\sigma}})$ -- i.e. the non-tensorial Christoffel symbol -- by assumption. By relaxing this convention, the geometry of the manifold can be extended to include not only the curvature $\tensor{R}{^\mu_{\nu\sigma\rho}}\equiv 2\tensor{\partial}{_{[\sigma|}}\tensor{\Gamma}{^\mu_{|\rho]\nu}}+\dots$, but also the \emph{torsion} $\tensor{T}{^\mu_{\nu\sigma}}\equiv 2\tensor{\Gamma}{^\mu_{[\nu\sigma]}}$ and \emph{non-metricity} $\tensor{Q}{_{\mu\nu\sigma}}\equiv\tensor{\nabla}{_\mu}\tensor{g}{_{\nu\sigma}}$. In this broader context, what is usually meant by `GR' is reached by the Einstein--Hilbert-like action $S_{\text{G}}\equiv\int\mathrm{d}^4x\sqrt{-g}L_{\text{G}}$, with
\begin{equation}
  L_{\text{G}}=-\frac{1}{2\kappa}R
  +\frac{1}{\kappa}\tensor{\lambda}{_\mu^{\rho\sigma}}\tensor{T}{^\mu_{\rho\sigma}}
  +\frac{1}{\kappa}\tensor{\hat{\lambda}}{_\mu^{\rho\sigma}}\tensor{Q}{^\mu_{\rho\sigma}}.
  \label{gr}
\end{equation}
The multiplers\footnote{Note some differences in our convention, regarding the dimensionality and density status of these~\cite{2019Univ....5..173B}.} $\tensor{\lambda}{_\mu^{\rho\sigma}}$ and $\tensor{\hat{\lambda}}{_\mu^{\rho\sigma}}$ constrain the geometry to be \emph{torsion-free} and \emph{metric}, while the dynamics are those of \emph{curvature}. At the time of writing, the attention of the community is drawn to the remarkable non-uniqueness of~\eqref{gr} as a realisation of GR in this broader, non-Riemannian context. By cycling multipliers onto different pairs in the curvature--torsion--non-metricity triad, and identifying suitable dynamical terms, one can construct the flat, metric \emph{teleparallel} equivalent of GR (TEGR) from pure torsion~\cite{Aldrovandi:2013wha}, and the flat, torsion-free \emph{symmetric} alternative (STEGR) from pure non-metricity~\cite{Nester:1998mp}. With~\eqref{gr}, these points in the space of non-Riemannian theories define the \emph{geometrical trinity of gravity}~\cite{2019Univ....5..173B}.

There is now activity to determine a preferred vertex of the trinity, and the viability of the surrounding non-Riemannian landscape. Various considerations must be balanced in a very large parameter space~\cite{BeltranJimenez:2019bnx,Blixt:2021anz,BeltranJimenez:2021kpj,BeltranJimenez:2020guo}. As we mentioned previously, the PGT may be thought of as the sector of the landscape in which $\tensor{Q}{^\mu_{\nu\sigma}}$ is suppressed, e.g. by a multiplier as in~\eqref{gr}. PGT is a convenient sector to study --- without prejudice to the ultimate r\^ole of non-metricity in constructing a viable gravity theory --- since it has a relatively self-contained history stemming from the Einstein--Cartan model~\cite{Trautman:2006fp}, and the gauge-theoretic interpretation pioneered by many authors~\cite{Hehl:1976kj,blagojevic2002gravitation,Obukhov:2006gea,1998RSPTA.356..487L}, beginning with Kibble~\cite{1961JMP.....2..212K}, Utiyama~\cite{PhysRev.101.1597} and also Sciama~\cite{RevModPhys.36.463}. 

The linear Hamiltonian structure of PGT is particularly well developed~\cite{blagojevic2002gravitation,1983PhRvD..28.2455B,1987PhRvD..35.3748B,2018PhRvD..98b4014B}. In the nonlinear structure, strong coupling phenomena are know to be \emph{abundant}, and in this regard PGT is feared to be representative of the broader non-Riemannian landscape. Critically however, the challenge of the Hamiltonian analysis has prevented this structure from being mapped in any comprehensive detail.
A few islands in the landscape were probed at the turn of the millenium~\cite{2002IJMPD..11..747Y}: minimal extensions to the Einstein--Cartan theory in which a single extra massive spin-parity ($J^P$) $1^+$, $1^-$ or $2^-$ torsion particle is present. In each case respectively, $1^-$, $1^+$ and $2^+$ modes were suggested to be strongly coupled. 
Since then, the PGT sector of the landscape has been treated with a sense of `\emph{hic sunt dracones}' -- i.e. avoided as potentially dangerous -- with apparently the only safe configurations being exclusive activation of the $0^+$ or $0^-$ scalar torsion particles~\cite{1999IJMPD...8..459Y}.
In light of our comments above, one can even afford to remain agnostic on the pathology of this strong coupling\footnote{It is important to understand that the main historical objection to the strongly coupled modes of the PGT has been their ghostly character, as inferred by an inspection of the signs of squared momenta in the Hamiltonian. These signs are fixed by the unitarity requirements of the desired, linearly active modes in each case, and in each case they are \emph{negative}~\cite{2002IJMPD..11..747Y}. This `\emph{catch-22}' is particular to the PGT, but, as cogently explained in~\cite{BeltranJimenez:2020lee}, there are principled reasons to be suspicious of stongly coupled surfaces, which appeal to neither the non-perturbative dynamics, nor the causality arguments mentioned above. Generically, a background which is strongly coupled had better not be one which is also seen in nature, since it cannot have been reached by any smooth trajectory through the phase space.}. It would seem, however, given the recent, promising developments in the non-Riemannian approach~\cite{2019Univ....5..173B,BeltranJimenez:2019bnx,Blixt:2021anz,BeltranJimenez:2021kpj,BeltranJimenez:2020guo}, that the extent of the phenomenon should still be understood, and general tools be developed to that end.

\vspace{15pt}

In this paper we present a computer algebra implementation of the nonlinear Hamiltonian analysis for a \emph{generalisation} of the full PGT, in which one may covariantly disable arbitrary irreps of the torsion and curvature by means of Lagrangian multiplier fields. 
This generalisation was put forward as an anti-strong-coupling measure in~\cite{mythesis}, and its Hamiltonian structure is elucidated in the companion paper~\cite{smooth}.
In the conventions of~\eqref{gr}, we may write generalised PGT as
\begin{align}
  \tensor{L}{_{\text{G}}}&=
  -\frac{\alp{0}}{2\kappa}R
  +\sum_{I=1}^{6}\Big(\tensor{\hat{\alpha}}{_I}\tensor{  R}{^{\mu\nu}_{\sigma\rho}}
				  +\tensor{\bar{\alpha}}{_I}\tensor{\lambda}{^{\mu\nu}_{\sigma\rho}}\Big)
				  \qprojlore[_{\mu\nu}^{\sigma\rho}_{\kappa\pi}^{\xi\zeta}]{I}\tensor{  R}{^{\kappa\pi}_{\xi\zeta}}\nonumber\\
				  &\ \ +\frac{1}{\kappa}\sum_{M=1}^{3}\Big(\tensor{\hat{\beta}}{_M}\tensor{ T}{^{\mu}_{\nu\sigma}}
				  +\tensor{\bar{\beta}}{_M}\tensor{\lambda}{^{\mu}_{\nu\sigma}}\Big)
				  \qprojlore[_{\mu}^{\nu\sigma}_{\pi}^{\xi\zeta}]{M}\tensor{  T}{^{\pi}_{\xi\zeta}}\nonumber\\
				  &\ \ +
\frac{1}{\kappa}\tensor{\hat{\lambda}}{_\mu^{\rho\sigma}}\tensor{Q}{^\mu_{\rho\sigma}}
,
\label{neocon2}
\end{align}
where the usual `quadratic' PGT is spanned by the ten parameters $\alp{0}$, $\{\alp{I}\}$, $\{\bet{M}\}$. The projections $\qprojlore[_{\mu\nu}^{\sigma\rho}_{\kappa\pi}^{\xi\zeta}]{I}$ and $\qprojlore[_{\mu}^{\nu\sigma}_{\pi}^{\xi\zeta}]{M}$ extract the $\soonethree$ field strength irreps.
The core of the implementation is (version 1.0.0 of~\cite{supp2}) the \emph{Hamiltonian Gauge Gravity Surveyor} (\HiGGS{}), a \Mathematica{} package grounded in the popular open-source \xAct{} tensor manipulation suite~\cite{Martin-Garcia:2007bqa,Martin-Garcia:2008yei,2008CoPhC.179..597M,Brizuela:2008ra,Pitrou:2013hga,Nutma:2013zea}.

The \HiGGS{} package is suitable for targeted use on a desktop computer. 
Since it is parallelised over Poisson brackets, \HiGGS{} also scales to clusters and supercomputers.
Development in this direction is with the aim of surveying the constraint structure of the non-Riemannian landscape at scale.
Modules from \HiGGS{}, in particular those concerned with bracket evaluation, can be used as a back-end in searching the parameter space for desirable canonical features --- as such features become better understood with time.

For the moment, we perform the brute-force `calibration' survey illustrated in~\cref{hpcsurvey}. In this run, which takes a little over $\SI{1}{\hour}$, the $1^+$, $1^-$ and $2^-$ Einstein--Cartan extensions are modified with all possible configurations of curvature- and torsion-disabling multipliers: 192 generalised Poincar\'e gauge theories in total. During the run \HiGGS{} obtains, for every theory, all the primary and secondary constraints which can be inferred from a knowledge of the literature, and then computes simple covariant expressions for the nonlinear Poisson brackets between all constraint pairs. All brackets identified in this survey can be found in the supplemental materials~\cite{supp}.

Examples are also provided of how to use \HiGGS{} to calculate constraint velocities when implementing the Dirac--Bergmann algorithm. Focus is on the minimal Einstein--Cartan extensions with strong coupling, and the viable $0^+$ and $0^-$ extensions. The unmodified Einstein--Cartan theory is also studied, and \HiGGS{} is used to show that it propagates only the two graviton polarisations. Finally, we will use the results of the initial survey to show how multipliers might conceivably be used to suppress strongly coupled fields. Note that the major undertaking of fully analysing the results, isolating and confirming any viable multiplier configurations, is left to future work.

The remainder of this paper is structured as follows. In~\cref{physicalproblem} we briefly set out our conventions for PGT, using the non-geometric, gauge-theoretic formulation. In~\cref{implementation} we describe the \HiGGS{} implementation, including general tools for the canonical manipulation curvature and torsion, up to Poisson brackets and higher-level functionality for the theory-specific calculation of constraints and velocities. Solutions for scaling the Hamiltonian analysis to high-performance computing (HPC) resources are also described. In~\cref{examples} we present examples of the algorithm, and the results of our initial survey in~\cref{hpcsurvey}. Conclusions follow in~\cref{conclusions}.






\subsection{Conventions}\label{physicalproblem}

Before proceeding, we introduce the gauge-theoretic formulation of Poincar\'e gauge theory, as used in~\cite{blagojevic2002gravitation,2016JMP....57i2505L,chapter2,chapter3,chapter4,mythesis,smooth}.
The gauge picture does not offer any special advantage over the geometric setup in~\eqref{gr}, but it is more consistent with previous Hamiltonian analyses in~\cite{blagojevic2002gravitation,1983PhRvD..28.2455B,1987PhRvD..35.3748B,Chen:2015vya,1999IJMPD...8..459Y,2002IJMPD..11..747Y,2018PhRvD..98b4014B}.

The geometric covariant derivative appearing in the definition $\tensor{Q}{_{\mu\nu\sigma}}\equiv\tensor{\nabla}{_\mu}\tensor{g}{_{\nu\sigma}}$, as it acts on a vector $\tensor{V}{^\mu}$, is written 
\begin{equation}
  \tensor{\nabla}{_\nu}\tensor{V}{^\mu}\equiv\tensor{\partial}{_\nu}\tensor{V}{^\mu}+\tensor{\Gamma}{^\mu_{\lambda\nu}}\tensor{V}{^\lambda}.
  \label{vdef}
\end{equation}
In the Poincar\'e gauge theory we enforce $\tensor{Q}{_{\mu\nu\sigma}}=0$ by assumption, though a multiplier could also be used. The general non-Riemannian connection 
\begin{equation}
  \tensor{\Gamma}{^\mu_{\nu\sigma}}=\tensor*{C}{^\mu_{\nu\sigma}}+\tensor{K}{^\mu_{\nu\sigma}}+\tensor{L}{^\mu_{\nu\sigma}},
\end{equation}
thus loses its \emph{disformation} part $\tensor{L}{^\mu_{\nu\sigma}}\equiv\frac{1}{2}\tensor{Q}{^\mu_{\nu\sigma}}-\tensor{Q}{_{(\nu|}^\mu_{|\sigma)}}=0$, but still conveys torsion through the \emph{contorsion} d.o.fs $\tensor{K}{^\mu_{\nu\sigma}}\equiv\frac{1}{2}\tensor{T}{^\mu_{\nu\sigma}}+\tensor{T}{_{(\nu|}^\mu_{|\sigma)}}$. Aside from the contorsion, the metric d.o.fs which source the Levi--Civita part are defined, as usual, by tangents to the coordinate curves ${\tensor{g}{_{\mu\nu}}\equiv\tensor{\bm{e}}{_\mu}\cdot{\bm{e}}{_\nu}}$.

We now move away from this geometric picture, to a `particle physics' setup where the underlying manifold is always flat Minkowski space $\check{\mathcal{  M}}$. The metric associated with (curvillinear) coorinate tangents is then ${\tensor{\gamma}{_{\mu\nu}}\equiv\tensor{\bm{e}}{_\mu}\cdot{\bm{e}}{_\nu}}$, and this metric is \emph{flat}. The coordinate basis is accompanied by a Lorentz basis, whose dot products give the Minkowskian metric components ${\tensor{\eta}{_{ij}}\equiv\tensor{\hat{\bm{e}}}{_i}\cdot{\hat{\bm{e}}}{_j}}$. The Lorentz basis can rotate locally under the proper, orthochronous Lorentz rotations, and is non-holonomic. Following on from~\eqref{vdef}, a vector $\tensor{\mathcal{  V}}{^i}$ with Lorentz indices has a covariant derivative
\begin{equation}
  \tensor{\mathcal{  D}}{_j}\tensor{\mathcal{  V}}{^i}\equiv\tensor{h}{_j^\mu}(\tensor{\partial}{_\mu}\tensor{\mathcal{  V}}{^i}+\tensor{A}{^{i}_{k\mu}}\tensor{\mathcal{  V}}{^k}),
  \label{poin_covd}
\end{equation}
where we define the (inverse of the) translational gauge field $\tensor{h}{_i^\mu}$ and the rotational gauge field ${\tensor{A}{^{ij}_\mu}\equiv\tensor{A}{^{[ij]}_\mu}}$ --- we will return to this derivative in~\cref{derivatives}.
These fields guarantee invariance under the general coordinate transformations (GCTs or passive diffeomorphisms) on $\check{\mathcal{  M}}$, and Lorentz rotations, and together they \emph{gauge} the Poincar{\'e} group $\poincare$. 

How to connect back to the geometric picture? The Minkowskian metric components can be recovered via the identities ${\tensor{b}{^i_\mu}\tensor{h}{_i^\nu}\equiv\tensor*{\delta}{_\mu^\nu}}$ and ${\tensor{b}{^i_\mu}\tensor{h}{_j^\mu}\equiv\tensor*{\delta}{_j^i}}$.
On the other hand, the usual system of clocks and rulers can be recovered using ${\tensor{g}{_{\mu\nu}}\equiv\tensor{\eta}{_{ij}}\tensor{b}{^i_\mu}\tensor{b}{^j_\nu}}$ and ${\tensor{g}{^{\mu\nu}}\equiv\tensor{\eta}{^{ij}}\tensor{h}{_i^\mu}\tensor{h}{_j^\nu}}$. The covariant measures on $\mathcal{  M}$ and $\check{\mathcal{  M}}$ are respectively $\sqrt{-g}$, i.e. the conventional ${g\equiv\det \tensor{g}{_{\mu\nu}}}$, and ${b\equiv h^{-1}\equiv\det\tensor{b}{^i_\mu}}$. Finally, the geometric field strength tensors -- referred to as the curvature and the torsion -- are provided by the formulae
\begin{subequations}
  \begin{align}
    \tensor{\mathcal{R}}{^{ij}_{kl}}&\equiv 2\tensor{h}{_k^\mu}\tensor{h}{_l^\nu}\big(\tensor{\partial}{_{[\mu|}}\tensor{A}{^{ij}_{|\nu]}}+\tensor{A}{^i_{m[\mu|}}\tensor{A}{^{mj}_{|\nu]}}\big),\label{riemanndef}
\\
    \tensor{\mathcal{T}}{^i_{kl}}&\equiv 2\tensor{h}{_k^\mu}\tensor{h}{_l^\nu}\big(\tensor{\partial}{_{[\mu|}}\tensor{b}{^i_{|\nu]}}+\tensor{A}{^i_{m[\mu|}}\tensor{b}{^m_{|\nu]}}\big).\label{torsiondef}
  \end{align}
\end{subequations}
The conversion of these back to the (numerical values of) the geometric components, is done by contraction with the translational gauge fields. The theory~\eqref{neocon2} meanwhile becomes
  \begin{align}
    \tensor{L}{_{\text{G}}}&=
    -\frac{1}{2}\alp{0}\planck^2\mathcal{R}
    \nonumber
    \\
    &\ \ 
    +\sum_{I=1}^{6}\Big(\tensor{\hat{\alpha}}{_I}\tensor{\mathcal{  R}}{^{ij}_{kl}}
				    +\tensor{\bar{\alpha}}{_I}\tensor{\lambda}{^{ij}_{kl}}\Big)
				    \projlore[_{ij}^{kl}_{nm}^{pq}]{I}\tensor{\mathcal{  R}}{^{nm}_{pq}}\nonumber\\
				    &\ \ +\planck^2\sum_{M=1}^{3}\Big(\tensor{\hat{\beta}}{_M}\tensor{\mathcal{  T}}{^{i}_{jk}}
				    +\tensor{\bar{\beta}}{_M}\tensor{\lambda}{^{i}_{jk}}\Big)
				    \projlore[_{i}^{jk}_{l}^{nm}]{M}\tensor{\mathcal{  T}}{^{l}_{nm}},
  \label{neocon}
  \end{align}
  where our metricity assumption dispenses with the need for the final term in~\eqref{neocon2}, and we use the Planck mass rather than the Einstein constant $\planck\equiv 1/\sqrt{\kappa}$.
  Once again, we reiterate that our use of the gauge theory setup over geometric alternatives (including those which use the tetrad and spin-connection over the metric and contorsion) is merely a matter of convenience in the present work.

In this article we will follow~\cite{Hohmann:2020muq} by using the following syntax highlighting for code listings: keywords for \Mathematica{} are typeset in \lstinline!green!, for \xAct{} in \lstinline!blue! and for \HiGGS{} in \lstinline!red!. 
This paper uses the `West coast' signature $(+,-,-,-)$.

\section{Implementation}\label{implementation}
In this section the implementation of the Hamiltonian analysis in the \HiGGS{} package is described.
Many aspects of our particular approach will be inefficient: the package is monolithic and expensive in terms of memory and maintenance.
At the time of writing, however, HPC is a \emph{cheap} resource. By carefully tuning only a few aspects of the implementation it is therefore possible to produce a product which surveys the theory space in a matter of hours.
It is important to emphasise that these `high-level' features of \HiGGS{} are currently limited to the generalised Poincar\'e gauge theory in~\eqref{neocon}. The question of scaling to arbitrary theories will be addressed in~\cref{conclusions}, where we will find that such a scaling is actually likely to \emph{reduce} the complexity of the package, so long as the \lstinline!PoissonBracket[]! function is upgraded so as to be `aware of' Leibniz's rule. Moreover, the `low-level' functionality, which is not specific to a Lagrangian, will be of standalone utility. In what follows, we assume a basic familiarity with the \xAct{} suite and Wolfram Language.

\subsection{Geometric setup}\label{geometricsetup}
We begin by describing the way in which the Riemann--Cartan geometry is implemented in \HiGGS{}. Whilst \xAct{} is perfectly capable of accommodating not only a Riemann--Cartan curvature, but also a torsion tensor, we prefer to adhere to the `particle physics' picture of gravitational gauge theories~\cite{lasenby-hobson-2016,blagojevic2002gravitation,1998RSPTA.356..487L,mythesis}, and set up all the physics on a \emph{flat} spacetime. The following equivalent \xAct{} commands are issued when the \HiGGS{} environment is initially built (see~\cref{preparingasciencesession}) with the command \lstinline!BuildHiGGS[]! 
\begin{lstlisting}[breaklines=true]
In[]:= DefManifold[M4, 4, IndexRange[{a, z}]];
DefMetric[-1, G[-m, -n], CD, {",", "\[PartialD]"}, PrintAs -> "\[Gamma]", FlatMetric -> True, SymCovDQ -> True];
\end{lstlisting}
This sets up a $D=4$ manifold \lstinline!M4! to represent $\check{\mathcal{M}}$, with a \emph{flat} negative-signature metric \lstinline!G[-m,-n]! to represent $\tensor{\gamma}{_{\mu\nu}}$, and \emph{flat} covariant derivative $\tensor{\check{\nabla}}{_\mu}$ represented by \lstinline!CD[-m][]!.
This offers several advantages, and foremost among these is an easy comparison with the very substantial body of literature on the Hamiltonian structure as mentioned in~\cref{physicalproblem}. We will also not be limited by the fact that the non-Riemannian features of \xAct{} are (presently) less flexible and comprehensive than those of a simple Minkowskian setup. This is to be expected, given that GR is the preferred effective theory of gravitation.

An essential feature of this setup, which goes beyond the particle picture of the literature, is the conflation of holonomic and non-holonomic tangent spaces: a single collection of indices \lstinline!a!, \lstinline!b!, \lstinline!c!, etc. is ascribed to \lstinline!TangentM4!, and these represent both the Lorentz indices $i$, $j$, $k$ and coordinate indices $\mu$, $\nu$, $\sigma$. This practice might be anathema to the field of differential geometry, but it is acceptable for pragmatic, computational purposes. The coordinates are assumed implicitly to be Cartesian, so that the components of the flat metric $\tensor{\gamma}{_{\mu\nu}}$ are Minkowskian (equal to those of $\tensor{\eta}{_{ij}}$), and moreover that a rotational gauge is chosen in which $\tensor{\bm e}{_0}=\tensor{\hat{\bm e}}{_0}$, $\tensor{\bm e}{_1}=\tensor{\hat{\bm e}}{_1}$, $\tensor{\bm e}{_2}=\tensor{\hat{\bm e}}{_2}$ and $\tensor{\bm e}{_3}=\tensor{\hat{\bm e}}{_3}$. As a consequence of this gauge-fixed setup, the covariance of quantities is not guaranteed internally, as it would be if all the features of \xAct{} (such as user-defined connections) were fully exploited in \HiGGS{}. It is instead possible to check the covariance by visually inspecting the final results for explicit gauge fields: in practice this turns out to be very easy, and the elimination of bare gauge fields in \HiGGS{} is reliable. 

It is important to note that whilst the gauge is fixed to conveniently overload the indices of \lstinline!TangentM4!, the tensor structure of PGT is wholly preserved. This is in contrast, for example, with the seminal paper~\cite{1983PhRvD..28.2455B} in which the \emph{time gauge} imposes $\tensor{h}{_a^0}=0$, thereby massively simplifying the various algebraic expressions. On the contrary, the algebraic expressions produced from \HiGGS{} are valid for \emph{any gauge}, once the various shared indices are interpreted as being either Greek or Roman. Care is taken in the definitions to avoid any ambiguity over this division of indices.

Following on from our discussion of indices, we note that the 1, 2, 3 spacelike indices $a$, $b$, $c$ and $\alpha$, $\beta$, $\gamma$ are also subsumed into \lstinline!a!, \lstinline!b!, \lstinline!c!. These may be extracted by means of the projection operator \lstinline!G3[a,b]!, which represents $\tensor{\gamma}{^{\alpha\beta}}$, and lies at the heart of the Arnowitt--Deser--Misner (ADM) split. 
The ADM or $3+1$ split uses a spacelike foliation which is characterised by timelike unit vector $\tensor{n}{_k}$, defined as
\begin{equation}
	\tensor{n}{_k}\equiv\tensor{h}{_k^0}/\sqrt{\tensor{g}{^{00}}},
	\label{foliation_master}
\end{equation}
where we recall that the (gravitational) metric components are recovered by $\tensor{g}{^{\mu\nu}}\equiv\tensor{h}{_i^\mu}\tensor{h}{_j^\nu}\tensor{\eta}{^{ij}}$.
Any vector with local Lorentz indices may then be decomposed into perpendicular and parallel components ${\tensor{\mathcal{  V}}{^i}=\tensor{\mathcal{  V}}{^\perp}\tensor{n}{^i}+\tensor{\mathcal{  V}}{^{\ovl i}}}$. An overbar is used to denote the parallel indices. There are then some identities $\smash{\tensor{b}{^{\ovl k}_\alpha}\tensor{h}{_{\ovl{l}}^\alpha}\equiv\tensor*{\delta}{_{\ovl l}^{\ovl k}}}$ and $\smash{\tensor{b}{^{\ovl k}_\alpha}\tensor{h}{_{\ovl{k}}^\beta}\equiv\tensor*{\delta}{_\alpha^\beta}}$, which follow from~\eqref{foliation_master}.
The gauge fields $\tensor{h}{_i^\mu}$, $\tensor{b}{^i_\mu}$ and $\tensor{A}{^{ij}_{\mu}}$ are denoted by \lstinline!H[-a,b]!, \lstinline!B[a,-b]! and \lstinline!A[a,b,-c]!. The timelike vector $\tensor{n}{^i}$ is represented by \lstinline!V[a]!, and we see that a variety of identities are then implied within the built \HiGGS{} environment
\begin{lstlisting}[breaklines=true]
In[]:= quantity = {H[-a, b], B[a, -b], A[a, b, -c], V[a], V[a]*V[-a], A[a, -a, -c], A[a, b, -c] + A[b, a, -c], H[-a, i]*B[a, -j], H[-a, i]*B[c, -i], G3[-a, -b]*G3[b, -d], G3[-a, a], B[a, -b]*G3[b, -c]*V[-a], CD[-a][G3[-c, b]]}
Out[]= |
\vspace{-4pt}
\begin{flushleft}
\includegraphics[width=\linewidth]{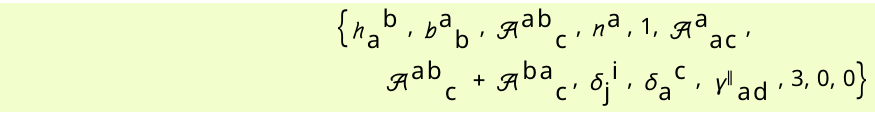}
\end{flushleft}
\vspace{-5pt}
|
In[]:= ToCanonical[quantity]
Out[]= |
\vspace{-4pt}
\begin{flushleft}
\includegraphics[width=\linewidth]{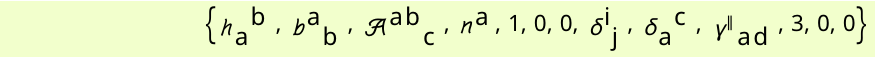}
\end{flushleft}
\vspace{-5pt}
|
\end{lstlisting}
Note that the covariant derivative \lstinline!CD[-i][]! represents, in our Cartesian gauge choice, the basic coordinate derivative $\tensor{\check{\nabla}}{_\mu}\to\tensor{\partial}{_\mu}$. 
The ADM \emph{lapse} function, and a \emph{shift} vector are next defined, according to 
\begin{equation}
N\equiv\tensor{n}{_k}\tensor{b}{^k_0}, \quad \tensor{N}{^\alpha}\equiv\tensor{h}{_{\ovl{k}}^\alpha}\tensor{b}{^{\ovl{k}}_0}.
\end{equation}
These functions carry information about the part $\tensor{b}{^{\ovl{k}}_0}$ of the translational gauge field (which, as we will shortly see, is \emph{non-physical}), wheres $\tensor{n}{^i}$ is independent of this quantity.
In \HiGGS{} the lapse is \lstinline!Lapse[]! and we use also the spatial measure \lstinline!J[]! for the quantity $J\equiv b/N$, where $b\equiv \det \tensor{b}{^i_\mu}$.
Some further identities are then
\begin{equation}
  \begin{gathered}
    \frac{\partial\tensor{n}{_l}}{\partial\tensor{b}{^k_\mu}}\equiv -\tensor{n}{_k}\tensor{h}{_{\overline{l}}^\mu}, \quad \frac{\partial\tensor{h}{_l^\nu}}{\partial\tensor{b}{^k_\mu}}\equiv -\tensor{h}{_k^\nu}\tensor{h}{_l^\mu},\quad
    \frac{\partial b}{\partial\tensor{b}{^k_\nu}}\equiv b\tensor{h}{_k^\nu}, \\ 
    \frac{\partial J}{\partial\tensor{b}{^k_\nu}}\equiv J\tensor{h}{_{\overline{k}}^\nu}, \quad \frac{\partial N}{\partial\tensor{b}{^k_\nu}}\equiv N\tensor{n}{_k}\tensor{h}{_{\perp}^\nu}.
  \end{gathered}
  \label{furtheridentities}
\end{equation}
These identities are also incorporated into the \HiGGS{} environment, which prefers to extract -- from all derivatives of quantities dependent on the translational gauge field -- the form \lstinline!CD[-a][B[-b,-c]]!. It is practical to replace all instances of \lstinline!CD[-a][H[-b,-c]]! accordingly, since the momentum \lstinline!BPi[-a,-b]! is defined according to \lstinline!B[-a,-b]!, and so we find 
\begin{lstlisting}[breaklines=true]
In[]:= quantity = ScreenDollarIndices[{J[], Lapse[], CD[-a][V[-j]], CD[-a][H[-j, n]], CD[-a][J[]], CD[-a][Lapse[]]}]
Out[]= |
\vspace{-4pt}
\begin{flushleft}
\includegraphics[width=\linewidth]{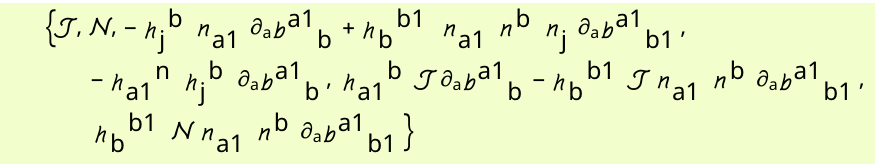}
\end{flushleft}
\vspace{-5pt}
|
\end{lstlisting}

Now that the geometric setup is in place, we introduce the canonical setup~\cite{blagojevic2002gravitation,Henneaux:1992ig,1813910}. The \emph{canonical} momenta are 
\begin{equation}
  \tensor{\pi}{_i^\mu}\equiv\frac{\partial bL_{\text{G}}}{\partial(\partial_0\tensor{b}{^i_{\mu}})}, \quad \tensor{\pi}{_{ij}^{\mu}}\equiv\frac{\partial bL_{\text{G}}}{\partial(\partial_0\tensor{A}{^{ij}_{\mu}})}.
  \label{canonicalmomenta}
\end{equation}
As with our earlier work~\cite{mythesis,smooth,chapter4}, we will \emph{neglect} the matter Lagrangian $L_\text{M}$.
The field strengths in~\eqref{riemanndef} and~\eqref{torsiondef} are independent of the velocities for $\tensor{b}{^k_0}$ and $\tensor{A}{^{ij}_0}$, so~\eqref{canonicalmomenta} imply $10$ primary constraints of the form
\begin{equation}
  \tensor{\varphi}{_k^0}\equiv\tensor{\pi}{_k^0}\approx 0, \quad \tensor{\varphi}{_{ij}^0}\equiv\tensor{\pi}{_{ij}^0}\approx 0.
  \label{sureprimaries}
\end{equation}
From~\eqref{sureprimaries} we arrive at the result mentioned above, that the conjugate field $\tensor{b}{^k_0}$ is non-physical; the same applies to $\tensor{A}{^{ij}_0}$. By ($\approx$), the \emph{weak} equality is denoted, i.e. not an approximation. The constraints~\eqref{sureprimaries} are referred to as the `sure' primary, first class (sPFC) constraints: they are a consequence of Poincar{\'e} symmetry, and they apply for all choices of the $\{\alp{I}\}$, $\{\bet{M}\}$, $\{\calp{I}\}$, $\{\cbet{M}\}$.
According to~\eqref{canonicalmomenta}, \HiGGS{} must support a field momentum for both \lstinline!B[a,b]! and \lstinline!A[a,b,c]!. The $\tensor{\pi}{_i^\mu}$, $\tensor{\pi}{_{ij}^\mu}$ and the `parallel' $\tensor{\hat{\pi}}{_i^{\ovl{j}}}$, $\tensor{\hat{\pi}}{_{ij}^{\ovl{k}}}$ are defined as follows
\begin{lstlisting}[breaklines=true]
In[]:= quantity = {BPi[-a, b], APi[-a, -b, c], BPiP[-a, -b], APiP[-a, -b, -c], BPiP[-a, -b]*V[b], APiP[-a, -b, -c]*V[c]}
Out[]= |
\vspace{-4pt}
\begin{flushleft}
\includegraphics[width=\linewidth]{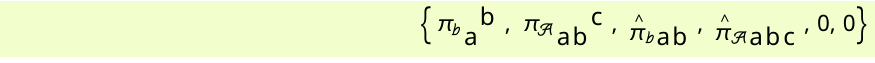}
\end{flushleft}
\vspace{-5pt}
|
In[]:= ScreenDollarIndices[ToCanonical[quantity /. PiPToPi]]
Out[]= |
\vspace{-4pt}
\begin{flushleft}
\includegraphics[width=\linewidth]{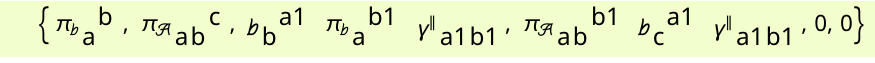}
\end{flushleft}
\vspace{-5pt}
|
\end{lstlisting}
The purpose of the parallel momenta is seen above: they are the physical parts not touched by the sPFCs, and defined according to $\smash{\tensor{\hat{\pi}}{_{i}^{\overline{k}}}\equiv\tensor{\pi}{_{i}^\alpha}\tensor{b}{^{k}_\alpha}}$ and $\smash{\tensor{\hat{\pi}}{_{ij}^{\overline{k}}}\equiv\tensor{\pi}{_{ij}^\alpha}\tensor{b}{^{k}_\alpha}}$. 

We will see in~\cref{low-levelfunctions} that the specific internal rule \lstinline!PiPToPi! should not often need to be used in practice, and has a more general alternative in the \lstinline!ToBasicForm[]! command which is provided officially by the package. In general, parallel and perpendicular quantities can be accessed with some projections
\begin{lstlisting}[breaklines=true]
In[]:= quantity = {PPerp[-a, b], PPara[-a, b]}
Out[]= |
\vspace{-4pt}
\begin{flushleft}
\includegraphics[width=\linewidth]{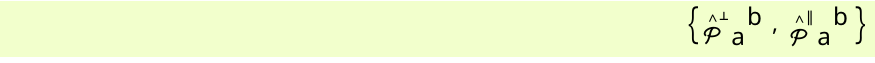}
\end{flushleft}
\vspace{-5pt}
|
In[]:= ScreenDollarIndices[ToCanonical[quantity /. PADMActivate]]
Out[]= |
\vspace{-4pt}
\begin{flushleft}
\includegraphics[width=\linewidth]{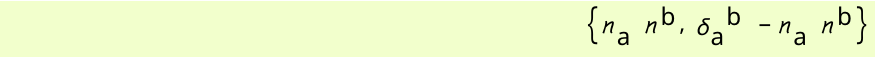}
\end{flushleft}
\vspace{-5pt}
|
\end{lstlisting}

The flat manifold of course has vanishing Riemannian curvature $\tensor{R}{^\mu_{\nu\sigma\lambda}}=0$, and so we must be careful not to use the \xAct{} quantities \lstinline!RiemannCD[-a,-b,-c,-d]!, \lstinline!RicciCD[-c,-d]! or \lstinline!RicciScalarCD[]!. Instead, the field strengths $\tensor{\mathcal{R}}{_{ijkl}}$ and $\tensor{\mathcal{T}}{_{ijk}}$ are given by their own tensors, and these expand to give the definitions in~\cref{riemanndef,torsiondef}
\begin{lstlisting}[breaklines=true]
In[]:= quantity = {R[-a, -b, -c, -d], R[a, -a, -c, -d], R[-a, -b, -c, c], T[a, -b, -c], T[a, -b, b]}
Out[]= |
\vspace{-4pt}
\begin{flushleft}
\includegraphics[width=\linewidth]{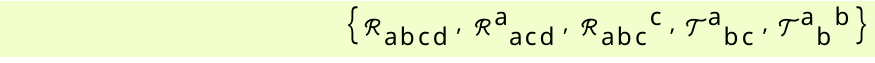}
\end{flushleft}
\vspace{-5pt}
|
In[]:= ScreenDollarIndices[ToCanonical[quantity /. ExpandStrengths]]
Out[]= |
\vspace{-4pt}
\begin{flushleft}
\includegraphics[width=\linewidth]{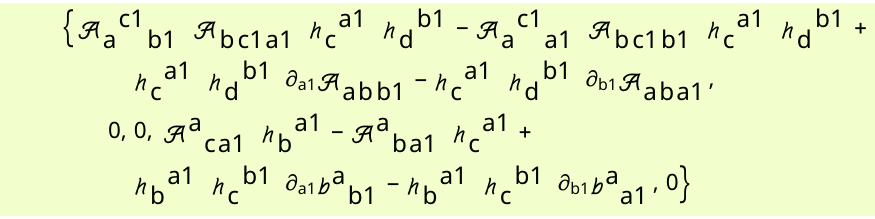}
\end{flushleft}
\vspace{-5pt}
|
\end{lstlisting}
The final set of fields which we introduce are the multipliers $\tensor{\lambda}{_{ijkl}}$ and $\tensor{\lambda}{_{ijk}}$ -- these are precisely the same shape as $\tensor{\mathcal{R}}{_{ijkl}}$ and $\tensor{\mathcal{T}}{_{ijk}}$, and we write them
\begin{lstlisting}[breaklines=true]
In[]:= quantity = {RLambda[-a, -b, -c, -d], RLambda[a, -a, -c, -d], RLambda[-a, -b, -c, c], TLambda[a, -b, -c], TLambda[a, -b, b]}
Out[]= |
\vspace{-4pt}
\begin{flushleft}
\includegraphics[width=\linewidth]{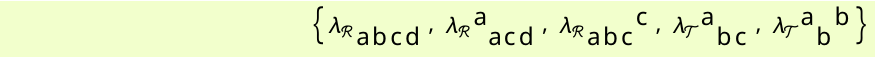}
\end{flushleft}
\vspace{-5pt}
|
In[]:= ScreenDollarIndices[ToCanonical[quantity]]
Out[]= |
\vspace{-4pt}
\begin{flushleft}
\includegraphics[width=\linewidth]{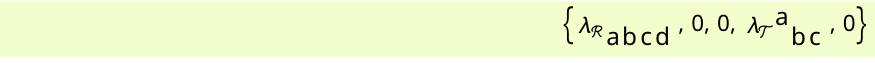}
\end{flushleft}
\vspace{-5pt}
|
\end{lstlisting}
It should be emphasised again that rules such as \lstinline!ExpandStrengths!, \lstinline!PADMActivate! and \lstinline!PiPToPi! should not often be needed, and are subsumed under the \lstinline!ToBasicForm[]! command.

\subsection{Irreducible decompositions}\label{irreducibledecompositions}

The Hamiltonian analysis invites decomposition into irresducible representations of the Lorentz group $\soonethree$, and (through the ADM split) the special orthogonal group $\othree$. The former is useful also in the Lagrangian picture, since the Lagrangian formulation of a theory is typically Lorentz-covariant, it is useful to split field representations into blocks which transform only among themselves. In the Hamiltonian case, the choice of slicing introduces a `preferred' timelike vector -- this is not unique, so covariance is not ultimately lost -- but the symmetry in the context of the slicing is reduced to the spatial rotations $\sothree$. In the case of rotations, the irreps of tensor fields correspond to states of definite spin and parity, allowing us to designate the parts as $J^P$.

\subsubsection{Lorentz group}\label{lorentzgroup}

We do not often encounter the Lorentz decomposition in the course of Hamiltonian calculations, but it is implemented across the higher-rank tensors, for example the `human-readable' decomposition into familiar irreps such as the \emph{Weyl}
\begin{align}
  \tensor{\mathcal{W}}{_{ijkl}}&\equiv\tensor{\mathcal{R}}{_{ijkl}}-\frac{1}{2}\big(\tensor{\eta}{_{ik}}\tensor{\mathcal{R}}{_{jl}}-\tensor{\eta}{_{il}}\tensor{\mathcal{R}}{_{jk}}-\tensor{\eta}{_{jk}}\tensor{\mathcal{R}}{_{il}}
  \nonumber\\
&\ \ +\tensor{\eta}{_{jl}}\tensor{\mathcal{R}}{_{ik}}\big)+\frac{1}{6}\mathcal{R},
\end{align}
the \emph{Ricci} and the \emph{Ricci scalar}
\begin{equation}
  \tensor{\mathcal{R}}{_{ij}}\equiv\tensor{\mathcal{R}}{^l_{ilj}}, 
  \quad 
  \tensor{\mathcal{R}}{}\equiv\tensor{\mathcal{R}}{^l_{l}}, 
\end{equation}
and the \emph{tensor} torsion, defined as the remainder of $\tensor{\mathcal{T}}{^i_{jk}}$ after removing the \emph{vector} (i.e. torsion contraction) and \emph{pseudovector} components
\begin{equation}
  \tensor{\mathcal{T}}{_{i}}\equiv\tensor{\mathcal{T}}{^l_{il}}, 
  \quad 
  \tensor[^*]{\mathcal{T}}{_{i}}\equiv\tensor{\epsilon}{_{ijkl}}\tensor{\mathcal{T}}{^{jkl}}, 
\end{equation}
-- see e.g.~\cite{10.1143/PTP.64.866,1999IJMPD...8..459Y,2002IJMPD..11..747Y}. These components may be accessed as follows
\begin{lstlisting}[breaklines=true]
In[]:= ScreenDollarIndices[ToCanonical[R[-a, -b, -c, -d] /. StrengthSO13Activate]]
Out[]= |
\vspace{-4pt}
\begin{flushleft}
\includegraphics[width=\linewidth]{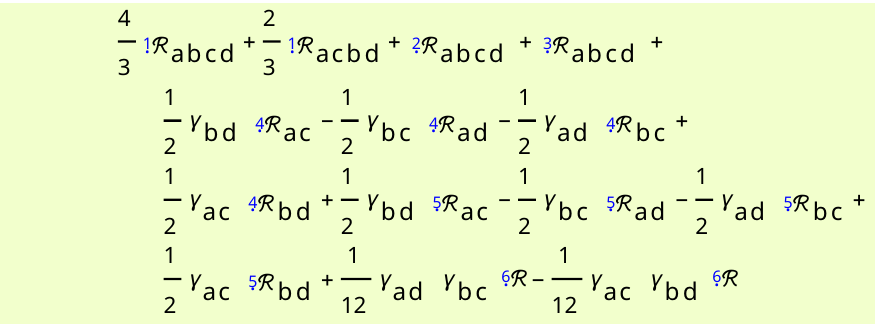}
\end{flushleft}
\vspace{-5pt}
|
In[]:= ScreenDollarIndices[ToCanonical[TLambda[-a, -b, -c] /. StrengthLambdaSO13Activate]]
Out[]= |
\vspace{-4pt}
\begin{flushleft}
\includegraphics[width=\linewidth]{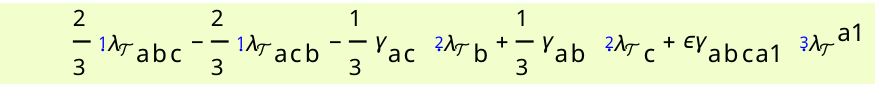}
\end{flushleft}
\vspace{-5pt}
|
\end{lstlisting}
More commonly, we might need access to the complete, orthonormal $\soonethree$ operators which appear in~\eqref{neocon}. These may be associated with the couplings $\{\alp{I}\}$, $\{\bet{M}\}$, $\{\calp{I}\}$, $\{\cbet{M}\}$ as follows, for the purposes of constructing a Lagrangian\footnote{Note that \HiGGS{} does not define a Planck mass $\planck$: all scales are absorbed into the couplings \lstinline!Bet1!, \lstinline!Bet2!, \lstinline!Bet3!, which are really $\planck^2\bet{1}$, etc., and likewise for the multiplier coefficients and \lstinline!Alp0!.}
\begin{lstlisting}[breaklines=true]
In[]:= quantity = (Alp1*PR1[-i, -k, -l, -m, a, b, c, d] + Alp2*PR2[-i, -k, -l, -m, a, b, c, d] + Alp3*PR3[-i, -k, -l, -m, a, b, c, d] + Alp4*PR4[-i, -k, -l, -m, a, b, c, d] + Alp5*PR5[-i, -k, -l, -m, a, b, c, d] + Alp6*PR6[-i, -k, -l, -m, a, b, c, d])*R[-a, -b, -c, -d]
Out[]= |
\vspace{-4pt}
\begin{flushleft}
\includegraphics[width=\linewidth]{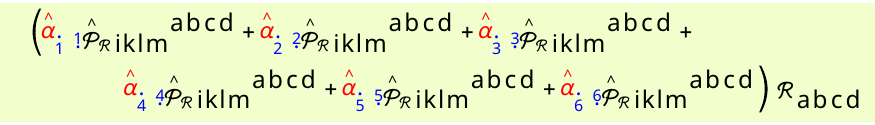}
\end{flushleft}
\vspace{-5pt}
|
In[]:= quantity = ScreenDollarIndices[ToCanonical[quantity /. PActivate /. {Alp1 -> 1, Alp2 -> 1, Alp3 -> 1, Alp4 -> 1, Alp5 -> 1, Alp6 -> 1}]]
Out[]= |
\vspace{-4pt}
\begin{flushleft}
\includegraphics[width=\linewidth]{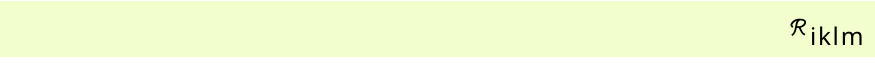}
\end{flushleft}
\vspace{-5pt}
|
In[]:= ScreenDollarIndices[ContractMetric[ToCanonical[T[i, k, l]*(Bet1*PT1[-i, -k, -l, a, b, c])*T[-a, -b, -c] + cAlp6*R[i, k, l, m]*PR6[-i, -k, -l, -m, a, b, c, d]*R[-a, -b, -c, -d] + T[i, k, l]*(cBet2*PT2[-i, -k, -l, a, b, c] + cBet3*PT3[-i, -k, -l, a, b, c])*TLambda[-a, -b, -c] /. PActivate]]]
Out[]= |
\vspace{-4pt}
\begin{flushleft}
\includegraphics[width=\linewidth]{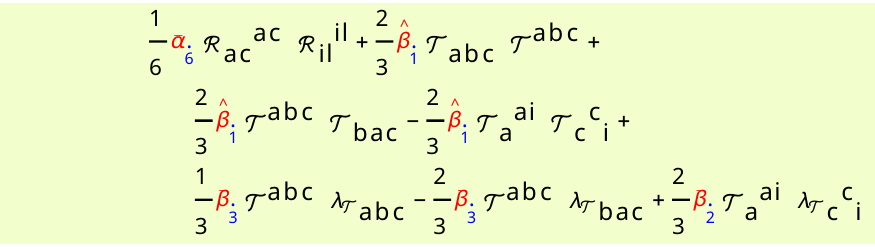}
\end{flushleft}
\vspace{-5pt}
|
\end{lstlisting}
From this we see how to access the $\projlore[_{ij}^{kl}_{nm}^{pq}]{I}$ and $\projlore[_{j}^{kl}_{m}^{pq}]{M}$ projections.

\subsubsection{Rotation group}\label{lorentzgroup}

The $\othree$ projections are similarly defined, but of course refer to the vector $\foli{i}$, not just the Lorentzian metric $\tensor{\eta}{_{ij}}$. The parallel momenta can be decomposed in this way: for the translational case
\begin{subequations}
\begin{align}
  \tensor{\hat{\pi}}{_{k\ovl{l}}}&=\tensor{\hat\pi}{_{\ovl{kl}}}+\tensor{n}{_k}\PiP[\ovl{l}]{B1m},\label{traper}\\
  \tensor{\hat{\pi}}{_{\ovl{kl}}}&=\frac{1}{3}\etad{\ovl{kl}}\PiP{B0p}+\PiP[\ovl{kl}]{B1p}+\PiP[\ovl{kl}]{B2p},
  \label{trapar}
\end{align}
\end{subequations}
where the second term in~\eqref{traper} is the $1^-$ vector mode, and the terms in~\eqref{trapar} are respectively the $0^+$ scalar, skew-symmetric $1^+$ vector and symmetric-traceless $2^+$ tensor modes. 
The rotational case is similarly decomposed as
\begin{subequations}
\begin{align}
  \tensor{\hat{\pi}}{_{kl\ovl{m}}}&=\tensor{\hat{\pi}}{_{\ovl{klm}}}+2\tensor{n}{_{[k}}\tensor{\hat{\pi}}{_{\perp\ovl{l}]\ovl{m}}},\\
  \tensor{\hat{\pi}}{_{\perp\ovl{kl}}}&=\frac{1}{3}\etad{\ovl{kl}}\PiP{A0p}+\PiP[\ovl{kl}]{A1p}+\PiP[\ovl{kl}]{A2p},\label{rotper}\\
  \tensor{\hat{\pi}}{_{\ovl{klm}}}&=\frac{1}{6}\epsd{\ovl{klm}}\PiP{A0m}+\PiP[[\ovl{k}]{A1m}\etad{{{\ovl{l}]\ovl{m}}}}+\frac{4}{3}\PiP[\ovl{klm}]{A2m},
  \label{rotpar}
\end{align}
\end{subequations}
where~\eqref{rotper} are the $0^+$, $1^+$ and $2^+$ modes and~\eqref{rotpar} are the $0^-$, $1^-$ and $2^-$ modes. Accordingly, we access these as follows

\begin{lstlisting}[breaklines=true]
In[]:= ScreenDollarIndices[ToCanonical[{APiP[-a, -b, -c], BPiP[-a, -b]} /. PiPToPiPO3]]
Out[]= |
\vspace{-4pt}
\begin{flushleft}
\includegraphics[width=\linewidth]{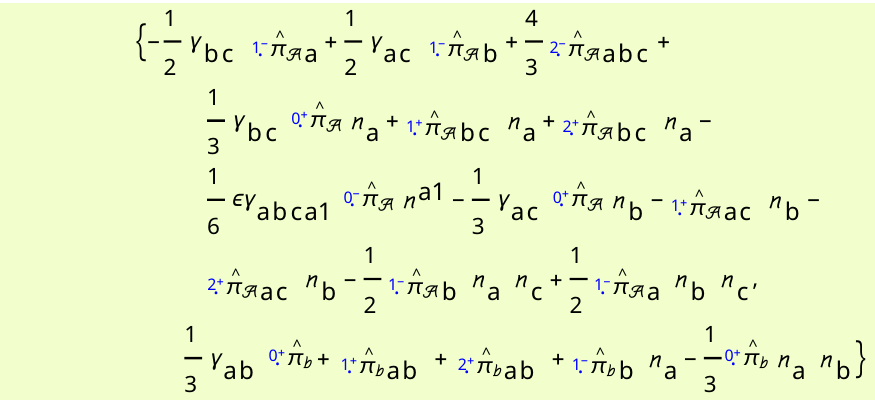}
\end{flushleft}
\vspace{-5pt}
|
\end{lstlisting}

The parallel parts of the field strengths can also be decomposed, since they are canonical, but we neglect the perpendicular parts which depend on the unphysical fields $\tensor{b}{^i_0}$ and $\tensor{A}{^{ij}_0}$, and also on non-canonical velocities. The field strength decomposition is
\begin{subequations}
\begin{align}
  \tensor{\mathcal{R}}{_{ijkl}}&=\tensor{\mathcal{R}}{_{ij\ovl{kl}}}+2\tensor{n}{_{[k|}}\tensor{\mathcal{R}}{_{ij\perp|\ovl{l}]}},\label{koffl}
 \\
 \tensor{\mathcal{T}}{_{ikl}}&=\tensor{\mathcal{T}}{_{i\ovl{kl}}}+2\tensor{n}{_{[k|}}\tensor{\mathcal{T}}{_{i\perp|\ovl{l}]}}.\label{foffl}
\end{align}
\end{subequations}
We notice that $\tensor{\mathcal{R}}{_{ij\ovl{kl}}}$ and $\tensor{\mathcal{R}}{_{ij\perp\ovl{l}}}$ both share all six $J^P$ representations present in the rotational momentum. In the parallel case, these are denoted by the $0^+$ scalar $\cR[]{A0p}$, $1^+$ dual vector $\cR[\ovl{ij}]{A1p}$, $2^+$ symmetric-traceless tensor $\cR[\ovl{ij}]{A2p}$, $0^-$ pseudoscalar $\cR[]{A0m}$, $1^-$ vector $\cR[\ovl{i}]{A1m}$ and $2^-$ tensor $\cR[\ovl{ijk}]{A2m}$. In general, the ($\langle\cdot \rangle$) brackets indicate the symmetric-traceless operation. 
The perpendicular irreps are denoted as the $0^+$ scalar $\ncR[]{A0p}$, $1^+$ dual vector $\ncR[\ovl{ij}]{A1p}$, $2^+$ symmetric-traceless tensor $\ncR[\ovl{ij}]{A2p}$, $0^-$ pseudoscalar $\ncR[]{A0m}$, $1^-$ vector $\ncR[\ovl{i}]{A1m}$ and $2^-$ tensor $\ncR[\ovl{ijk}]{A2m}$.
The situation for the torsion is slightly different, because of the reduced number of components. The parallel $\tensor{\mathcal{T}}{_{i\ovl{kl}}}$ contains the $0^-$ pseudoscalar $\cT[]{A0m}$, $1^+$ dual vector $\cT[\ovl{ij}]{B1p}$, $1^-$ vector $\cT[\ovl{i}]{B1m}$ and $2^-$ tensor $\cT[\ovl{ijk}]{A2m}$. The perpendicular $\tensor{\mathcal{T}}{_{i\perp\ovl{l}}}$ contains the $0^+$ scalar $\ncT[]{B0p}$, $1^+$ dual vector $\ncT[\ovl{ij}]{B1p}$, $1^-$ vector $\ncT[\ovl{i}]{B1m}$ and $2^+$ tensor $\ncT[\ovl{ij}]{B2p}$ -- just as with the translational momentum.

In \HiGGS{}, only the canonical, parallel parts of these tensors are decomposed. The multipliers share the same tensor structure, and both their parallel and perpendicular parts are decomposed: this is because all multiplier fields are assumed to be canonical. The resultant decomposition is
\begin{lstlisting}[breaklines=true]
In[]:= quantity = ScreenDollarIndices[ToCanonical[{R[-a, -b, -c, -d], RLambda[-a, -b, -c, -d], T[-a, -b, -c], TLambda[-a, -b, -c]} /. StrengthDecompose /. StrengthLambdaDecompose]]
Out[]= |
\vspace{-4pt}
\begin{flushleft}
\includegraphics[width=\linewidth]{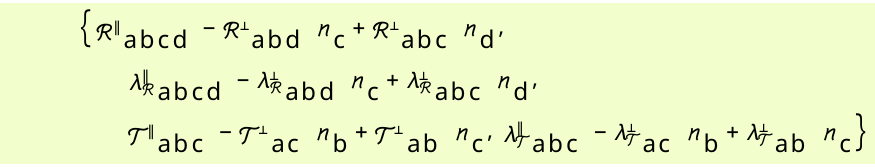}
\end{flushleft}
\vspace{-5pt}
|
In[]:= Simplify[ScreenDollarIndices[ContractMetric[ToCanonical[quantity[[1]] /. StrengthPToStrengthPO3]]]]
Out[]= |
\vspace{-4pt}
\begin{flushleft}
\includegraphics[width=\linewidth]{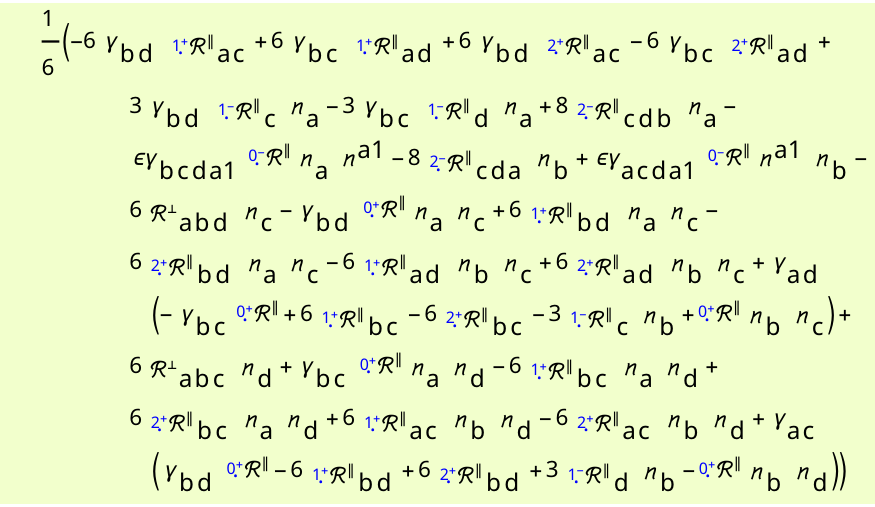}
\end{flushleft}
\vspace{-5pt}
|
In[]:= Simplify[ScreenDollarIndices[ContractMetric[ToCanonical[quantity[[4]] /. StrengthLambdaPToStrengthLambdaPO3 /. StrengthLambdaPerpToStrengthLambdaPerpO3]]]]
Out[]= |
\vspace{-4pt}
\begin{flushleft}
\includegraphics[width=\linewidth]{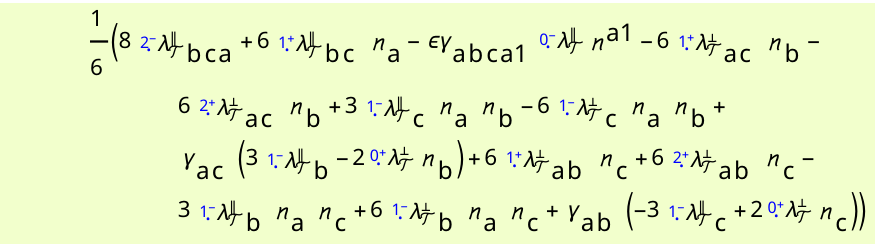}
\end{flushleft}
\vspace{-5pt}
|
\end{lstlisting}
Note that our typeset convention for the $\sothree$ multiplier irreps will be simply to recycle the field strength expressions above, substituting the symbols $\mathcal{R},\ \mathcal{T}\to \lambda$. This is a general rule unless stated otherwise: i.e., tensors whose (Lorentz-invariant) index structure are identical will use the same $\sothree$ irrep notation, but a different symbol.

It is possible, if needed, to recover the explicit `human-readable' projection operators used in the above calculations, such as appear in $\tensor[^A]{\hat{\pi}}{_{\acu{l}}}\equiv\projorthhum[_{\acu{l}}^{ij}_{\ovl{k}}]{A}\tensor{\hat{\pi}}{_{ij}^{\ovl{k}}}$ and $\tensor[^E]{\hat{\pi}}{_{\acu{l}}}\equiv\projorthhum[_{\acu{l}}^{i}_{\ovl{k}}]{E}\tensor{\hat{\pi}}{_{i}^{\ovl{k}}}$, where the $J^P$ sectors are represented by $A$ and $E$ indices according to our conventions in~\cite{smooth}; the $\acu{l}$ accent denotes variable indices, and these projections are defined as obtaining precisely the irreps in~\cref{traper,trapar,rotper,rotpar}. To reflect the various conventions used in the literature, these projections are sometimes defined differently for spin representations contained within parts of the field strengths and momenta whose tensor structures are identical -- for example $\tensor{\hat{\pi}}{_{\ovl{ijk}}}$ and $\tensor{\mathcal{T}}{_{\ovl{ijk}}}$. We therefore use different symbol names for momenta and field strength projections (or those of multipliers), and find e.g. for the selection
\begin{lstlisting}[breaklines=true]
In[]:= quantity = ScreenDollarIndices[ContractMetric[ToCanonical[{PB0p[e, f]*PBPara[-e, -f, a, c]*BPiP[-a, -c], PB1m[-n, f]*PBPerp[-f, a, c]*BPiP[-a, -c], PA2p[-n, -m, e, f]*PAPerp[-e, -f, a, b, c]*APiP[-a, -b, -c], PR0p[e, f, g, h]*RP[-e, -f, -g, -h], PR0m[e, f, g]*RPerp[-e, -f, -g], PR1p[-n, -m, e, f, g, h]*RLambdaP[-e, -f, -g, -h], PT1p[-a, -b, c, d]*TPerp[-c, -d], PT2m[-a, -b, -c, d, e, f]*TP[-d, -e, -f]} /. PADMPiActivate]]]
Out[]= |
\vspace{-4pt}
\begin{flushleft}
\includegraphics[width=\linewidth]{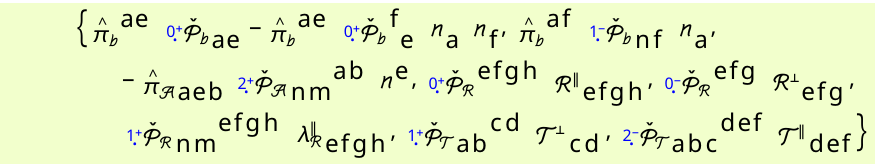}
\end{flushleft}
\vspace{-5pt}
|
In[]:= quantity = quantity /. PO3PiActivate /. PO3TActivate /. PO3RActivate /. PiPToPiPO3 /. StrengthPToStrengthPO3 /. StrengthPerpToStrengthPerpO3 /. StrengthLambdaPToStrengthLambdaPO3 /. StrengthLambdaPerpToStrengthLambdaPerpO3; quantity = Simplify[ScreenDollarIndices[ContractMetric[ToCanonical[quantity]]]]; quantity = Simplify[ScreenDollarIndices[ContractMetric[ToCanonical[quantity]]]]; 
Out[]= |
\vspace{-4pt}
\begin{flushleft}
\includegraphics[width=\linewidth]{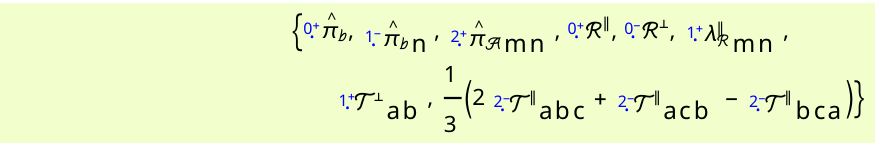}
\end{flushleft}
\vspace{-5pt}
|
\end{lstlisting}
Note above that it is sometimes necessary to repeatedly apply routines in order to recover the desired form. In prinicple, the \HiGGS{} commands are constructed so as to be idempotent, but broadly in \Mathematica{} and \xAct{} repeated commands are sometimes helpful when multiple functions are nested.

\subsection{Derivatives}\label{derivatives}

While the gauge-fixed Minkowskian setup initially makes for easy development, it sometimes means that we have to `reinvent the wheel' in order to access machinery for which there is already a very sophisticated implementation in \xAct{}. A clear example of this is given by the way in which \HiGGS{} handles gauge-covariant derivatives in the Poincar\'e gauge theory. Generalising from~\eqref{poin_covd}, we recall from~\cite{2016JMP....57i2505L,mythesis,chapter2,blagojevic2002gravitation} that for some matter field $\varphi$ we have 
\begin{equation}
  \tensor{\mathcal{  D}}{_i}\varphi\equiv \tensor{h}{_i^\mu}\Big( \tensor{\partial}{_\mu}+\frac{1}{2}\tensor{A}{^{kl}_\mu}\tensor{\Sigma}{_{kl}}\cdot \Big)\varphi,
  \quad
  \tensor{D}{_\mu}\varphi\equiv \tensor{b}{^i_\mu}\tensor{\mathcal{  D}}{_i}\varphi,
  \label{pgtcd}
\end{equation}
where $\tensor{\Sigma}{_{ij}}$ are the Lorentz group generators specific to the representation of $\varphi$. This construction could be implemented using the \xAct{} command \lstinline!DefCovD[]!, in such a way that the $\tensor{\Sigma}{_{ij}}$ generators are automatically calculated for tensorial $\varphi$ representations of the kind that arise in the Hamiltonian analysis of the matter-free theory. In particular, the derivative $\tensor{D}{_\mu}\varphi$ is geometrically interpreted as $\tensor{\nabla}{_\mu}\varphi$ as it appears in~\eqref{vdef}. Rather than following this route, \HiGGS{} defines a pair of first derivatives (broadly corresponding to the two definitions in~\eqref{pgtcd}) for every canonical quantity which might be of interest. This process is not very efficient or flexible, but it is sufficient when explicit covariant derivatives are scarce. Indeed, the most common occurance of the gauge covariant derivative is through its commutator
\begin{equation}
  2\tensor{\mathcal{  D}}{_{[i}}\tensor{\mathcal{  D}}{_{j]}}\varphi=\Big(\frac{1}{2}\tensor{\mathcal{  R}}{^{kl}_{ij}}\tensor{\Sigma}{_{kl}}\cdot
  -\tensor{\mathcal{  T}}{^k_{ij}}\tensor{\mathcal{  D}}{_k}\Big)\varphi,
\end{equation}
in the form of the field strengths: as we saw in~\cref{geometricsetup}, these have a \emph{separate} implementation. Explicit gradients will tend to arise mostly at the \emph{end} of the calculations for which \HiGGS{} is designed. Gradients of the field strengths cannot usually be fed back into the algorithm anyway, since they would require an implementation of the second order Euler--Lagrange equations in order to be processed.

Based on the understanding that we are always dealing with arbitrarily-indexed (possibly mixed) tensorial representations, \HiGGS' two preferred forms of the derivative are $\tensor{D}{_\mu}\tensor{\varphi}{_{\acu{u}\acu{\mu}}}$ and $\tensor*{\delta}{^{\acu{\ovl{u}}}_{\acu{\ovl{v}}}}\tensor*{\delta}{^{\acu{\ovl{w}}}_{\acu{\ovl{s}}}}\tensor{h}{_{\acu{w}}^{\acu{\mu}}}\tensor{\mathcal{D}}{_{\ovl{k}}}\tensor{\varphi}{_{\acu{u}\acu{\mu}}}$. The former is straightforward\footnote{Note that we use an accent to indicate an arbitrary number of indices, following from~\cite{2020arXiv200502228L,smooth}.}, and the latter is the parallel projection of the gradient on all indices. Picking a couple of $\othree$ irreps at random, we find for example
\begin{lstlisting}[breaklines=true]
In[]:= quantity = {DPiPA2m[-z, -a, -b, -c], CD[-z][PiPA2m[-a, -b, -c]], DRP2p[-z, -a, -b], CD[-z][RP2p[-a, -b]]}
Out[]= |
\vspace{-4pt}
\begin{flushleft}
\includegraphics[width=\linewidth]{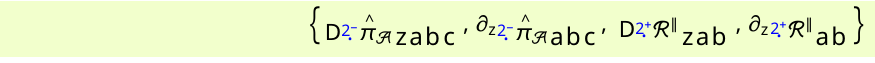}
\end{flushleft}
\vspace{-5pt}
|
In[]:= quantity = ScreenDollarIndices[ContractMetric[ToCanonical[quantity /. DRPDeactivate /. DPiPDeactivate]]]
Out[]= |
\vspace{-4pt}
\begin{flushleft}
\includegraphics[width=\linewidth]{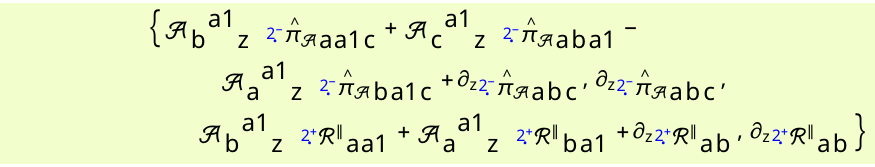}
\end{flushleft}
\vspace{-5pt}
|
In[]:= quantity = ScreenDollarIndices[ContractMetric[ToCanonical[quantity /. DRPActivate /. DPiPActivate]]]
Out[]= |
\vspace{-4pt}
\begin{flushleft}
\includegraphics[width=\linewidth]{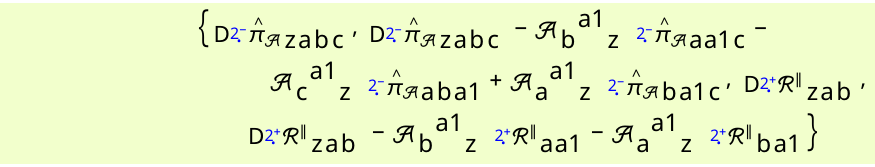}
\end{flushleft}
\vspace{-5pt}
|
\end{lstlisting}
One cannot generally pass from $\tensor{D}{_\mu}\tensor{\varphi}{_{\acu{u}\acu{\mu}}}$ to a parallel derivative, unless only the spacelike indices are involved, e.g. $\tensor{D}{_\alpha}\tensor{\varphi}{_{\acu{u}\acu{\mu}}}$. Since \HiGGS{} tries to achive parallel derivatives wherever possible, this serves as one of the internal checks on the canonical status of a quantity, making sure that any unacceptable time derivatives would appear explicitally in the output. Accordingly, continuing from above
\begin{widetext}
\begin{lstlisting}[breaklines=true]
In[]:= quantity = ScreenDollarIndices[ContractMetric[ToCanonical[(G3[-y, z]*#1 & ) /@ quantity[[{1, 3}]]]]]
Out[]= |
\vspace{-4pt}
\begin{flushleft}
\includegraphics[width=\linewidth]{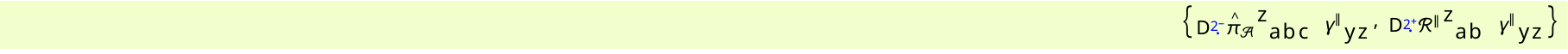}
\end{flushleft}
\vspace{-5pt}
|
In[]:= quantity = ScreenDollarIndices[ContractMetric[ToCanonical[quantity /. DpRPActivate /. DpPiPActivate]]]
Out[]= |
\vspace{-4pt}
\begin{flushleft}
\includegraphics[width=\linewidth]{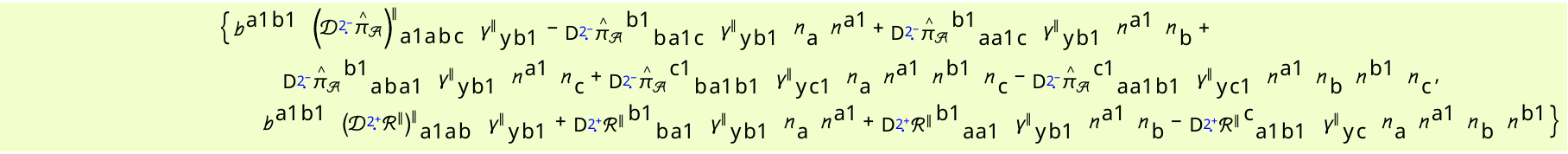}
\end{flushleft}
\vspace{-5pt}
|
In[]:= quantity = ScreenDollarIndices[ContractMetric[ToCanonical[quantity /. DpRPDeactivate /. DpPiPDeactivate]]]
Out[]= |
\vspace{-4pt}
\begin{flushleft}
\includegraphics[width=\linewidth]{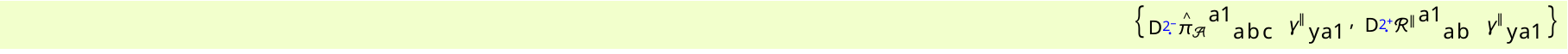}
\end{flushleft}
\vspace{-5pt}
|
\end{lstlisting}
\end{widetext}
We see in the penultimate expression above the two parallel derivatives. If an expression containing gradients is covariant, velocity-independent and parallel, then the various other terms should cancel among themselves. As we have mentioned, covariance also requires that gauge fields do not explicitally appear, but are implicit in covariant quantities. Overall, the work of packaging an expression into a canonical, covariant form, parallel if possible and free from unphysical fields, is done by the \lstinline!ToNesterForm[]! command, which we introduce in~\cref{low-levelfunctions}.

As a final comment on the use of derivatives, we recall that the formulae~\cref{riemanndef,torsiondef} imply a pair of Bianchi identities~\cite{10.1143/PTP.64.866,blagojevic2002gravitation} as follows
\begin{subequations}
	\begin{align}
		\tensor{\epsilon}{^{\rho\mu\lambda\nu}}\tensor{D}{_\mu}\left(\tensor{b}{^i_\lambda}\tensor{b}{^j_\nu}\tensor{\mathcal{T}}{^s_{ij}}\right)&\equiv\tensor{\epsilon}{^{\rho\mu\lambda\nu}}\tensor{b}{^k_\mu}\tensor{b}{^i_\lambda}\tensor{b}{^j_\nu}\tensor{\mathcal{R}}{^s_{kij}},\label{bianchi1}\\
		\tensor{\epsilon}{^{\rho\mu\lambda\nu}}\tensor{D}{_\lambda}\left(\tensor{b}{^k_\mu}\tensor{b}{^l_\nu}\tensor{\mathcal{R}}{^{ij}_{kl}}\right)&\equiv 0.\label{bianchi2}
	\end{align}
\end{subequations}
These can be verified using the tools already introduced, but the process will be easier with the \lstinline!ToBasicForm[]! and \lstinline!ToNesterForm[]! commands.

\subsection{Low-level functions}\label{low-levelfunctions}

Throughout~\cref{geometricsetup,irreducibledecompositions} we have used many rules, such as \lstinline!DpRPDeactivate!, \lstinline!DPiPActivate!, etc., which --- whilst they are not confined to \lstinline!xAct`HiGGS`Private`! and so are available to the user --- should not often be needed within a \HiGGS{} science session. Instead these rules are wrapped into two `official' functions: \lstinline!ToNesterForm[]!, which strives to collect expressions\footnote{The chosen naming refers to the fact that the irrep conventions and notation of collected expressions tends to align most closely with those of a collection of articles -- very useful during development -- for which Nester is a common author, see e.g.~\cite{Chen:2015vya,Hehl:1976kj,1999IJMPD...8..459Y,2002IJMPD..11..747Y,2008PhRvD..78b3522S,2009JCAP...10..027C,2011PhRvD..83b4001B,2011JPhCS.330a2005H,2011IJMPD..20.2125H,2015arXiv151201202H,1998AcPPB..29..961C}.}, and \lstinline!ToBasicForm[]!, which expands them. Naturally \lstinline!ToNesterForm[]! is the more complicated of the two. It is, in some sense, the extension of the \lstinline!ToCanonical[]! command from \xAct{} into \HiGGS{}.

\subsubsection{Module: `ToBasicForm'}

We begin by breaking some expressions which we know to be covariant, using \lstinline!ToBasicForm[]!
\begin{lstlisting}[breaklines=true]
In[]:= quantity = ToBasicForm[{T[i, -j, -k], PiPB0p[], PiPA1p[-i, -j]}]
Out[]= |
\vspace{-4pt}
\begin{flushleft}
\includegraphics[width=\linewidth]{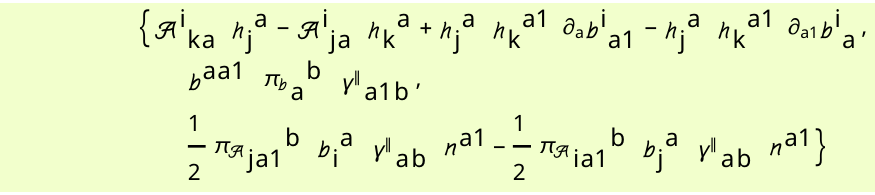}
\end{flushleft}
\vspace{-5pt}
|
\end{lstlisting}
We see that the results are expressed in terms of the bare gauge fields, non-parallel momenta, independent ADM quantities and coordinate derivatives. If a Poisson bracket were to be manually evaluated between the fields in \lstinline!quantity!, it would first be necessary to perform the expansion provided by \lstinline!ToBasicForm[]!, before calculating variational derivatives and multiplying. For this reason, the \lstinline!PoissonBracket[]! command which we introduce shortly relies on \lstinline!ToBasicForm[]! as an initial step. In the case above, all the brackets would be relatively straightforward to evaluate by hand: to see why the strong coupling problem demands an implementation such as \HiGGS{}, let us break open a simple momentum gradient
\begin{widetext}
\begin{lstlisting}[breaklines=true]
In[]:= ToBasicForm[DpPiPB1m[-k, -i]]
Out[]= |
\vspace{-4pt}
\begin{flushleft}
\includegraphics[width=\linewidth]{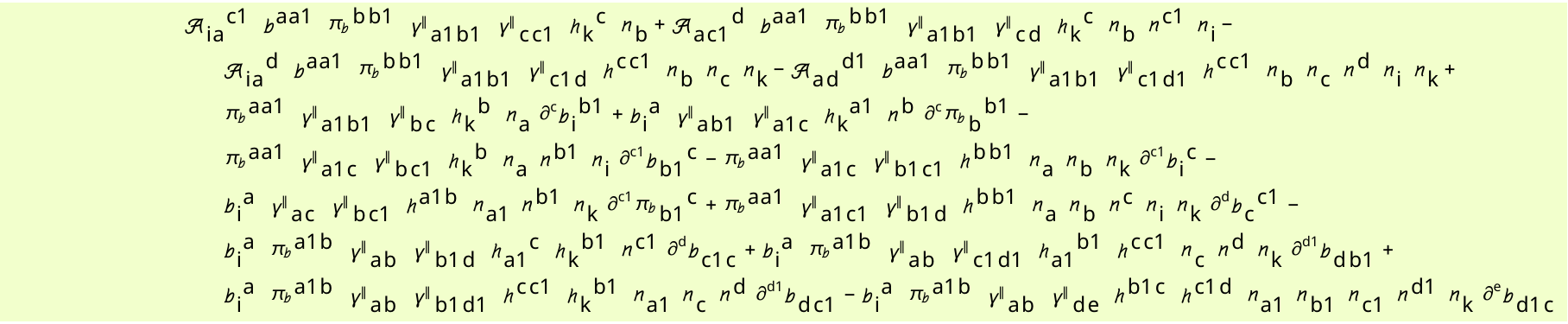}
\end{flushleft}
\vspace{-5pt}
|
\end{lstlisting}
\end{widetext}
The gradient of a more complicated irrep such as \lstinline!DpPiPA2m[-i,-j,-k]! would span many pages after an application of \lstinline!ToBasicForm[]!, rendering manual evaluation of brackets impractical, yet we will see in~\cref{examples} that \emph{there is nothing to stop these terms arising in the analysis}.

Now is a convenient time to verify the second Bianchi identity in~\eqref{bianchi2}. 
Setting up the derivative, and using the gauge-fixed $\tensor{\epsilon}{^{\mu\nu\sigma\lambda}}$ which follows from $\tensor{\gamma}{_{\mu\nu}}$ -- \lstinline!epsilonG[a,b,c,d]! -- we find
\begin{lstlisting}[breaklines=true]
In[]:= quantity2 = ToBasicForm[epsilonG[r, m, l, n]*(CD[-l][B[k, -m]*B[q, -n]*R[i, j, -k, -q]] + A[i, -x, -l]*B[k, -m]*B[q, -n]*R[x, j, -k, -q] + A[j, -x, -l]*B[k, -m]*B[q, -n]*R[i, x, -k, -q])]; 
Out[]= |
\vspace{-4pt}
\begin{flushleft}
\includegraphics[width=\linewidth]{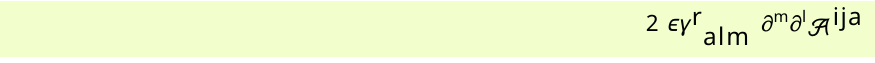}
\end{flushleft}
\vspace{-5pt}
|
\end{lstlisting}
Evidently, this quantity vanishes, and so what we are seeing is a side-effect of the \HiGGS{} geometric setup: \xAct{} does not know that \lstinline!CD[-i][]! is the partial derivative, so we need a final step
\begin{lstlisting}[breaklines=true]
In[]:= quantity2 = (ToCanonical[(1/2)*(#1 + CommuteCovDs[#1, CD, {l, m}])] & )[quantity2]; 
Out[]= |
\vspace{-4pt}
\begin{flushleft}
\includegraphics[width=\linewidth]{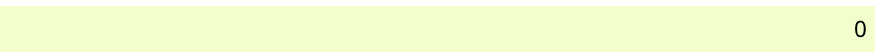}
\end{flushleft}
\vspace{-5pt}
|
\end{lstlisting}

\subsubsection{Module: `ToNesterForm'}

Let us now undo the breaking, using \lstinline!ToNesterForm[]!. 
In general, much of the functionality of \lstinline!ToNesterForm[]! has to do with fully incorporating the known primary and secondary constraints of the theory during the course of simplification -- so as to leave no extra conditions for the user to worry about. Before a theory has been defined (see~\cref{high-levelfunctions}) using \lstinline!DefTheory[]!, we will have to suppress this activity by passing the option \lstinline!"ToShell"->False!, yielding
\begin{lstlisting}[breaklines=true]
In[]:= quantity = ToNesterForm[quantity, "ToShell" -> False]
Out[]= |
\vspace{-4pt}
\begin{flushleft}
\includegraphics[width=\linewidth]{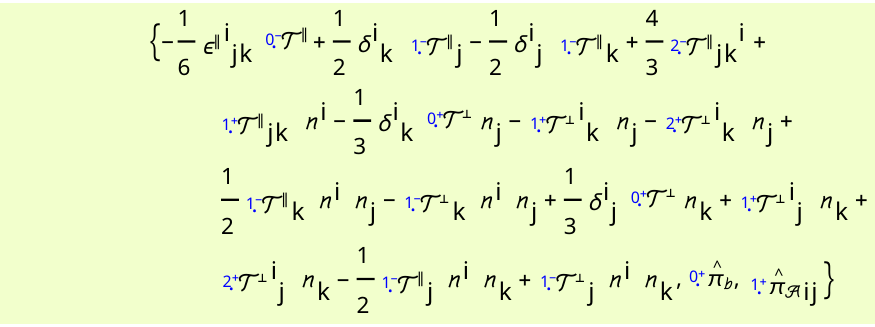}
\end{flushleft}
\vspace{-5pt}
|
\end{lstlisting}
Note that whilst the results \emph{are} covariant, we do not get back the same form from \lstinline!ToNesterForm[]! as that which was provided to \lstinline!ToBasicForm[]!, instead we recover the $\othree$ expansion. The reason for this is that reducible quantities such as \lstinline!T[-a,-b,-c]! are usually quickly broken up during the course of the Hamiltonian analysis: the output of \lstinline!ToNesterForm[]! is tuned so as to be useful in that context. Moreover, we note that this operation will not now work in reverse, since input such as \lstinline!TP2m[-a,-b,-c]//ToBasicForm! will not return a broken expression. There is no special reason behind this: \lstinline!ToBasicForm[]! is not a sophisticated function, it essentially imposes a list of internal rules on its argument, and the expansion of $\othree$ field strength irreps, as with those of momenta, would be perfectly straightforward to implement in the source if and when needed.

When \lstinline!ToNesterForm[]! is passed a non-covarint quantity, it is unable to return a covariant result. Nonetheless, it tries, returning for the innocuous gradient $\tensor{\partial}{_\mu}\tensor{b}{^i_\nu}$
\begin{widetext}
\begin{lstlisting}[breaklines=true]
In[]:= ToNesterForm[CD[-i][B[-a, -b]], "ToShell" -> False]
Out[]= |
\vspace{-4pt}
\begin{flushleft}
\includegraphics[width=\linewidth]{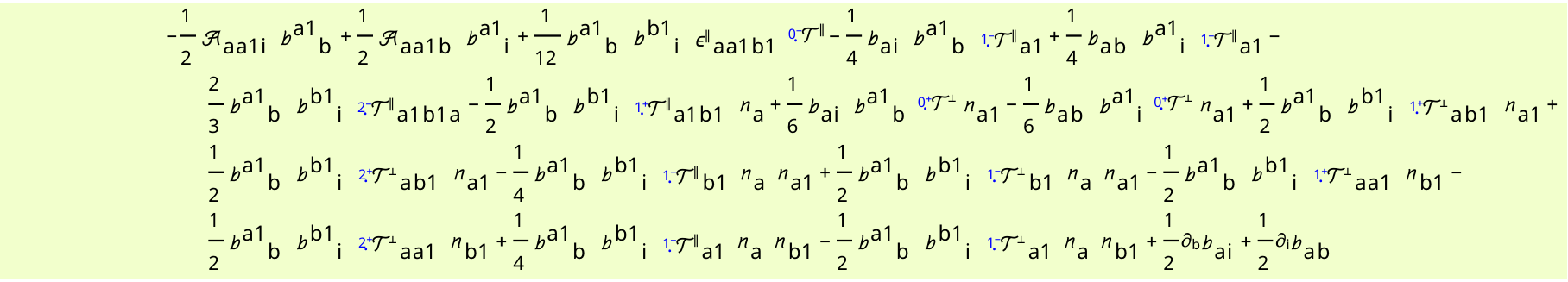}
\end{flushleft}
\vspace{-5pt}
|
\end{lstlisting}
\end{widetext}
As with this case, it is usually easy to identify nonphysical expressions which indicate a human error, through the appearance of bare gauge fields. 
If needed, this covariance check is easy to automate through an output search with the \Mathematica{} \lstinline!Head[]! function for an unwanted \HiGGS{} quantity, such as \lstinline!A! or \lstinline!B!.
We can also observe in the output above part of the route taken by \lstinline!ToNesterForm[]!. Gauge field gradients are converted, where possible, to field strengths and covariant derivatives.
The residual spin connection terms are then extensively manipulated in an attempt to cancel them.
Leftover asymmetric derivatives of the translational gauge field can sometimes be ascribed to covariant quantities through~\eqref{furtheridentities}, for example the following gradient cannot be expressed through the torsion alone
\begin{lstlisting}[breaklines=true]
In[]:= ToNesterForm[PPara[-y, v]*H[-v, w]*G3[-w, m]*PPara[-q, b1]*H[-b1, a1]*G3[-a1, l]*(G3[-l, n]*V[-k]*CD[-m][B[k, -n]] + G3[-l, n]*B[j, -n]*A[k, -j, -m]*V[-k]) /. PADMActivate, "ToShell" -> False]
Out[]= |
\vspace{-4pt}
\begin{flushleft}
\includegraphics[width=\linewidth]{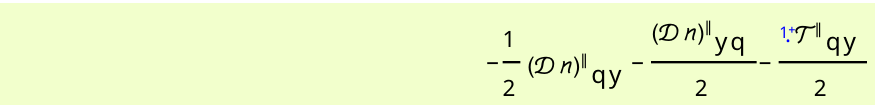}
\end{flushleft}
\vspace{-5pt}
|
\end{lstlisting}
Note that the above result could equally be written as the single term \lstinline!-DpV[-q,-y]!, but \lstinline!ToNesterForm[]! takes the opportunity to separate out the antisymmetric part.

Before moving on, we return to verify the first Bianchi identity in~\eqref{bianchi2}. We can access the canonical (i.e. velocity-independent) part of this identity by projecting~\cref{bianchi2} with $\tensor{b}{^\perp_\mu}$, or equivalently using in place of $\tensor{\epsilon}{^{\mu\nu\lambda\sigma}}$ the foliation equivalent $\epsu{\ovl{ijk}}$, which is given in \HiGGS{} by \lstinline!Eps[i,j,k]!,
\begin{lstlisting}[breaklines=true]
In[]:= quantity = ToBasicForm[Eps[u, v, w]*H[-u, m]*H[-v, l]*H[-w, n]*(CD[-m][B[i, -l]*B[j, -n]*T[s, -i, -j]] + A[s, -x, -m]*B[i, -l]*B[j, -n]*T[x, -i, -j]) /. PADMActivate]; 
Out[]= |
\vspace{-4pt}
\begin{flushleft}
\includegraphics[width=\linewidth]{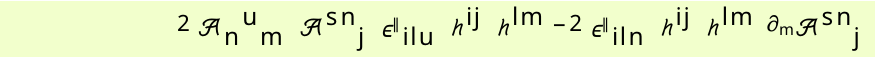}
\end{flushleft}
\vspace{-5pt}
|
In[]:= quantity = (ToCanonical[V[-r]*(1/2)*(#1 + CommuteCovDs[#1, CD, {j, l}])] & )[quantity]; 
Out[]= |
\vspace{-4pt}
\begin{flushleft}
\includegraphics[width=\linewidth]{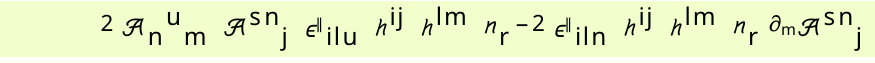}
\end{flushleft}
\vspace{-5pt}
|
In[]:= quantity = ToNesterForm[quantity, "ToShell" -> False]; 
Out[]= |
\vspace{-4pt}
\begin{flushleft}
\includegraphics[width=\linewidth]{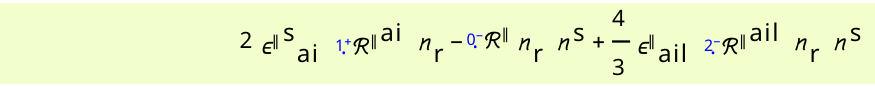}
\end{flushleft}
\vspace{-5pt}
|
\end{lstlisting}
The output here is then equal to the $\othree$ irreps contained within $\epsu{\ovl{kij}}\tensor{\mathcal{R}}{^s_{\ovl{kij}}}$, as expected.

\subsubsection{Module: `PoissonBracket'}\label{poissonbracket}

The \lstinline!PoissonBracket[]! command is the third `official' function provided by the \HiGGS{} package. 
The Poisson bracket appearing in this article is defined for general functionals $\mathcal{A}$ and $\mathcal{B}$ of the gravitational fields and their conjugate momenta
\begin{equation}
	\begin{aligned}
	\Big\{\mathcal{A},\mathcal{B}\Big\}\equiv 
	\int\mathrm{d}^3x
	\Bigg[
	\frac{\delta\mathcal{A}}{\delta\tensor{b}{^{i}_{\mu}}}
	\frac{\delta\mathcal{B}}{\delta\tensor{\pi}{_{i}^{\mu}}}
	+
	\frac{\delta\mathcal{A}}{\delta\tensor{A}{^{ij}_{\mu}}}
	\frac{\delta\mathcal{B}}{\delta\tensor{\pi}{_{ij}^{\mu}}}
	\\
	-
	\frac{\delta\mathcal{A}}{\delta\tensor{\pi}{_{i}^{\mu}}}
	\frac{\delta\mathcal{B}}{\delta\tensor{b}{^{i}_{\mu}}}
	-
	\frac{\delta\mathcal{A}}{\delta\tensor{\pi}{_{ij}^{\mu}}}
	\frac{\delta\mathcal{B}}{\delta\tensor{A}{^{ij}_{\mu}}}
	\Bigg],
	\end{aligned}
	\label{superpoisson}
\end{equation}
with a natural extension of the formula when multiplier fields are admitted.
The formula~\eqref{superpoisson} may appear no more daunting than a commonplace action variation, but in practice $\mathcal{A}$ and $\mathcal{B}$ are frequently \emph{local tensors} rather than nonlocal scalars. Locality signifies that the underlying functionals contain Dirac distribuions, themselves subject to the total derivatives of the generalised Euler--Lagrange equations. The full ramifications of covariantly removing these Dirac gradients are detailed in~\cite{mythesis,smooth}, and some special cases are discussed in electrodynamics on the lightcone~\cite{blagrec1} and noncritical string theory~\cite{blagrec2}. Currently, \HiGGS{} is able to accommodate the \emph{first order} Euler--Lagrange formalism in \lstinline!PoissonBracket[]!. First order brackets, evaluated by inserting the spatial dependence into~\eqref{superpoisson}, produce four terms of the form
\begin{align}
		\Big\{\mathcal{A}(\bm{x}_1),&\ \mathcal{B}(\bm{x}_2)\Big\}\equiv 
		\nonumber
		\\
	\int\mathrm{d}^3x
	\big[&
		J_1(\bm{x})
		\delta^3(\bm{x}-\bm{x}_1)
		\delta^3(\bm{x}-\bm{x}_2)
		\nonumber
		\\
		&+
		\tensor{J}{_2^\alpha}(\bm{x})
		\delta^3(\bm{x}-\bm{x}_1)
		\tensor{\partial}{_\alpha}\delta^3(\bm{x}-\bm{x}_2)
		\nonumber
		\\
		&+
		\tensor{J}{_3^\alpha}(\bm{x})
		\tensor{\partial}{_\alpha}\delta^3(\bm{x}-\bm{x}_1)
		\delta^3(\bm{x}-\bm{x}_2)
		\nonumber
		\\
		&+
		\tensor{J}{_4^{\alpha\beta}}(\bm{x})
		\tensor{\partial}{_\alpha}\delta^3(\bm{x}-\bm{x}_1)
		\tensor{\partial}{_\beta}\delta^3(\bm{x}-\bm{x}_2)
	      \big],
	\label{coefflist}
	      \\
	      &\nonumber
\end{align}
where the $J_1$, $\tensor{J}{_2^\alpha}$, $\tensor{J}{_3^\alpha}$ and $\tensor{J}{_4^{\alpha\beta}}$ can be determined by certain formulae.
Note that by our conventions in~\cite{chapter4,mythesis,smooth}, we will in future denote by $\delta^3$ the equal-time Dirac function ${\delta^3(\bm{x}_1-\bm{x}_2)}$.
Without any special instructions, \lstinline!PoissonBracket[]! returns a \lstinline!List! of the four Dirac coefficients in~\eqref{coefflist}. This behaviour is tied into calls to \lstinline!PoissonBracket[]! from within \lstinline!Velocity[]!, which we introduce in~\cref{high-levelfunctions}. 
Note that \lstinline!PoissonBracket[]! contains calls to \lstinline!ToNesterForm[]!, and so works to exhaust transformations which can be applied to the output by virtue of the known primary and secondary constrints. In this sense, it depends on the theory introduced by \lstinline!DefTheory[]! and so for the time being we must again pass the option \lstinline!"ToShell"->False!. We begin with a very simple bracket, whose output can be understood in terms of our previous \lstinline!ToNesterForm[]! result for \lstinline!DpV[-a,-b]!
\begin{lstlisting}[breaklines=true]
In[]:= PoissonBracket[PiPB2p[-a, -b], TP1p[-c, -d], "ToShell" -> False]; 
Out[]= |
\vspace{-4pt}
\begin{flushleft}
\includegraphics[width=\linewidth]{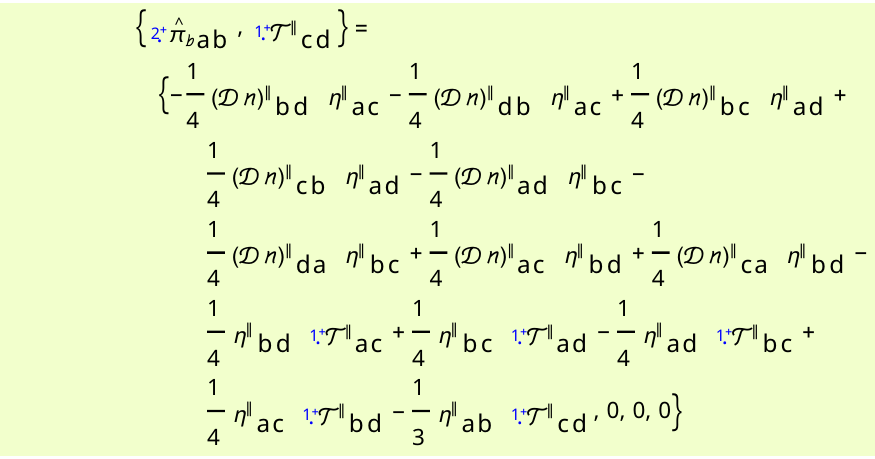}
\end{flushleft}
\vspace{-5pt}
|
\end{lstlisting}
The bracket is not `surficial', in the sense that the latter three entries vanish and the nonvanishing part of the bracket is a compact, covariant expression.
If the latter three coefficients are nonvanishing however, explicit covariance in this expression will be lost, as we can see by modifying the previous example
\begin{widetext}
\begin{lstlisting}[breaklines=true]
In[]:= PoissonBracket[PiPB2p[-a, -b], TP1m[-c], "ToShell" -> False]; 
Out[]= |
\vspace{-4pt}
\begin{flushleft}
\includegraphics[width=\linewidth]{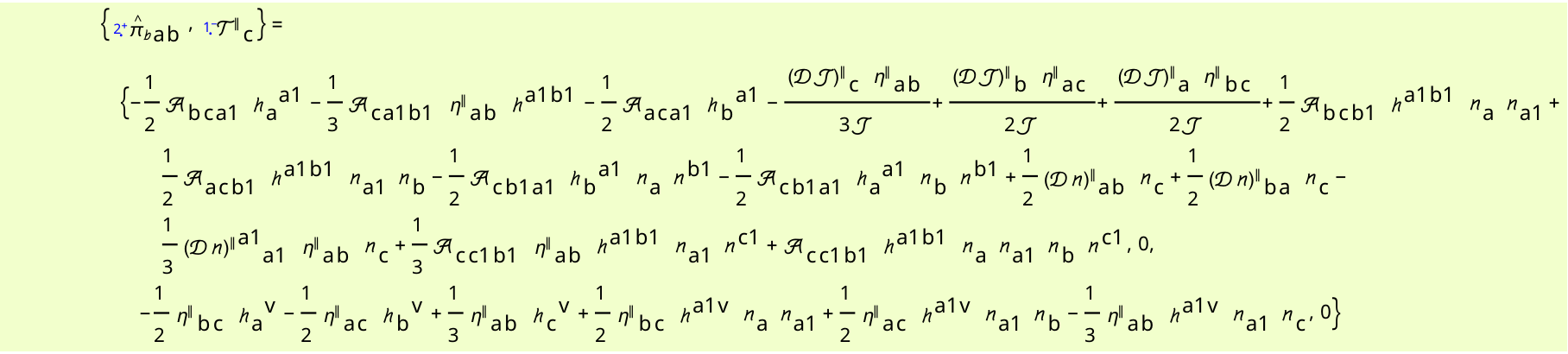}
\end{flushleft}
\vspace{-5pt}
|
\end{lstlisting}
\end{widetext}
In such `surficial' cases the default output of \lstinline!PoissonBracket[]! will not be helpful for visual inspection. To resolve this we can recall, again from~\cite{mythesis}, that~\eqref{coefflist} can be alternatively expressed as
\begin{equation}
  \begin{aligned}
  \int&\mathrm{d}^3x_2\big\{\fA(\bm{x}_1),\fB(\bm{x}_2) \big\}\fC(\bm{x}_2)\equiv
	\tensor{\mathcal{J}}{_1_{\acu{v}}}(\bm{x}_1)\fC(\bm{x}_1)
	\\
	&
	+\tensor{\mathcal{J}}{_2_{\acu{v}}^\alpha}(\bm{x}_1)\tensor{D}{_\alpha}\fC(\bm{x}_1)
	+\tensor{\mathcal{J}}{_3_{\acu{v}}^{\alpha\beta}}(\bm{x}_1)\tensor{D}{_\alpha}\tensor{D}{_\beta}\fC(\bm{x}_1).
\label{defpoi}
\end{aligned}
\end{equation}
The three-component \lstinline!List! output corresponding to~\eqref{defpoi} can be produced by passing the option \lstinline!"Surficial"->True!. Trying again with this option, we obtain
\begin{lstlisting}[breaklines=true]
In[]:= PoissonBracket[PiPB2p[-i, -j], TP1m[-l], "ToShell" -> False, "Surficial" -> True]; 
Out[]= |
\vspace{-4pt}
\begin{flushleft}
\includegraphics[width=\linewidth]{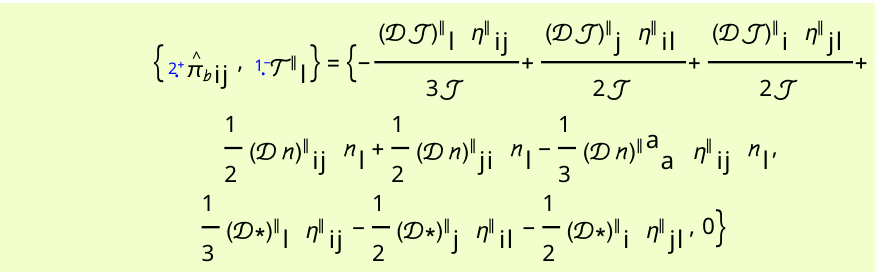}
\end{flushleft}
\vspace{-5pt}
|
\end{lstlisting}
We see that the result is indeed covariant, and fairly simple. We will come back to this `surficial' case in~\cref{yo-nesterunittests}, where we attempt to recover the historical results in~\cite{1999IJMPD...8..459Y,2002IJMPD..11..747Y}.

\subsection{High-level functions}\label{high-levelfunctions}

As we mentioned in~\cref{low-levelfunctions}, much of the work done by \lstinline!ToNesterForm[]! has to do with the imposition of the theory shell so as to simplify the argument. The particular shell used is not specifically that of primary vs secondary constraints, but it is restricted to the constraints of which we have prior knowledge from the literature, and does not include new constraints discovered in the course of a \HiGGS{} session. 
In particular, we rely on the so-called \emph{if-constraint} structure which was discovered by Blagojevi\'c and Nikoli\'c in~\cite{1983PhRvD..28.2455B}. Depending on the Lagrangian parameters in~\eqref{neocon}, the number and type of primary constraints may be radically different: these contingent primaries are called primary if-constraints (PiC). There are similarly secondary (SiC) and tertiary (TiC) quantities, etc.
Returning to~\cite{smooth}, the PiCs of the theory~\eqref{neocon} take the form
\begin{subequations}
  \begin{align}
  \tensor[^A]{\varphi}{_{\acu{v}}}&\equiv\frac{1}{J}\tensor[^A]{\hat{\pi}}{_{\acu{v}}}
  +2\alp{0}\planck^2\projorthhum[_{\acu{v}\perp\ovl{k}}^{\ovl{k}}]{A}
  -8\calpm[A]{\perp\perp}\projorthhum[_{\acu{v}}_{jk}^{\ovl{m}}]{A}\tensor{\lambda}{^{jk}_{\perp\ovl{m}}}
  \nonumber\\
  &\ \ \ -4\projorthhum[_{\acu{v}}_{jk}^{\ovl{lm}}]{A} 
  \Big(\calpm[A]{\perp\parallel}\tensor{\lambda}{^{jk}_{\ovl{lm}}}+2\alpm[A]{\perp\parallel}\tensor{\mathcal{  R}}{^{jk}_{\ovl{lm}}}\Big),
  \label{fulpic}\\
  \tensor[^E]{\varphi}{_{\acu{v}}}&\equiv\frac{1}{J}\tensor[^E]{\hat{\pi}}{_{\acu{v}}}
  -4\planck^2\cbetm[E]{\perp\perp}\projorthhum[_{\acu{v}}_{j}^{\ovl{m}}]{E}\tensor{\lambda}{^{j}_{\perp\ovl{m}}}
  \nonumber\\
  &\ \ \ -2\planck^2\projorthhum[_{\acu{v}}_{j}^{\ovl{lm}}]{E} 
  \Big(\cbetm[E]{\perp\parallel}\tensor{\lambda}{^{j}_{\ovl{lm}}}+2\betm[E]{\perp\parallel}\tensor{\mathcal{  T}}{^{j}_{\ovl{lm}}}\Big),
  \label{fulpictor}
\end{align}
\end{subequations}
where we defined e.g. ${\alpm[A]{\parallel\parallel}\equiv\sum_{I}\projmatrix[AI]{\parallel\parallel}\alp{I}}$, with the matrix $\projorthhum[{_{\acu{p}}^{lm}_{\ovl{nq}}}]{A}\projlore[{_{lm}^{\ovl{nq}}_{ij}^{\ovl{rk}}}]{I}\equiv \projmatrix[AI]{\parallel\parallel}\projorthhum[{_{\acu{p}}_{ij}^{\ovl{rk}}}]{A}$. In the handling of PiCs, it is extremely useful to refer to the functions
\begin{align}
\mu(x) \equiv \left\{\begin{array}{lr}
 x^{-1}, & \text{for } x\neq 0\\
 0, & \text{for } x=0,
\end{array}\right.\quad
\nu(x) \equiv 1-|\text{sgn}(\mu(x))|,
\end{align}
and we note that the PiC functions defined in~\cref{fulpic,fulpictor} are only constrained when ${\nu(\alpm[A]{\perp\perp})=1}$ or ${\nu(\betm[E]{\perp\perp})=1}$.
The structure of~\cref{fulpic,fulpictor} is quite useful in that it allows momenta to substituted for parallel field strengths. These substitutions are among those performed every time \lstinline!ToNesterForm[]! is called, unless the option \lstinline!"ToShell"->False! is passed as above in~\cref{low-levelfunctions}.

\subsubsection{Module: `DefTheory'}\label{deftheory}

In order to discover these PiCs, we must first set up the shell using \lstinline!DefTheory[]!. Let us consider the simple example of Einstein--Cartan theory, i.e. the substantial restriction of~\eqref{neocon} to the simple Einstein--Hilbert term
\begin{equation}
	L_{\text{G}}=-\frac{1}{2}\alp{0}\planck^2\mathcal{R}.
	\label{ehl}
\end{equation}
We implement the theory~\eqref{ehl} by passing to the \lstinline!DefTheory[]! command the system of equations which deactivates all the $\{\alp{A}\}$, $\{\bet{E}\}$, $\{\calp{A}\}$, $\{\cbet{E}\}$, while leaving $\alp{0}$ untouched. We want to store our knowledge of the shell once it has been obtained, and so we pass the label \lstinline!"Export"->"EinsteinCartan"!, which will be used to construct a filename. The input is
\begin{lstlisting}[breaklines=true]
In[]:= DefTheory[{Alp1 == 0, Alp2 == 0, Alp3 == 0, Alp4 == 0, Alp5 == 0, Alp6 == 0, Bet1 == 0, Bet2 == 0, Bet3 == 0, cAlp1 == 0, cAlp2 == 0, cAlp3 == 0, cAlp4 == 0, cAlp5 == 0, cAlp6 == 0, cBet1 == 0, cBet2 == 0, cBet3 == 0}, "Export" -> "EinsteinCartan", "Order" -> Infinity]; 
Out[]= |
\vspace{-10pt}
|
** DefTheory: Found the following primary if-constraints:
|
\vspace{-4pt}
\begin{flushleft}
\includegraphics[width=\linewidth]{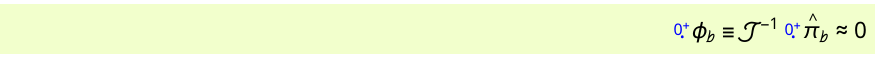}
\end{flushleft}
\vspace{-5pt}
|
|
\vspace{-4pt}
\begin{flushleft}
\includegraphics[width=\linewidth]{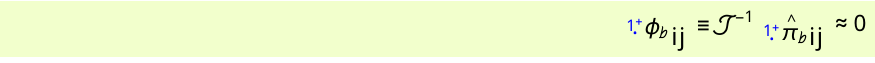}
\end{flushleft}
\vspace{-5pt}
|
|
\vspace{-4pt}
\begin{flushleft}
\includegraphics[width=\linewidth]{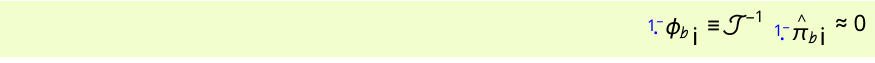}
\end{flushleft}
\vspace{-5pt}
|
|
\vspace{-4pt}
\begin{flushleft}
\includegraphics[width=\linewidth]{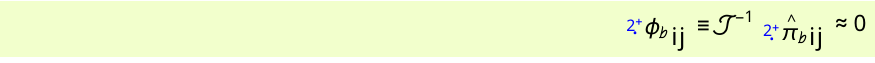}
\end{flushleft}
\vspace{-5pt}
|
|
\vspace{-4pt}
\begin{flushleft}
\includegraphics[width=\linewidth]{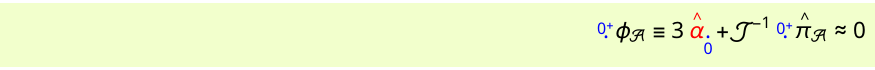}
\end{flushleft}
\vspace{-5pt}
|
|
\vspace{-4pt}
\begin{flushleft}
\includegraphics[width=\linewidth]{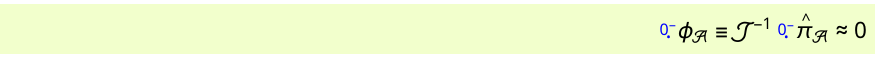}
\end{flushleft}
\vspace{-5pt}
|
|
\vspace{-4pt}
\begin{flushleft}
\includegraphics[width=\linewidth]{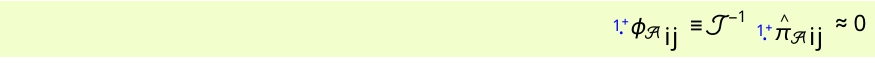}
\end{flushleft}
\vspace{-5pt}
|
|
\vspace{-4pt}
\begin{flushleft}
\includegraphics[width=\linewidth]{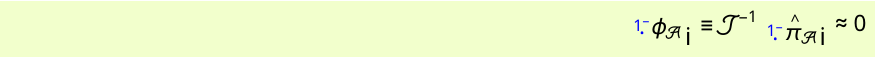}
\end{flushleft}
\vspace{-5pt}
|
|
\vspace{-4pt}
\begin{flushleft}
\includegraphics[width=\linewidth]{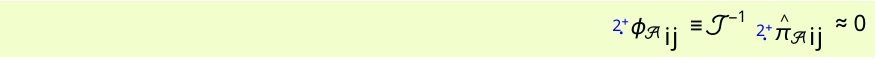}
\end{flushleft}
\vspace{-5pt}
|
|
\vspace{-4pt}
\begin{flushleft}
\includegraphics[width=\linewidth]{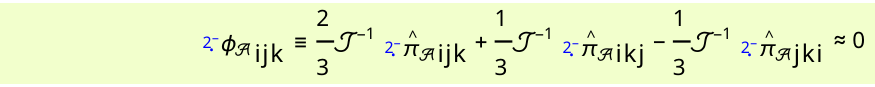}
\end{flushleft}
\vspace{-5pt}
|
|
\vspace{-10pt}
|
** DefTheory: Found the following secondary perpendicular if-constraints:
|
\vspace{-10pt}
|
** DefTheory: Found the following secondary parallel if-constraints:
|
\vspace{-10pt}
|
** DefTheory: Found the following secondary singular if-constraints:
|
\vspace{-10pt}
|
** DefTheory: The super-Hamiltonian is:
|
\vspace{-4pt}
\begin{flushleft}
\includegraphics[width=\linewidth]{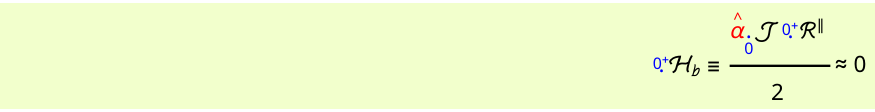}
\end{flushleft}
\vspace{-5pt}
|
|
\vspace{-10pt}
|
** DefTheory: The linear super-momentum is:
|
\vspace{-4pt}
\begin{flushleft}
\includegraphics[width=\linewidth]{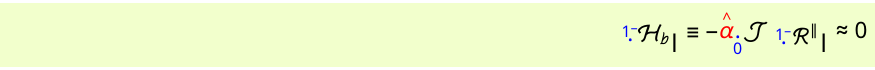}
\end{flushleft}
\vspace{-5pt}
|
|
\vspace{-10pt}
|
** DefTheory: The 1- part of the angular super-momentum is:
|
\vspace{-4pt}
\begin{flushleft}
\includegraphics[width=\linewidth]{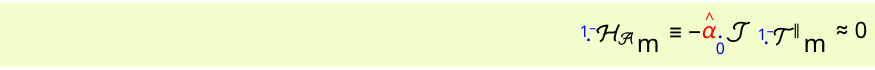}
\end{flushleft}
\vspace{-5pt}
|
|
\vspace{-10pt}
|
** DefTheory: The 1+ part of the angular super-momentum is:
|
\vspace{-4pt}
\begin{flushleft}
\includegraphics[width=\linewidth]{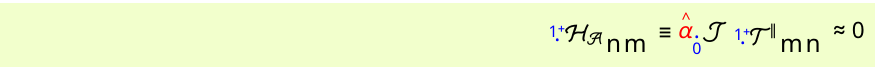}
\end{flushleft}
\vspace{-5pt}
|
|
\vspace{-10pt}
|
 ** DefTheory: Exporting the binary at svy/EinsteinCartan.thr.mx
\end{lstlisting}
In the output above we can see the listing of the PiCs. It is clear from~\eqref{ehl} and also from the form of the constraints in~\cref{fulpic,fulpictor}, why we get this specific list. All the coefficients associated with the $\soonethree$ irreps of $\tensor{\mathcal{R}}{_{ijkl}}$ and $\tensor{\mathcal{T}}{_{ijk}}$ are vanishing, and so all the PiC functions become constraints. Moreover, these constraints are very simple in their form: they are all pure momenta, with the single exception being in the case of the $0^+$ roton constraint, which contains a constant
\begin{equation}
	\pic[]{A0p}=3\alp{0}\planck^2+\frac{1}{J}\PiP[]{A0p}\approx 0.
	\label{funnyconstraint}
\end{equation}

Following this listing of the PiCs, there are references to perpendicular, parallel and singular SiCs, none of which are present in the Einstein--Cartan theory. We will return to these in~\cref{simplehpcsurvey}, but for now we note that they follow from the imposition of multipliers in~\eqref{neocon}, and their presence is fully understood in~\cite{smooth}, just as the presence of the PiCs of the basic Poincar\'e gauge theory is understood in~\cite{1983PhRvD..28.2455B,2018PhRvD..98b4014B}. This is the intended scope of \lstinline!DefTheory[]!: to elucidate not only the primary constraints, but all of that part of the constraint structure which is already known from the literature. Every if-constraint identified by \lstinline!DefTheory[]! is used during the \lstinline!DefTheory[]! call in the construction of very large internal rule sets \lstinline!HiGGS§StrengthPShellToStrengthPO3!, \lstinline!HiGGS§PiPShellToPiPPO3!, \lstinline!HiGGS§TheoryCDPiPToCDPiPO3! and \lstinline!HiGGS§TheoryPiPToPiPO3!. These rules are applied -- by default -- during \lstinline!ToNesterForm! calls. Most of the if-constraints that can arise, as detailed in~\cite{smooth}, are of the form $\frac{1}{J}\tensor[^A]{\hat{\pi}}{_{\acu{u}}}+\cdots\approx 0$ or $\frac{1}{J}\tensor[^E]{\hat{\pi}}{_{\acu{u}}}+\cdots\approx 0$, and so the shell is defined by \emph{replacing} all possible instances of the momenta in favor of other quantities. There are also the `parallel' SiCs, which deactivate irreps within the canonical parts of the field strength tensors in~\cref{koffl,foffl} -- these are straightforwardly implemented.

Following the listing of the if-constraints, the output above details the structure of the canonical Hamiltonian, which from~\cite{smooth} is written
\begin{equation}
  \tensor{\mathcal{H}}{_{\text{C}}}\equiv N\tensor{\mathcal{H}}{_\perp}+\tensor{N}{^\alpha}\tensor{\mathcal{H}}{_\alpha}-\frac{1}{2}\tensor{A}{^{ij}_0}\tensor{\mathcal{H}}{_{ij}}+\tensor{\partial}{_\alpha}\tensor{\mathscr{D}}{^\alpha}.
  \label{canonicalhamiltonian}
\end{equation}
This \emph{Dirac} form~\cite{1987PhRvD..35.3748B,PhysRevD.30.2508} is useful because it expresses part of the Hamiltonian as a linear combination of the nonphysical fields $N$, $\tensor{N}{^\alpha}$ and $\tensor{A}{^{ij}_{0}}$. The physical (and canonical) coefficients of these undetermined fields are\footnote{Note that $\tensor{\mathscr{D}}{^\alpha} \equiv \tensor{b}{^i_0}\tensor{\pi}{_i^\alpha}+\frac{1}{2}\tensor{A}{^{ij}_0}\tensor{\pi}{_{ij}^\alpha}$.}
\begin{subequations}
\begin{align}
  \tensor{\mathcal{H}}{_\perp} &\equiv\tensor{\hat{\pi}}{_i^{\overline{k}}}\tensor{\mathcal{  T}}{^i_{\perp\overline{k}}}+\frac{1}{2}\tensor{\hat{\pi}}{_{ij}^{\overline{k}}}\tensor{\mathcal{  R}}{^{ij}_{\perp\overline{k}}}-JL_{\text{G}}-\tensor{n}{^k}\tensor{D}{_\alpha}\tensor{\pi}{_k^\alpha},\label{total_hamiltonian}\\
  \tensor{\mathcal{H}}{_\alpha} & \equiv\tensor{\pi}{_i^\beta}\tensor{T}{^i_{\alpha\beta}}+\frac{1}{2}\tensor{\pi}{_{ij}^\beta}\tensor{R}{^{ij}_{\alpha\beta}}-\tensor{b}{^k_\alpha}\tensor{D}{_\beta}\tensor{\pi}{_k^\beta},\label{total_hamiltonian_start}\\
  \tensor{\mathcal{H}}{_{ij}}& \equiv 2\tensor{\pi}{_{[i}^\alpha}\tensor{b}{_{j]\alpha}}+\tensor{D}{_\alpha}\tensor{\pi}{_{ij}^\alpha},\label{total_hamiltonian_int}
\end{align}
\end{subequations}
and they form the `sure' secondary FC constraints (sSFCs)
\begin{equation}
  \tensor{\mathcal{H}}{_\perp}\approx 0, \quad \tensor{\mathcal{H}}{_\alpha}\approx 0, \quad \tensor{\mathcal{H}}{_{ij}}\approx 0.
  \label{suresecondaries}
\end{equation}
We see that these 10 constraints are divided up under $\othree$ to give the final four entries in the output above, with $\tensor{\mathcal{H}}{_{ij}}=\tensor{\mathcal{H}}{_{\ovl{ij}}}+2\foli{{[}i|}\tensor{\mathcal{H}}{_{\perp|\ovl{j}{]}}}$. The (linearised) values returned for $\tensor{\mathcal{H}}{_\perp}$ and $\quad \tensor{\mathcal{H}}{_\alpha}$ seem sensible in the context of the constraint in~\eqref{funnyconstraint} -- we can see from this that \HiGGS{} has begun imposing the PiC shell. What about the angular super-momentum? Let us verify by hand that the answer is correct. The first term in~\eqref{total_hamiltonian_int} vanishes on the PiC shell, because \HiGGS{} has told us that $\PiP{B0p}\approx\PiP{B1p}\approx\PiP{B1m}\approx\PiP{B2p}\approx0$. For the second term, we find from~\eqref{funnyconstraint} after a few lines that
\begin{equation}
	\tensor{\pi}{_{kl}^\beta}\approx-2\alp{0}\planck^2 J\foli{{[}k}\tensor{h}{_{\ovl{l}{]}}^\beta},
\end{equation}
from which we determine
\begin{equation}
	\begin{aligned}
		\tensor{\mathcal{H}}{_{kl}}&\approx -2\alp{0}\planck^2J\bigg[
		\tensor{h}{_{\ovl{m}}^\beta}\left(\tensor{\partial}{_\alpha}\tensor{b}{^m_\alpha}\right)\foli{{[}k}\tensor{h}{_{\ovl{l}{]}}^\alpha}
		\\
		&\ \ \ \ \ \ \ \ \ \ \ \ 
		+\tensor{D}{_\alpha}\left(\foli{{[}k}\tensor{h}{_{\ovl{l}{]}}^\alpha}\right)
	\bigg]
		\\
		&
		=-2\alp{0}\planck^2J\bigg(\cT[\ovl{kl}]{B1p}+2\foli{{[}k}\cT[\ovl{l}{]}]{B1m}\bigg).
	\end{aligned}
\end{equation}
These are indeed the $1^+$ and $1^-$ parts of the torsion, as \HiGGS{} is claiming.

\subsubsection{Module: `StudyTheory' and `Velocity'}\label{amalgum}

The greater part of the analysis -- i.e. that which is not necessarily encoded in the literature -- is requested by means of the \lstinline!StudyTheory[]! command. We will see in~\cref{parallelisationandhpc} that this command is just a high-level wrapper for parallelising Poisson matrices and constraint velocities, and also for \lstinline!DefTheory[]! as discussed in~\cref{deftheory}. We will see how to use \lstinline!StudyTheory[]! for a batch of theories in~\cref{yo-nesterunittests}. For now we note that having called \lstinline!DefTheory[]! above, we can pass the option \lstinline!"Import"->True! to the command
\begin{lstlisting}[breaklines=true]
In[]:= JobsBatch = {{"EinsteinCartan", {Alp1 == 0, Alp2 == 0, Alp3 == 0, Alp4 == 0, Alp5 == 0, Alp6 == 0, Bet1 == 0, Bet2 == 0, Bet3 == 0, cAlp1 == 0, cAlp2 == 0, cAlp3 == 0, cAlp4 == 0, cAlp5 == 0, cAlp6 == 0, cBet1 == 0, cBet2 == 0, cBet3 == 0}}}; 
JobsBatch~StudyTheory~("Import" -> True);
\end{lstlisting}
and -- since the calculation is expensive -- run it through a Wolfram Language package file. The evaluation of the Poisson matrix is fairly straightforward, and made up from calls to \lstinline!PoissonBracket[]! as discussed in~\cref{poissonbracket}. It is better then to turn to \lstinline!Velocity[]!. 

The linearised velocity of (e.g.) some PiC $\tensor[^B]{\varphi}{_{\acu{v}}}$ is calculated using the formula
\begin{equation}
  \tensor[^B]{\dot{\varphi}}{_{\acu{v}}}(\bm{x}_1)=\int\mathrm{d}^3x_2\left\{\tensor[^B]{\varphi}{_{\acu{v}}}(\bm{x}_1),\tensor{\mathcal{H}}{_{\text{T}}}(\bm{x}_2)\right\},
\label{svel}
\end{equation}
where from~\eqref{canonicalhamiltonian} we need evaluate only the commutator with $\mathcal{H}_{\perp}$, which is re-expressed in~\cite{smooth} on the PiC shell as
\begin{equation}
\begin{aligned}
  \tensor{\mathcal{H}}{_\perp}&\equiv\frac{J}{64}\sum_{A}\ctmp[A]{\perp}\mu(\alpm[A]{\perp\perp})\tensor[^A]{\varphi}{_{\acu{v}}}\tensor[^A]{\varphi}{^{\acu{v}}}\\
  &\ \ \ +\frac{J}{16\planck}\sum_{E}\ctmp[E]{\perp}\mu(\betm[E]{\perp\perp})\tensor[^E]{\varphi}{_{\acu{v}}}\tensor[^E]{\varphi}{^{\acu{v}}}+\frac{1}{2}\alp{0}\planck^2\cR[]{A0p}\\
    &\ \ \ -J\sum_{I}\Big(\alp{I}\tensor{\mathcal{  R}}{^{ij}_{\ovl{kl}}}+\calp{I}\tensor{\lambda}{^{ij}_{kl}}\Big)\projlore[_{ij}^{kl}_{nm}^{\ovl{pq}}]{I}\tensor{\mathcal{  R}}{^{nm}_{\ovl{pq}}}\\
    &\ \ \ -J\planck^2\sum_{M}\Big(\bet{M}\tensor{\mathcal{  T}}{^{i}_{\ovl{kl}}}+\cbet{M}\tensor{\lambda}{^{i}_{kl}}\Big)\projlore[_{i}^{kl}_{n}^{\ovl{pq}}]{M}\tensor{\mathcal{  T}}{^{n}_{\ovl{pq}}}\\
	&\ \ \ -\tensor{n}{^k}\tensor{D}{_\alpha}\tensor{\pi}{_k^\alpha}.\label{grandperp}
\end{aligned}
\end{equation}
A particular problem now is that~\eqref{grandperp} has a \emph{quadratic} structure. If we apply \lstinline!PoissonBracket[]! to evaluate a velocity for some PiC $\{\tensor[^B]{\varphi}{_{\acu{v}}},\mathcal{H}_{\perp}\}$, we will actually be calling \lstinline!ToBasicForm[]! directly on $\mathcal{H}_{\perp}$. This will generally result in a very expensive computation, which we avoid in \lstinline!Velocity[]! by manually applying the Leibniz rule. Accordingly in the \HiGGS{} source there \emph{is} a collection of prepared formulae for generic expressions such as
\begin{equation}
	\begin{aligned}
		\left\{\tensor[^B]{\varphi}{_{\acu{v}}},\mathcal{H}_{\perp}\right\}&=\frac{J}{32}\sum_{A}\ctmp[A]{\perp}\mu(\alpm[A]{\perp\perp})\left\{\tensor[^B]{\varphi}{_{\acu{v}}},\tensor[^A]{\varphi}{_{\acu{u}}}\right\}\tensor[^A]{\varphi}{^{\acu{u}}}+...,
		\\
		&=\frac{1}{32}\sum_{A}\ctmp[A]{\perp}\mu(\alpm[A]{\perp\perp})\Bigg[J\tensor{J}{_{1\acu{v}\acu{u}}}\tensor[^A]{\varphi}{^{\acu{u}}}+
		\\
		&\ \ \ \ \ \
		+\tensor{J}{_{2\acu{v}\acu{u}}^\alpha}\tensor{\partial}{_\alpha}\left(J\tensor[^A]{\varphi}{^{\acu{u}}}\right)
		-\tensor{\partial}{_\alpha}\left(J\tensor{J}{_{3\acu{v}\acu{u}}^\alpha}\tensor[^A]{\varphi}{^{\acu{u}}}\right)
		\\
		&\ \ \ \ \ \ \ \
		-\tensor{\partial}{_\alpha}\left(\tensor{J}{_{4\acu{v}\acu{u}}^{\alpha\beta}}\tensor{\partial}{_\beta}\left(J\tensor[^A]{\varphi}{^{\acu{u}}}\right)\right)\Bigg]+...,
	\label{template}
	\end{aligned}
\end{equation}
which collectively act as a template for each velocity. For each $\othree$ index $A$ shown in the sum in the second equality in~\eqref{template}, we need the four portions of a single Poisson bracket, as given in~\eqref{coefflist} and returned as a \lstinline!List! by a call to \lstinline!PoissonBracket[]!. Accordingly, the whole of~\eqref{grandperp} is broken into \emph{blocks}, each of four terms, which follow from a single bracket, and these are evaluated in parallel as discussed in~\cref{parallelisationandhpc}.

As we mentioned in~\cref{poissonbracket}, the formula~\eqref{coefflist} is not explicitally covariant, and hence the appearance in~\eqref{template} of partial derivatives. Ultimately, a call to \lstinline!ToNesterForm[]! is needed to restore explicit covariance to the quadratic expression~\eqref{template}. This setup is not so good: the use of block formulae restricts us to calculating velocities over the given $\mathcal{H}_{\perp}$ -- it is also susceptible to human error, and concludes with an expensive simplification process.
We propose that any future iterations of \HiGGS{} implement velocities based two extensions to \lstinline!PoissonBracket[]!:
\begin{enumerate}
	\item The operands should be expressed as sums of products of covariant quantities, and the Leibniz rule be used to distribute the operation. 
	\item The operation should return the covariant form given in \eqref{defpoi}, so that covariance is maintained at all steps.
\end{enumerate}
An improvement of \lstinline!PoissonBracket[]! along these lines should be straightforward to implement, and would also be equally emenable to parallelisation: we defer its development to future work.


\subsubsection{Module: `ViewTheory'}\label{viewtheory}

The final module \lstinline!ViewTheory[]! is used to produce a human-readable summary of the results from \lstinline!StudyTheory[]!. Since we already have a summary of the literature constraint structure from the \lstinline!DefTheory[]! call in~\cref{deftheory}, we pass for brevity our earlier theory name with the option \lstinline!"Literature"->False!, producing
\begin{lstlisting}[breaklines=true]
In[]:= ViewTheory["EinsteinCartanVelocities", "Literature" -> False]; 
Out[]= |
\vspace{-10pt}
|
 ** DefTheory: Incorporating the binary at svy/EinsteinCartanVelocities.thr.mx
|
\vspace{-4pt}
\begin{flushleft}
\includegraphics[width=\linewidth]{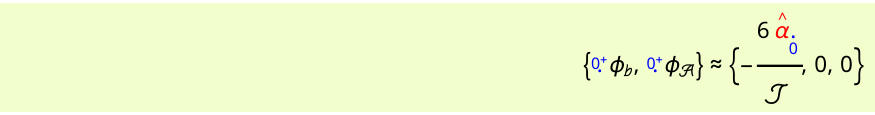}
\end{flushleft}
\vspace{-5pt}
|
|
\vspace{-4pt}
\begin{flushleft}
\includegraphics[width=\linewidth]{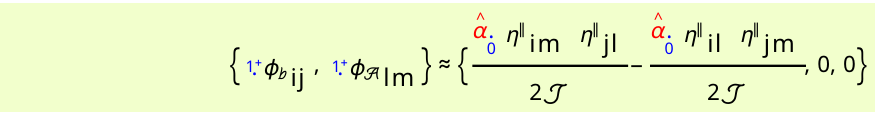}
\end{flushleft}
\vspace{-5pt}
|
|
\vspace{-4pt}
\begin{flushleft}
\includegraphics[width=\linewidth]{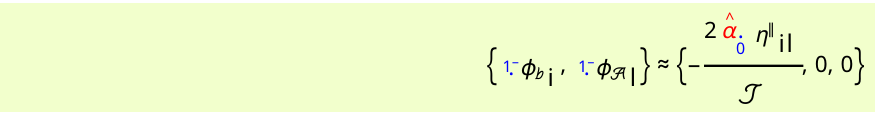}
\end{flushleft}
\vspace{-5pt}
|
|
\vspace{-4pt}
\begin{flushleft}
\includegraphics[width=\linewidth]{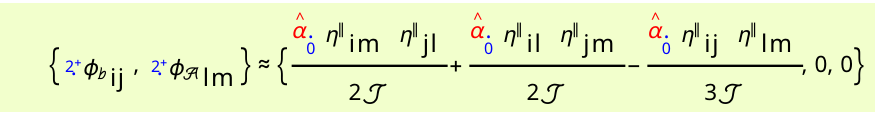}
\end{flushleft}
\vspace{-5pt}
|
|
\vspace{-4pt}
\begin{flushleft}
\includegraphics[width=\linewidth]{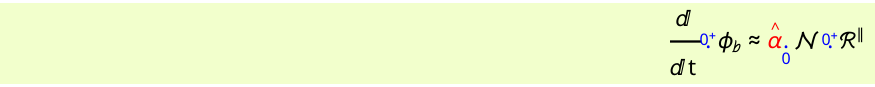}
\end{flushleft}
\vspace{-5pt}
|
|
\vspace{-4pt}
\begin{flushleft}
\includegraphics[width=\linewidth]{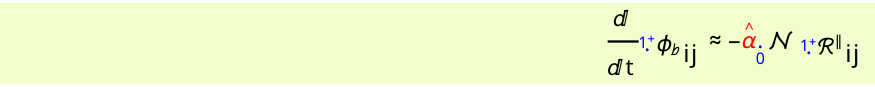}
\end{flushleft}
\vspace{-5pt}
|
|
\vspace{-4pt}
\begin{flushleft}
\includegraphics[width=\linewidth]{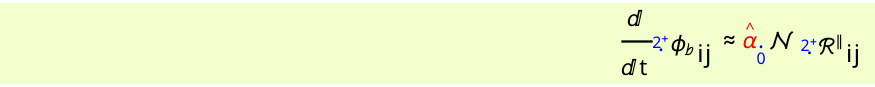}
\end{flushleft}
\vspace{-5pt}
|
|
\vspace{-4pt}
\begin{flushleft}
\includegraphics[width=\linewidth]{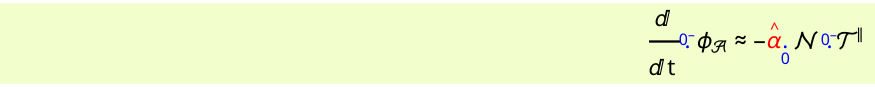}
\end{flushleft}
\vspace{-5pt}
|
|
\vspace{-4pt}
\begin{flushleft}
\includegraphics[width=\linewidth]{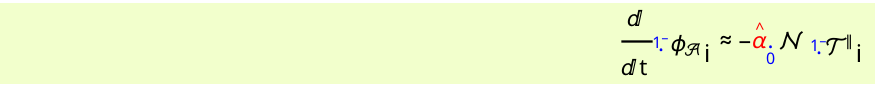}
\end{flushleft}
\vspace{-5pt}
|
|
\vspace{-4pt}
\begin{flushleft}
\includegraphics[width=\linewidth]{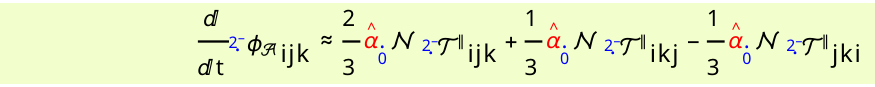}
\end{flushleft}
\vspace{-5pt}
|
\end{lstlisting}
We see how the results are stored in the theory binary at \lstinline!svy/EinsteinCartan.thr.mx!. There are \emph{four} nonvanishing Poisson brackets, all between the translational and rotational pairs of the same $\othree$ irrep, $\pic[]{B0p}$ vs $\pic[]{A0p}$, $\pic[\ovl{ij}]{B1p}$ vs $\pic[\ovl{ij}]{A0p}$, $\pic[\ovl{i}]{B1m}$ vs $\pic[\ovl{i}]{A1m}$, and $\pic[\ovl{ij}]{B2p}$ vs $\pic[\ovl{ij}]{A2p}$.
These are the \emph{conjugate pairs}, whose commutators are are proportional to mass parameters in the theory~\cite{1983PhRvD..28.2455B} -- here mediated by the Einstein--Hilbert term $\alp{0}$. The involved PiCs are second class (SC). 

What about the remaining PiCs $\pic[]{A0m}$ and $\pic[]{A2m}$? These $0^-$ and $2^-$ irreps\footnote{Note that the velocity of $\pic[\ovl{ijk}]{A2m}$ is not simplified: \HiGGS{} inherits this inabiliy from \xAct{}, as no efficient algorithm is known for simplifying multi-term symmetries~\cite{Nutma:2013zea} such as that of the $2^-$ sector.} are not represented in the translational sector. Previous analyses tell us that their relevant commutators will be with their own secondaries~\cite{1983PhRvD..28.2455B}. Accordingly, we turn to the velocities, which are also provided in the above output. Unlike the if-constraints, the linearised sSFC Hamiltonian constraints are not automatically implemented when \lstinline!"ToShell"->True! is passed: we therefore note that since $\cRl[]{A0p}\approx\cTl[\ovl{i}]{B1p}\approx 0$, from the output of \lstinline!DefTheory[]! above, both rotational and translational (Hamiltonian) multipliers for the $0^+$ and $1^-$ sectors will also vanish at linear order. The same is true of the translational $1^+$ and $2^+$ Hamiltonian multipliers, but the rotational counterparts will be linearly proportional to $\cRl[\ovl{ij}]{A1p}$ and $\cRl[\ovl{ij}]{A2p}$. Returning to the `lonely' $0^+$ and $2^-$ sectors, we can go right ahead and calculate the expected brackets mentioned above. We first find
\begin{lstlisting}[breaklines=true]
In[]:= PoissonBracket[PhiA0m[], TP0m[], "ToShell" -> True]; 
Out[]= |
\vspace{-4pt}
\begin{flushleft}
\includegraphics[width=\linewidth]{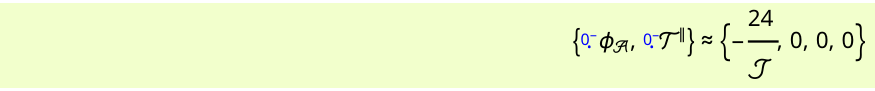}
\end{flushleft}
\vspace{-5pt}
|
\end{lstlisting}
As expected, this commutator survives at linear order: one could determine also the $0^-$ multiplier, but it is enough to notice that both $\pic[]{A0m}$ and its linearised secondary\footnote{The ($\flat$) symbol denotes linearisation near Minkowski spacetime.} $\sicl[]{A0m}\equiv\cTl[]{A0m}\approx 0$ also become SC. Moving on, we find
\begin{lstlisting}[breaklines=true]
In[]:= PoissonBracket[PhiA2m[-i, -j, -k], TP2m[-l, -m, -n], "ToShell" -> True]; 
Out[]= |
\vspace{-4pt}
\begin{flushleft}
\includegraphics[width=\linewidth]{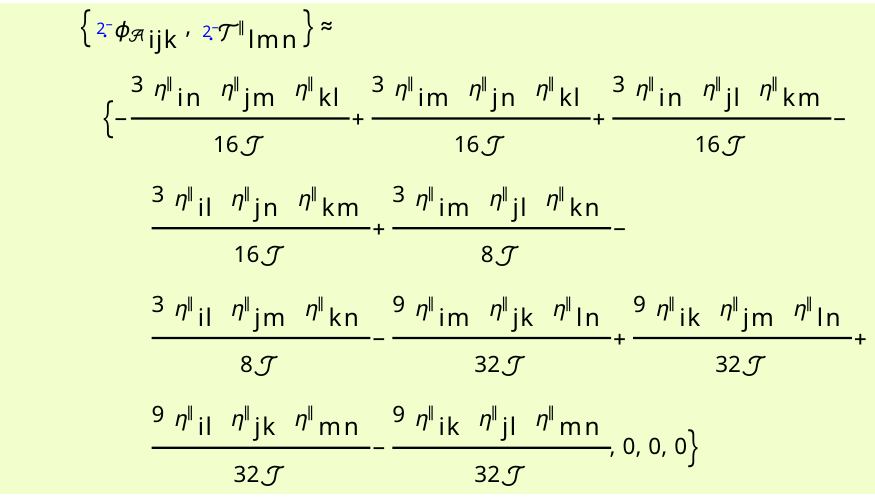}
\end{flushleft}
\vspace{-5pt}
|
\end{lstlisting}
The same, then, is true of $\pic[\ovl{ijk}]{A2m}$ and its linearised secondary $\sicl[\ovl{ijk}]{A2m}\equiv\cTl[\ovl{ijk}]{A2m}\approx 0$, and so the algorithm terminates.

What can \HiGGS{} tell us about the physics of Einstein--Cartan theory? The PGT contains $2\times(24+16)=80$ na\"ive d.o.fs in its gauge fields. The non-physicality of the lapse and shift (the Poincar\'e gauge symmetry), remove $2\times 10$ d.o.fs through the sure, primary (sPFC) constraints; a further $20$ d.o.f are removed via the sSFCs. We further learn from \HiGGS{} about the number and class (all SC) of the if-constraints. The final d.o.f count is thus
\begin{equation}
  \begin{aligned}
  \frac{1}{2}(&80-2\times 10-2\times 10\\
  &-(1+1+3+3+3+3+5+5)\\
&-(1+1+5+5))=2,
\label{carbon}
\end{aligned}
\end{equation}
i.e. the massless graviton.

\subsection{Parallelisation and HPC}\label{parallelisationandhpc}

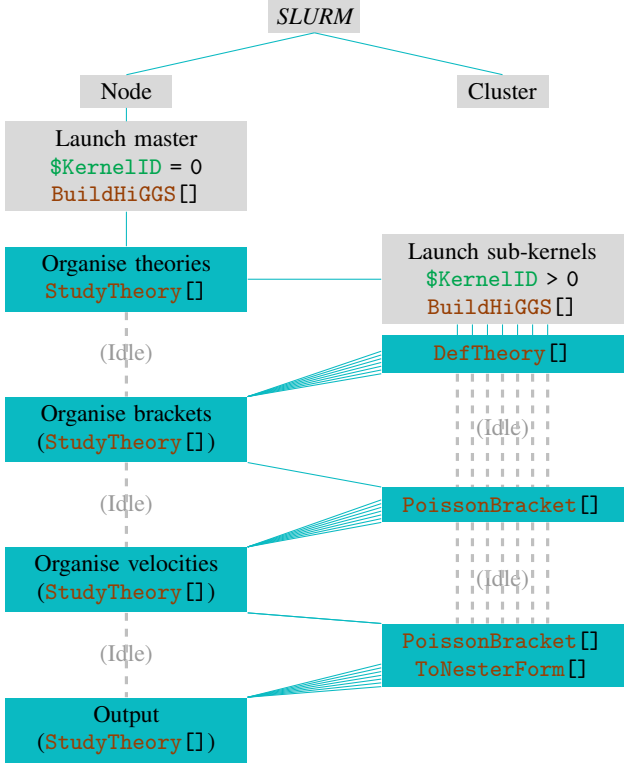
\begin{figure}
\tikzstyle{hpc} = [rectangle, minimum width=1cm, minimum height=0cm,text centered, fill=gray!30,anchor = center,text width = 1cm]
\tikzstyle{idl} = [rectangle, minimum width=1cm, minimum height=0cm,text centered, fill=white!50,anchor = center,text width = 1cm,opacity=0.4]
\tikzstyle{lau} = [rectangle, minimum width=3cm, minimum height=0cm,text centered, fill=gray!30,anchor = center,text width = 3cm]
\tikzstyle{org} = [rectangle, minimum width=3cm, minimum height=0cm,text centered, fill=Aquamarine!90,anchor = center,text width = 3cm]
\tikzstyle{cal} = [rectangle, minimum width=3cm, minimum height=0cm,text centered, fill=Aquamarine!90,anchor = center,text width = 3cm]
\tikzstyle{arrow} = [Aquamarine!90,thin]
\tikzstyle{idle} = [gray!50,dashed,very thick]
\begin{tikzpicture}[node distance=1cm]
  \node (slurm) [hpc] {\SLURM{}};
	\node (node) [hpc, below left = 1cm and 2.5cm of slurm.center,anchor=center] {Node};
	\node (cluster) [hpc, right = 5cm of node.center,anchor=center] {Cluster};
	\node (laumas) [lau, below = 1cm of node.center,anchor=center] {Launch master \lstinline!WL§KernelID = 0! \lstinline!BuildHiGGS[]!};
	\node (orgdef) [org, below = 1.5cm of laumas.center,anchor=center] {Organise theories \lstinline!StudyTheory[]!};
	\node (orgidl) [idl, below = 1cm of orgdef.center,anchor=center] {(Idle)};
	\node (lausla) [lau, right = 5cm of orgdef.center,anchor=center] {Launch sub-kernels \lstinline!WL§KernelID > 0! \lstinline!BuildHiGGS[]!};
	\node (caldef) [cal, below = 1cm of lausla.center,anchor=center] {\lstinline!DefTheory[]!};
	\node (orgidl) [idl, below = 1cm of caldef.center,anchor=center] {(Idle)};
	\node (orgppm) [org, below = 2cm of orgdef.center,anchor=center] {Organise brackets (\lstinline!StudyTheory[]!)};
	\node (orgidl) [idl, below = 1cm of orgppm.center,anchor=center] {(Idle)};
	\node (calppm) [cal, below = 2cm of caldef.center,anchor=center] {\lstinline!PoissonBracket[]!};
	\node (orgidl) [idl, below = 1cm of calppm.center,anchor=center] {(Idle)};
	\node (orgvel) [org, below = 2cm of orgppm.center,anchor=center] {Organise velocities (\lstinline!StudyTheory[]!)};
	\node (orgidl) [idl, below = 1cm of orgvel.center,anchor=center] {(Idle)};
	\node (calvel) [cal, below = 2cm of calppm.center,anchor=center] {\lstinline!PoissonBracket[]! \lstinline!ToNesterForm[]!};
	\node (orgend) [org, below = 2cm of orgvel.center,anchor=center] {Output (\lstinline!StudyTheory[]!)};
	\draw [arrow] (slurm.south) -- (cluster.north);
	\draw [arrow] (slurm.south) -- (node.north);
	\draw [arrow] (node) -- (laumas);
	\draw [arrow] (laumas) -- (orgdef);
	\draw [arrow] (orgdef) -- (lausla);
	\draw [arrow] (orgppm.south east) -- (calppm.north west);
	\draw [arrow] (orgvel.south east) -- (calvel.north west);
	\draw [arrow] (orgvel.south east) -- (calvel.north west);
	\draw [idle] (orgdef) -- (orgppm);
	\draw [idle] (orgppm) -- (orgvel);
	\draw [idle] (orgvel) -- (orgend);
	\begin{scope}
		\foreach \i in {-3,...,3}{%
			\draw [arrow] ([xshift=\i * 0.2 cm]lausla.south) -- ([xshift=\i * 0.2 cm]caldef.north);
			\draw [idle] ([xshift=\i * 0.2 cm]caldef.south) -- ([xshift=\i * 0.2 cm]calppm.north);
			\draw [idle] ([xshift=\i * 0.2 cm]calppm.south) -- ([xshift=\i * 0.2 cm]calvel.north);}
		\foreach \i in {0,...,6}{%
			\draw [arrow] ([yshift=\i * 0.05 cm]caldef.south west) -- (orgppm.north east);
			\draw [arrow] ([yshift=\i * 0.05 cm]calppm.south west) -- (orgvel.north east);
			\draw [arrow] ([yshift=\i * 0.05 cm]calvel.south west) -- (orgend.north east);}
	\end{scope}
\end{tikzpicture}
\caption{\label{flowchart} The parallelisation in \HiGGS{} is not sophisticated, but it allows for pragmatic scaling of the analysis to surveys. A large batch of theories can be broken via \lstinline!SBATCH! directives over computation nodes, with a reasonable group size of one theory per core per node. Each node is controlled by a master kernel, which coordinates its portion of the survey through a single call to \lstinline!StudyTheory[]!. Three phases of work are then delegated to the sub-kernels. First, the constraint shell is partially reconstructed from the literature knowledge of the theory. The commutators between all constraints of all theories are then evaluated in parallel. Finally, the velocities of the constraints are decomposed into ther constituent Poisson brackets: these are evaluated and simplified in parallel before returning to the master kernel for synthesis of the results.}
\end{figure}

The current version of \HiGGS{} is not only designed for the local or desktop operations conducted in~\crefrange{geometricsetup}{high-levelfunctions}. The intended use is the evaluation of quantities useful to the Hamiltonian analysis, in an environment where multiple parallel kernels are available. As mentioned in~\cref{introduction}, there are some generic features of the Hamiltonian analysis which lend themselves very well to parallelisation. 

Na\"ively, we notice that once the primary constraints have been identified, the resulting constraint chains (i.e. the recursive consistency conditions of each primary) can notionally be evaluated in parallel. This seems a natural route down which to parallelise, but in reality chains might seem to continue indefinitely in a kernel with knowledge only of the primary constraint shell. Velocities are the most expensive quantities, and so it seems prudent to pause after each is calculated, to see if the corresponding acceleration is strictly necessary. However, when we parallelise over chains we also find that the velocities, accelerations and jerks rapidly desynchronise: this is to be expected given the varying complexities of the $\othree$ irreps which underlie each chain. Accordingly, pre-velocity checks would require e.g. a centralised knowledge of the complete shell to be kept on the master kernel, to which sub-kernels could refer as needed. We could even imagine a setup where one sub-kernel is interrupted in mid-evaluation on the basis of a report from another.

While it is tempting to develop something sophisticated along these lines, the route turns out not to be practical. In practice, the number of chains is typically very few per theory (and strictly not more than ten in the case of the Poincar\'e gauge theory without multipliers). The desynchronisation effect is then so severe that most kernels would sit idle whilst waiting for the highest-spin chain to complete. More importantly, the question of whether chain $A$ need continue based on a new constraint from chain $B$, is not a trivial problem in computer algebra. In the case of the PiCs, each constraint has as one term a unique part of the momentum: as mentioned in~\cref{deftheory} this lends itself to a replacement rule which reliably implements the PiC shell, but in more general cases it is less clear how to proceed.

Ultimately, the experience of~\cite{1999IJMPD...8..459Y,2002IJMPD..11..747Y} suggests that evaluation of the accelerations is not usually necessary anyway. Once the brackets between the PiCs are known, it is often possible to determine from a visual inspection whether any chains are worth continuing. Velocities are expensive because of the structure of the Hamiltonian: $\mathcal{H}_\perp$ is a sum of terms which are quadratic in nontrivial covariant quantities such as the $\tensor[^A]{\varphi}{_{\acu{v}}}$ and $\projorthhum[_{\acu{v}}_{nm}^{\ovl{pq}}]{A}\tensor{\mathcal{  R}}{^{nm}_{\ovl{pq}}}$. Calculation of an overall bracket with $\mathcal{H}_\perp$ thus requires very many sub-brackets.

The quadratic structure of $\mathcal{H}_\perp$ underlies other inefficiencies in the current \HiGGS{} implementation, affecting all terms except for the final $-\tensor{n}{^k}\tensor{D}{_\alpha}\tensor{\pi}{_k^\alpha}$.
As discussed already in~\cref{amalgum}, the way in which \HiGGS{} calculates velocities is based on~\eqref{superpoisson} rather than~\eqref{coefflist}. As a result, the calculation of a bracket while finding the velocity of $\fA(\bm{x}_2)$, involving a quadratic term $\fB(\bm{x}_2)\fC(\bm{x}_2)$ in $\mathcal{H}_\perp$ produces -- in the first instance -- a non-covariant \emph{quadratic} expression which must be covariantised internally by expensive calls to \lstinline!ToNesterForm[]!. This could be readily improvable with development, and in many cases we find that the extra penalty incurred by the quadratic covariantisation remains comparable to the cost of the prerequisite brackets. 

Since brackets turn out to be the fundamental unit of the analysis, it is over brackets that \HiGGS{} is parallelised. Brackets are cumbersome for humans to evaluate and covariantise: these tasks are relatively easy for computers, now that the infrastructure of~\crefrange{geometricsetup}{high-levelfunctions} is in place. On the other hand, shells made from constraints whose format is arbitrary are less easy for computers to use. Once covariantised however, such constraints often comprise straightforward tensor equations which humans can readily manipulate. This suggests a pragmatical division of labour in which \HiGGS{} returns an organised structure of covariantised brackets -- formerly the primary Poisson matrix (PPM)~\cite{2002IJMPD..11..747Y,chapter4,mythesis} -- and velocities for human inspection. If velocities are expensive becuase they require multiple brackets, we can break them up accordingly, along with the quadratic covariantisation steps: in this way we can reduce chain desynchronisation.

The final structure of a \HiGGS{} survey is as follows;
\begin{enumerate}
  \item A list of theories is passed to \lstinline!StudyTheory[]! in the master kernel. This function runs \lstinline!DefTheory[]!, for each theory, in its own parallel sub-kernel. The literature knowledge of the constraint shell of each theory is cached as a binary.
  \item Within the same \lstinline!StudyTheory[]! call, the master kernel imports the shells and prepares a combined \lstinline!List! of brackets which need to be evaluated for the Poisson matrices. The \lstinline!List! elements are delegated to all available parallel sub-kernels -- as many of which are launched as necessary. Brackets are transferred to the next available sub-kernel using the Wolfram Symbolic Transfer Protocol (\WSTP{}). The \WSTP{} overhead is marginal, since each sub-kernel only needs to transfer data after its \lstinline!PoissonBracket[]! call\footnote{The use of \lstinline!Print[]! for debugging and development purposes means that this is not strictly true, and there is also an overhead from the system timing wrappers which we use to produce plots such as~\cref{hpcsurvey}. These features are not actually needed when running a survey.}.
  \item Within the same \lstinline!StudyTheory[]! call, the master kernel decomposes each linearised velocity into blocks dependent on a single bracket. For the velocity of some PiC $\tensor[^B]{\varphi}{_{\acu{v}}}$ the blocks are;
		\begin{equation}
		\begin{aligned}
			&\left\{ \tensor[^B]{\varphi}{_{\acu{v}}}, \tensor[^A]{\varphi}{_{\acu{u}}} \right\},\quad \forall A: \nu(\alpm[A]{\perp\perp})=1, 
			\\
			&\left\{ \tensor[^B]{\varphi}{_{\acu{v}}}, \tensor[^E]{\varphi}{_{\acu{u}}} \right\},\quad \forall E:  \nu(\betm[E]{\perp\perp})=1,
			\\
			&\left\{ \tensor[^B]{\varphi}{_{\acu{v}}}, \tensor{\mathcal{R}}{_{ij\ovl{kl}}} \right\},
			\quad
			\left\{ \tensor[^B]{\varphi}{_{\acu{v}}}, \tensor{\mathcal{T}}{_{i\ovl{kl}}} \right\},
			\\	
			&\left\{ \tensor[^B]{\varphi}{_{\acu{v}}}, J \right\},
			\quad
			\left\{ \tensor[^B]{\varphi}{_{\acu{v}}}, -\tensor{n}{^k}\tensor{D}{_\alpha}\tensor{\pi}{_k^\alpha} \right\}.
		\end{aligned}
		\end{equation}
		The blocks are delegated to the sub-kernels again using \WSTP{}. After the \lstinline!PoissonBracket[]! call, each sub-kernel pre-processes its block using \lstinline!ToNesterForm[]! before passing the result back to the master kernel.
	      \item Within the same \lstinline!StudyTheory[]! call the blocks are recombined into velocities, and all results are cached in a final binary.
\end{enumerate}
We illustrate this process in~\cref{flowchart}.
In the parallel \HiGGS{} environment of any node, one has at any one time a collection of theories, each of which has associated with it a complex and growing structure of constraints and commutators. This environment suggests an object-oriented approach. While \Mathematica{} allows for object-oriented programming using e.g. \lstinline!Association[]! (see also~\cite{oop}), the scope of \HiGGS{} is simple enough that we can retain a procedural approach. 
In fact, the shell structure of any one theory is referenced fairly infrequently, matching the rate of \lstinline!PoissonBracket[]! calls (see~\cref{hpcsurvey} for an illustration of this). This means that while \WSTP{} is necessary for the scheduling of evaluations, the information needed to switch between theories within a sub-kernel can be sourced by using relay system of binary files. Consequently, very minimal use is made by \HiGGS{} of \lstinline!DistributeDefinitions[]! to transfer theory-specific data between kernels.
In a more serious implementation of the Dirac--Bergmann algorithm, we would envisage more extensive use of \WSTP{} and the object-oriented approach.

How efficient is the above approach? Pending an implementation of the Leibniz rule, the efficacy of~\cref{flowchart} as it applies to the whole constraint algorithm is hard to gauge. The full run is only implemented in~\cref{yo-nesterunittests} for a handful of previously studied theories (without multipliers). Averaged over those cases, and for the non-fundamental reasons outlined above, a minority of the \lstinline!Velocity[]! calls contain serial tasks which turn out to dominate the workload. The truly parallel fraction of the workload (including the initial evaluation of all constraint brackets) is then as low as $p\sim 10^{-2}$. Modelling the whole implementation in~\cref{flowchart} as an Amdahl task~\cite{10.1145/327070.327215}, the $n$-core speedup
\begin{equation}
  S(n)=\frac{1}{1-p+\frac{p}{n}},
  \label{Amdahl}
\end{equation}
would appear very limited. We get a fairer understanding, however, from the `calibration' survey set out in~\cref{simplehpcsurvey}. In~\cref{effic} we run a portion of this survey -- spanning only $2^6=64$ modified gravity theories -- on a cluster with a vairable number of cores per node. We take this to be a task of `reasonable size' when using \HiGGS{} to learn about the canonical structure of a theory. Since the survey does not contain \lstinline!Velocity[]! calls, $S(n)$ will be more sensitive to the relevant trade-off between \WSTP{} overhead and bracket evaluation. There are relatively few brackets, and this portion of the survey is now dominated by the (fast but serial) \lstinline!BuildHiGGS[]! and \lstinline!DefTheory[]! calls: the benchmarking is thus quick to run, though it still gives an impression of low efficiency due to~\eqref{Amdahl}.
The normalising time is based on five cores per node: roughly equal to the per-node number of theories.
We expect $S(n)\equiv \langle t(5)\rangle/\langle t(n)\rangle$ to be concave and sublinear, but we do not see it peak in our use-case. 
No Hamiltonian analysis tools have previously been made that could provide a more meaningful benchmark\footnote{See however an \xAct{} implementation of the $3+1$ Baumgarte--Shapiro--Shibata--Nakamura (BSSN) formulation of the bimetric gravity field equations~\cite{Torsello:2019wyp}.}, so rather than obtaining more comprehensive statistics we only want to observe here that, when the analysis is reduced to its constituent Poisson brackets, there is a clear benefit from parallelisation.

\begin{table}
  \caption{\label{effic} Approximate benchmarks for parallel brackets. The spin-parity $1^+$ theory in~\eqref{simple_spin_1p} is augmented with all configurations of the multipliers $\{\calp{3},\calp{4},\calp{6},\cbet{1},\cbet{2},\cbet{3}\}$ -- see~\eqref{neocon} -- and all the Poisson brackets are obtained. Either four or five theories are allocated per node, wallclock time $\langle t(n)\rangle$ is averaged over all 14 nodes, but apparent $\sigma$ remains uniformly high (i.e. suspicious) due to a paucity of expensive brackets. For survey structure and specifications of the Peta4 cluster, see~\cref{simplehpcsurvey}.}
\begin{center}
\begin{tabularx}{\linewidth}{X|X|l|l}
\hline\hline
CPU per node $n$ & Wallclock time $\langle t(n)\rangle$/s & Speedup $S(n)$ & Efficiency $5\times S(n)/n$\\
\hline
$32$ & $\SI{550(100)}{\nothing}$ & $\SI{2.97(85)}{\nothing}$ & $\SI{0.46(13)}{\nothing}$\\
$28$ & $\SI{565(68)}{\nothing}$ & $\SI{2.88(92)}{\nothing}$& $\SI{0.51(13)}{\nothing}$\\
$24$ & $\SI{590(100)}{\nothing}$ & $\SI{2.73(76)}{\nothing}$& $\SI{0.57(16)}{\nothing}$ \\
$20$ & $\SI{640(120)}{\nothing}$ & $\SI{2.53(73)}{\nothing}$& $\SI{0.63(18)}{\nothing}$ \\
$16$ & $\SI{720(160)}{\nothing}$ & $\SI{2.26(71)}{\nothing}$& $\SI{0.71(22)}{\nothing}$ \\
$12$ & $\SI{835(81)}{\nothing}$ & $\SI{1.95(46)}{\nothing}$& $\SI{0.81(19)}{\nothing}$ \\
$8$ & $\SI{1110(310)}{\nothing}$ & $\SI{1.46(52)}{\nothing}$& $\SI{0.91(32)}{\nothing}$ \\
\hline\hline
\end{tabularx}
\end{center}
\end{table}
\section{Examples}\label{examples}
Having introduced the implementation in~\cref{implementation}, we now give some basic examples. These will include a reproduction of the analysis in~\cite{1999IJMPD...8..459Y,2002IJMPD..11..747Y} and a simple HPC survey which extends those same `minimal' theories with the use of multipliers.
\subsection{Preparing a science session}\label{preparingasciencesession}
To prepare a science session, we begin by loading the package into a fresh \Mathematica{} kernel 
\begin{lstlisting}[breaklines=true]
In[]:= <<xAct`HiGGS`;
\end{lstlisting}
If the \HiGGS{} sources have been correctly placed with respect to the \xAct{} directory tree, a copyright greeter should be displayed, indicating that the context \lstinline!xAct`HiGGS`! and its dependancy contexts \lstinline/xAct`xTensor`/, \lstinline/xAct`xPerm`/, \lstinline/xAct`xCore`/, \lstinline/xAct`xTras`/ have been loaded. 

However, loading the package does not yet introduce the physics: the kernel is still in a fresh \xAct{} session, without even a differential manifold. To construct the very many physical definitions needed for science we must \emph{build} the package using the command
\begin{lstlisting}[breaklines=true]
In[]:= BuildHiGGS[];
\end{lstlisting}
This begins an execution in \lstinline!xAct`HiGGS`Private`! of \lstinline!xAct/HiGGS/HiGGS_sources.m!, most of which is taken up with symbol definitions, and in particular calls to \lstinline!DefTensor[]! and \lstinline|MakeRule[]|. To shorten the process, the most expensive definitions (including those for $\soonethree$ projection operators) have been stored in binary files under \lstinline!xAct/HiGGS/bin/build/*.mx! --- those binaries are imported at this time\footnote{Note that the \HiGGS{} binaries were compiled on a 64-bit machine: they can, if needed, be recompiled on alternative architecture via minor changes to the source as indicated in \lstinline!xAct/HiGGS/HiGGS.nb!.}.
In an active front end, the output cells displaying the progress of this build process are periodically deleted, and should finally be replaced by the following message (adjusting for memory)
\begin{lstlisting}[breaklines=true]
Out[]=   ** BuildHiGGS: The HiGGS environment is now ready to use and is occupying 63184600 bytes in RAM.
\end{lstlisting}
The context \lstinline!xAct`HiGGS`! should now be populated with many physical quantities, and as a result the session alone is very memory-intensive. We can view some of these quantities through the \xAct{} variables \lstinline!xAct§Tensors! and \lstinline[breaklines = false]!xAct§ConstantSymbols!.
From this point we may perform the various operations in~\cref{implementation}, or proceed directly to science.

\subsection{Yo--Nester unit tests}\label{yo-nesterunittests}

As a first application, and to verify that \HiGGS{} has been correctly calibrated, we recapitulate the historical analysis in~\cite{1999IJMPD...8..459Y,2002IJMPD..11..747Y}.

The theory in which the $0^+$ mode is active, as considered in~\cite{1999IJMPD...8..459Y}, is given by imposing the following constraints on~\eqref{neocon}
\begin{equation}
\begin{gathered}
	\alp{1}=\alp{2}=\alp{3}=\alp{4}=\alp{5}=2\bet{1}+\bet{2}=\bet{1}+2\bet{3}=0,\\
	\alp{0}\neq 0, \quad \alp{6}\neq 0.
	\label{spin_0p}
\end{gathered}
\end{equation}
All the constraint couplings $\{\calp{I}\}$ and $\{\cbet{M}\}$ are of course assumed to vanish, since they are only recently proposed in~\cite{mythesis,smooth}.
No other `simple' linear identities are assumed among the couplings, and this avoids collision with other theories. The other theory in~\cite{1999IJMPD...8..459Y}, in which the $0^-$ mode is instead allowed to propagate, is defined by
\begin{equation}
\begin{gathered}
	\alp{1}=\alp{2}=\alp{4}=\alp{5}=\alp{6}=2\bet{1}+\bet{2}=\bet{1}+2\bet{3}=0,\\
	\alp{0}\neq 0, \quad \alp{3}\neq 0.
	\label{spin_0m}
\end{gathered}
\end{equation}

In~\cite{2002IJMPD..11..747Y} the analysis was extended to `higher-spin' modes. The theory with only the $1^+$ mode propagating is
\begin{equation}
\begin{gathered}
	\alp{1}=\alp{2}=\alp{3}=\alp{4}=\alp{6}=\bet{1}=\bet{2}=0,\\
	\alp{0}\neq 0, \quad \alp{5}\neq 0, \quad \bet{3}\neq 0.
	\label{simple_spin_1p}
\end{gathered}
\end{equation}
The theory with only the $1^-$ mode propagating is
\begin{equation}
\begin{gathered}
	\alp{1}=\alp{2}=\alp{3}=\alp{4}=\alp{6}=\bet{1}=\bet{3}=0,\\
	\alp{0}\neq 0, \quad \alp{5}\neq 0, \quad \bet{2}\neq 0,
	\label{simple_spin_1m}
\end{gathered}
\end{equation}
and the theory with only the $2^-$ mode propagating is
\begin{equation}
\begin{gathered}
	\alp{2}=\alp{3}=\alp{4}=\alp{5}=\alp{6}=\bet{1}=\bet{2}=\bet{3}=0,\\
	\alp{0}\neq 0, \quad \alp{1}\neq 0.
	\label{simple_spin_2m}
\end{gathered}
\end{equation}
There is also tested in~\cite{2002IJMPD..11..747Y} a pair of more complex theories in which the $0^-$ and $2^-$ modes are both active at the same time. These are given respectively by
\begin{equation}
\begin{gathered}
	\alp{1}=\alp{3}=\alp{4}=\alp{5}=\alp{6}=\bet{1}=\bet{2}=\bet{3}=0,\\
	\alp{0}\neq 0, \quad \alp{2}< 0,
	\label{simple_spin_0-2m_a}
\end{gathered}
\end{equation}
and
\begin{equation}
\begin{gathered}
	\alp{2}=\alp{4}=\alp{5}=\alp{6}=\bet{1}=\bet{2}=\bet{3}=0,\\
	\alp{0}\neq 0.
	\label{simple_spin_0-2m_b}
\end{gathered}
\end{equation}

The seven configurations~\crefrange{spin_0p}{simple_spin_0-2m_b} can of course be processed in parallel. We set up a list called \lstinline!JobsBatch!, which stores the coupling conditions and assigns a string label to each theory

\begin{lstlisting}[breaklines=true]
In[]:= JobsBatch={};
In[]:= JobsBatch~AppendTo~{"spin_0p", {Alp1 == 0, Alp2 == 0, Alp3 == 0, Alp4 == 0, Alp5 == 0, 2Bet1 + Bet2 == 0, Bet1 + 2 Bet3 == 0, cAlp1 == 0, cAlp2 == 0, cAlp3 == 0, cAlp4 == 0, cAlp5 == 0, cAlp6 == 0, cBet1 == 0, cBet2 == 0, cBet3 == 0}};
\end{lstlisting}
Note that definition in terms of equalities is sufficient: the many strict inequations fixing the absence of other special linear conditions are always assumed by \HiGGS{} to be implicit, and signs associated with other inequalities do not affect the constraint structure we wish to probe (though they may well affect the unitarity).

Assuming we have similarly entered the parameters for all theories above, the major undertaking of evaluating all seven PPMs can be initiated with the command 
\begin{lstlisting}[breaklines=true]
In[]:= JobsBatch~StudyTheory~("Import"->True);
\end{lstlisting}
The option \lstinline!"Import"->True! assumes that \lstinline!DefTheory[]! has already been run on all seven cases. In that case binaries such as \lstinline[breaklines = false]!svy/spin_0p.thr.mx! etc., will have been created to store a summary of our prior knowledge of each constraint chain.

Even in the case of the small batch in \lstinline!JobsBatch!, the calculation is barely viable using sub-HPC resources. We perform the run on a dedicated data processing server with $\SI{129}{\giga\byte}$ memory and eight available $\SI{2.90}{\giga\hertz}$ Intel\textsuperscript\textregistered\ Xeon\textsuperscript\textregistered\ E5-2690 0 CPU cores, corresponding to a maximum of eight parallel \Mathematica{} kernels. This computation lasts $\sim\SI{14}{\hour}$, but as mentioned in~\cref{parallelisationandhpc} almost all of this time is spent on the inefficient evaluation of velocities.

Once the calculations of the \lstinline!StudyTheory[]! call are complete, we recall from~\cref{viewtheory} that the results are displayed in human-readable form via the \lstinline!ViewTheory[]! command. We will not provide full analysis here, but focus on the exemplar $1^+$ case. Skipping the breakdown of PiCs and neglecting the velocities, all the nonlinear brackets are
\begin{lstlisting}[breaklines=true]
In[]:= ViewTheory["simple__spin_1p", "Literature" -> False, "Velocities" -> False]; 
Out[]= |
\vspace{-10pt}
|
 ** DefTheory: Incorporating the binary at svy/simple__spin_1p.thr.mx
|
\vspace{-4pt}
\begin{flushleft}
\includegraphics[width=\linewidth]{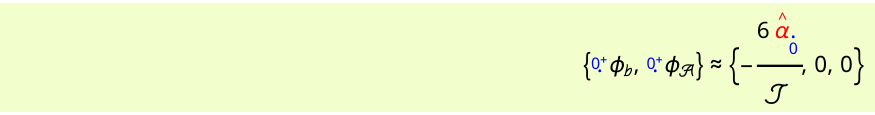}
\end{flushleft}
\vspace{-5pt}
|
|
\vspace{-4pt}
\begin{flushleft}
\includegraphics[width=\linewidth]{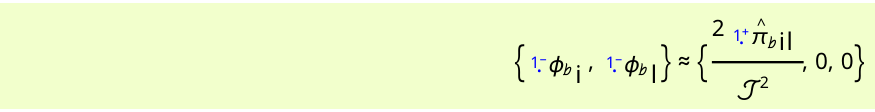}
\end{flushleft}
\vspace{-5pt}
|
|
\vspace{-4pt}
\begin{flushleft}
\includegraphics[width=\linewidth]{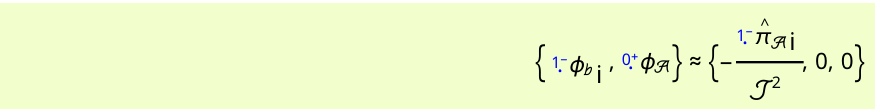}
\end{flushleft}
\vspace{-5pt}
|
|
\vspace{-4pt}
\begin{flushleft}
\includegraphics[width=\linewidth]{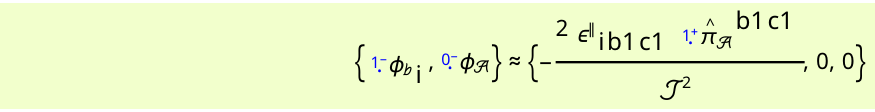}
\end{flushleft}
\vspace{-5pt}
|
|
\vspace{-4pt}
\begin{flushleft}
\includegraphics[width=\linewidth]{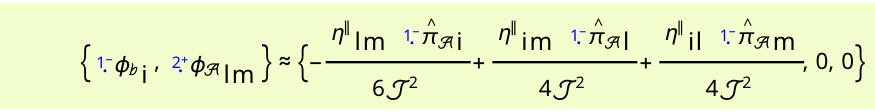}
\end{flushleft}
\vspace{-5pt}
|
|
\vspace{-4pt}
\begin{flushleft}
\includegraphics[width=\linewidth]{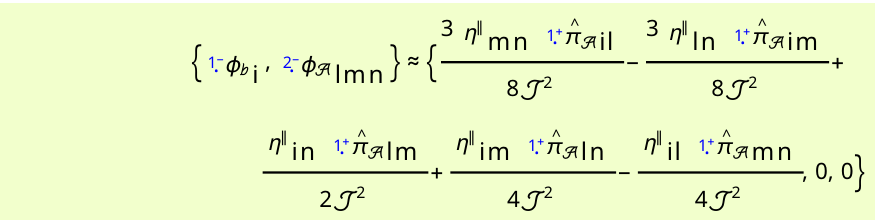}
\end{flushleft}
\vspace{-5pt}
|
|
\vspace{-4pt}
\begin{flushleft}
\includegraphics[width=\linewidth]{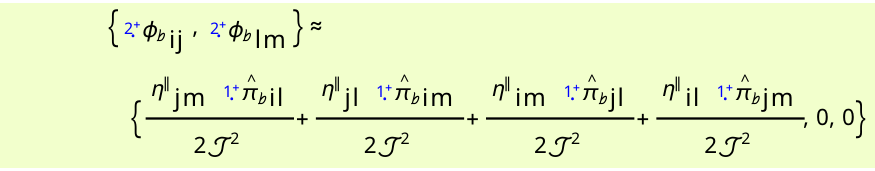}
\end{flushleft}
\vspace{-5pt}
|
|
\vspace{-4pt}
\begin{flushleft}
\includegraphics[width=\linewidth]{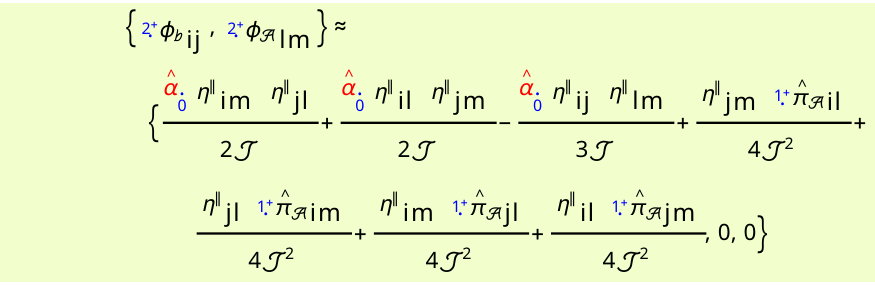}
\end{flushleft}
\vspace{-5pt}
|
|
\vspace{-4pt}
\begin{flushleft}
\includegraphics[width=\linewidth]{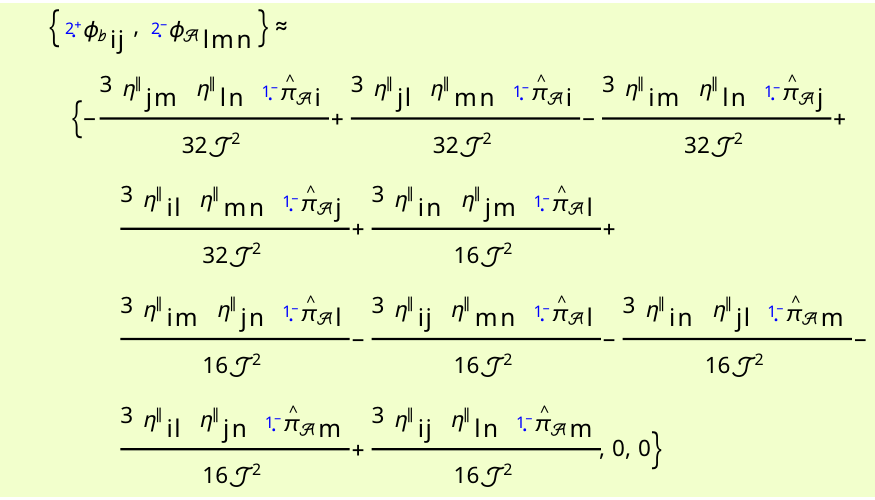}
\end{flushleft}
\vspace{-5pt}
|
\end{lstlisting}
This information (which by itself is obtained within the first minutes of the \lstinline!StudyTheory[]! run) encodes the whole nonlinear primary Poisson matrix (PPM) of the theory: it may be compared with the expressions carefully obtained in~\cite{2002IJMPD..11..747Y}. We see that the brackets $\{\pic[\ovl{i}]{B1m},\pic[\ovl{l}]{B1m}\}$, $\{\pic[\ovl{i}]{B1m},\pic[]{A0p}\}$, $\{\pic[\ovl{i}]{B1m},\pic[]{A0m}\}$, $\{\pic[\ovl{i}]{B1m},\pic[\ovl{lm}]{A2p}\}$, $\{\pic[\ovl{i}]{B1m},\pic[\ovl{lmn}]{A2m}\}$, $\{\pic[\ovl{ij}]{B2p},\pic[\ovl{lm}]{B2p}\}$ and $\{\pic[\ovl{ij}]{B2p},\pic[\ovl{lmn}]{A2m}\}$ are \emph{strictly} nonlinear: they vanish in the linear theory for which the $1^+$ mode is moving.
The effect of this discontinuity in the constraint structure, when moving from the linear to the nonlinear theory, is well decribed in~\cite{2002IJMPD..11..747Y}. In short, a counting analogous to that performed in~\eqref{carbon} shows that the three extra d.o.f of the massive $1^-$ mode are nonlinearly activated: \HiGGS{} thus tells us that a vector torsion mode is \emph{strongly coupled}.

The results of the remaining cases in \lstinline!JobsBatch! are provided in the supplemental materials~\cite{supp}. We find that, up to terms which vanish due to multi-term symmetries (i.e. which cannot be eliminated in the \xAct{} architecture), the nonlinear brackets agree with those found in~\cite{1999IJMPD...8..459Y,2002IJMPD..11..747Y}. There is a possible exception in the case of the general bracket $\{\pic[\ovl{ij}]{B1p},\pic[\ovl{l}]{B1m}\}$, which we find to be `surficial' in ways illustrated already by the bracket in~\cref{poissonbracket}. This subtlety appears only to touch the scalar mode results in~\cite{1999IJMPD...8..459Y}: it seems worthwhile to investigate -- in future work -- if and how this might affect the final d.o.f counting. For brevity we will omit the velocities here\footnote{The few velocities which are known from~\cite{1999IJMPD...8..459Y} corroborate our results, however we note that \lstinline!ToNesterForm[]! actually \emph{fails} to fully covariantise the velocity of $\pic[\ovl{i}]{B1m}$ in the simple $0^-$ case. This `bug' is not known to occur elsewhere, and certainly not during the evaluation of nonlinear brackets: the main feature of \HiGGS{}. We suggest that a fully nonlinear `Leibniz rule' implementation of velocities, as recommended above, would greatly reduce the risk of such effects by demanding less of \lstinline!ToNesterForm[]! in each call.}, since even the covariantised, linear expressions can be very cumbersome. Examples are provided in~\cref{appendix} for the minimal $1^+$ case above: these illustrate how gradients of the field strengths can arise as the algorithm progresses, necessitating a future extension of \HiGGS{} to the second-order Euler--Lagrange formalism.

\subsection{Simple HPC survey}\label{simplehpcsurvey}

The purpose of this section is to further extend our knowledge of the theories examined in~\cref{yo-nesterunittests}, by running \HiGGS{} on basic HPC resources, so as to chart the effects of introducing multiplier fields. 
The findings of this survey, and any viable multiplier configurations, will be presented in future work.
The starting point will be the `minimal' Poincar\'e gauge theories set out in~\crefrange{spin_0p}{simple_spin_0-2m_b}. Of these, we know that~\eqref{spin_0p} and~\eqref{spin_0m} are the traditional cases in which massive $0^+$ and $0^-$ scalars propagate safely alongside the usual $2^+$ graviton. Of the cases with active higher-spin modes, we know that~\eqref{simple_spin_0-2m_a} and~\eqref{simple_spin_0-2m_b} propagate both $0^-$ and $2^-$ modes, and \emph{may} be safe\footnote{The uncertainty in these cases is based on the fact that the rank of the PPM may still change on non-Minkowskian surfaces in the phase space. As mentioned in~\cref{introduction}, whether this is actually a physical problem should be determined by closer analysis.}. 

The minimal `problem' cases are~\cref{simple_spin_1p,simple_spin_1m,simple_spin_2m}. These are ostensibly simple linearised theories which propagate extra $1^+$, $1^-$ and $2^-$ modes respectively, but which appear to suffer from mode activation in the fully nonliear regime. This mode activation is identified by counting the canonical degrees of freedom in the Hamiltonian analysis: its association with a particular $J^P$ sector is not shown explicitly, but inferred by examining the unconstrained PiC functions in each case. These are as follows; 
\begin{itemize}
  \item In the minimal $1^+$ theory~\eqref{simple_spin_1p}, $\pic[\ovl{ij}]{B1p}$, $\pic[\ovl{ij}]{A1p}$ and $\pic[\ovl{i}]{A1m}$ are unconstrained, the $1^-$ mode is thought to be activated.
  \item In the minimal $1^-$ theory~\eqref{simple_spin_1m}, $\pic[]{B0p}$, $\pic[\ovl{i}]{B1m}$, $\pic[\ovl{ij}]{A1p}$ and $\pic[\ovl{i}]{A1m}$ are unconstrained, the $1^+$ mode is thought to be activated.
  \item In the minimal $2^-$ theory~\eqref{simple_spin_2m}, $\pic[\ovl{ij}]{A2p}$ and $\pic[\ovl{ijk}]{A2m}$ are unconstrained, the $2^+$ mode is thought to be activated.
\end{itemize}
In~\cite{smooth} a potential avenue was outlined for preventing mode activation by means of the multiplier fields set out in~\eqref{neocon}. For any conventional Poincar\'e gauge theory, there are $2^6\times 2^3$ possible multiplier configurations. Any pair of configurations differs, if not by the constrained $J^P$ sectors, then by the `singular' or `parallel' nature of a constrained sector. These differences may have quite unpredictable consequences in the analysis. In order to efficiently explore the space of multiplier configurations, it is therefore important to have all the commutators between the known constraints to hand.

\subsubsection{Scope of survey}\label{scope}

By adding various multipliers, there are $3\times 2^6\times 2^3=1536$ separate theories which can be constructed from the minimal PGTs. However, some of these can be ruled out immediately on phenomenological grounds. In~\cite{smooth} we focussed on the modification of~\eqref{simple_spin_1p} by allowing $\cbet{2}\neq 0$. In the Lagrangian picture, this maps to a pair of constraints on the $1^-$ part of the torsion tensor, suppressing velocities or equating them to gradients. The main phenomenological constraint on the minimal extensions is the requirement that they contain the Einstein--Cartan dynamics, for which the gravitational field still is described by the Riemann--Cartan curvature. For this reason, we suspect that it will generally be safer to impose the $\{\cbet{M}\}$ than the $\{\calp{I}\}$.

How dangerous is it to suppress parts of the rotational sector? To get an idea, we consider how the two d.o.fs in Einstein--Cartan gravity might be manifest in a gravitational wave-type solution. 
Rather than studying the Einstein--Cartan waves directly, we instead introduce novel solutions to the special (purely quadratic, i.e. $\alp{0}=0$) PGT from~\cite{chapter2,chapter3,chapter4,mythesis} which describe null pp-waves on the Minkowski background~\cite{another_lasenby}.
The wave solutions are formulated in the \emph{Brinkmann} gauge, rather than the more popular transverse-traceless (TT) setup~\cite{Lasenby:2019gmi}. To introduce the Brinkmann gauge we start with a Cartesian coordinate system, and the rotation gauge is first chosen as in the \HiGGS{} environment, so that the local Lorentz and coordinate bases are aligned. We then define `perpendicular' and `null' vectors as 
\begin{equation}
\begin{gathered}
    \phantom{\tensor{\bm{e}}{_t}}
  \tensor{\bm{e}}{_t}\equiv\tensor{\hat{\bm{e}}}{_0},\quad \tensor{\bm{e}}{_x}\equiv\tensor{\hat{\bm{e}}}{_1},\quad \tensor{\bm{e}}{_y}\equiv\tensor{\hat{\bm{e}}}{_2},\quad \tensor{\bm{e}}{_z}\equiv\tensor{\hat{\bm{e}}}{_3}, \\
  \tensor{\bm{e}}{_{\perp}}\equiv \cos (\theta)\tensor{\bm{e}}{_{x}}+\sin (\theta)\tensor{\bm{e}}{_{y}},\quad \tensor{\bm{e}}{_{+}}\equiv\tensor{\bm{e}}{_{t}}+\tensor{\bm{e}}{_{z}}.
    \label{rotgag}
\end{gathered}
\end{equation}
We will denote the `wave coordinate' as $\tau\equiv t-z$; the wave amplitude is taken in all solutions to be a smooth and compact scalar function $\mathcal{  A}\equiv\mathcal{  A}(\tau)$.
The Brinkmann gauge is complementary to the TT gauge in the sense that it confines waves to the time and longitudinal components of the metric perturbation.
The wave will therefore alter the definition of the unit timelike vector $\tensor{n}{^i}$, which defines the foliation\footnote{For this reason the Brinkmann gauge will not be a natural choice for the canonical analysis.}, and unit spacelike vector $\tensor{l}{_i}$ which defines the direction of travel, but not the polarisation vector $\tensor{\varepsilon}{_{\ovl{i}}}$. Restricting to the case of weak waves, we will then have $\tensor{n}{_i}\equiv \tensor{h}{_i^t}/\sqrt{|\tensor{g}{^{tt}}|}= \tensor{(\bm{e}^t)}{_i}+\mathcal{  O}(\mathcal{  A})$, $\tensor{l}{_i}\equiv \tensor{h}{_i^z}/\sqrt{|\tensor{g}{^{zz}}|}=\tensor{(\bm{e}^z)}{_i}+\mathcal{  O}(\mathcal{  A})$ and $\tensor{\varepsilon}{_{\ovl{i}}}\equiv \cos(\theta)\tensor{(\bm{e}^{x})}{_i}+\sin(\theta)\tensor{(\bm{e}^{y})}{_i}$.
While the Brinkmann gauge is opposed to the TT gauge at the level of the metric, this turns out to be somewhat reversed at the level of the field strengths. Accordingly, it is useful to define the `TT symmetric-traceless' operation on the indices of a general TT tensor $\tensor{X}{_{\ovl{ij}}}$ as $\tensor{X}{_{\llangle\ovl{ij}\rrangle}}\equiv\tensor{X}{_{(\ovl{ij})}}-\frac{1}{2}\tensor{X}{^{\ovl{k}}_{\ovl{k}}}( \etad{\ovl{ij}}-\tensor{l}{_{\ovl{i}}}\tensor{l}{_{\ovl{j}}} )$, where we recall the original `symmetric-traceless' operator $\tensor{X}{_{\langle\ovl{ij}\rangle}}\equiv\tensor{X}{_{(\ovl{ij})}}-\frac{1}{3}\tensor{X}{^{\ovl{k}}_{\ovl{k}}}\etad{\ovl{ij}}$ from~\cite{chapter4}.

The new exact solution which concerns us describes a wave in the Riemann--Cartan curvature, with vanishing torsion and two d.o.fs quantified by a polarisation vector. It has the components
\begin{subequations}
  \begin{align}
    \cR[\ovl{ij}]{A2p}&=\mathcal{  A}\tensor{\varepsilon}{_{\llangle\ovl{i}}}\tensor{\varepsilon}{_{\ovl{j}\rrangle}}+\mathcal{  O}(\mathcal{  A}^2),\label{whale}\\
    \ncR[\ovl{ij}]{A2p}&=\mathcal{  A}\tensor{\varepsilon}{_{\llangle\ovl{i}}}\tensor{\varepsilon}{_{\ovl{j}\rrangle}}+\mathcal{  O}(\mathcal{  A}^2),\\
    \cR[\ovl{ijk}]{A2m}&=\mathcal{  A}\tensor{\varepsilon}{_{\llangle\ovl{k}}}\tensor{\varepsilon}{_{[\ovl{i}\rrangle}}\tensor{l}{_{\ovl{j}]}}+\mathcal{  O}(\mathcal{  A}^2),\\
    \ncR[\ovl{ijk}]{A2m}&=-\mathcal{  A}\tensor{\varepsilon}{_{\llangle\ovl{k}}}\tensor{\varepsilon}{_{[\ovl{i}\rrangle}}\tensor{l}{_{\ovl{j}]}}+\mathcal{  O}(\mathcal{  A}^2).\label{rc26}
  \end{align}
\end{subequations}
These components turn out to be identical~\cite{another_lasenby} to those of the Riemann tensor in the presence of the vacuum pp-waves known from GR. Recall that~\cref{simple_spin_1p,simple_spin_1m,simple_spin_2m} were initially set up as a modifications of Einstein--Cartan theory. Since the Einstein--Cartan theory differs from GR only by a contact torsion interaction, and since the linear $1^+$, $1^-$ and $2^+$ modes are massive, it would seem strange if the the null pp-wave solution~\crefrange{whale}{rc26} is not also \emph{mandatory} in the minimal extensions. Referring back to~\cite{smooth}, we find that we should then always take $\calp{1}=0$, since we would otherwise encounter the Lagrangian constraints
\begin{equation}
    \calp{1}\neq 0\Rightarrow \cR[\ovl{ij}]{A2p}+\ncR[\ovl{ij}]{A2p}
    \approx \cR[\ovl{ijk}]{A2m}-\ncR[\ovl{ijk}]{A2m}\approx 0,
\end{equation}
which both force the wave amplitude $\mathcal{A}\to 0$.
This already halves the volume of our parameter space, and so we make no attempt exhaustively determine the various other phenomenological constraints. 

Other limitations on the allowed multipliers come from the linearised particle spectrum. For the $1^+$ theory with $\calp{1}=0$ only the group $\{\calp{3},\calp{4},\calp{6},\cbet{1},\cbet{2},\cbet{3}\}$ does not immediately constrain $\PiP[\ovl{ij}]{A1p}$. Similarly for the $1^-$ theory the space is $\{\calp{2},\calp{3},\calp{6},\cbet{1},\cbet{2},\cbet{3}\}$. For the $2^-$ theory, we consider the group $\{\calp{3},\calp{5},\calp{6},\cbet{1},\cbet{2},\cbet{3}\}$. Rightly, we ought to allow $\calp{4}\neq 0$ in this case, especially since the momentum $\PiP[\ovl{ij}]{A2p}$ of the (anticipated) strongly coupled $2^+$ mode would then be disabled: just for this initial survey, however, the complexity of the $2^-$ calculations is such that restricting to $\calp{4}=0$ results in a significant economy.

The aim is then to obtain covariant expressions for all the possible commutators among all known if-constraints for all $3\times 2^3\times 2^3=192$ theories stipulated above. The current \HiGGS{} setup is supposed to be able to do this, and produce binaries of the results suitable for use in a database. The requisite calculations would take years on a single desktop computer core, so we distribute them over 14 nodes of the Peta4 supercomputer -- the CPU cluster component of the heterogeneous CSD3 facility. Each node has two $\SI{2.60}{\giga\hertz}$ Intel\textsuperscript\textregistered\ Xeon\textsuperscript\textregistered\ Skylake 6142 CPUs, each having 16 cores with $\SI{6}{\giga\byte}$ of memory apiece, amounting to 448 processors. As described in~\cref{parallelisationandhpc}, \HiGGS{} runs on a master kernel within each node. The master kernel delegates the analysis of a (randomly allocated) batch of theories, and has a total of 32 parallel sub-kernels at its disposal. 

An illustration of the survey was shown already in~\cref{hpcsurvey}. The node count of 14 is a service-level restriction on Peta4 rather than limitation of the implementation. 
We can see from a visual inspection that not all theories are equally expensive: some multiplier configurations engender more constraints and more brackets, while others more heavily involve the higher-spin sectors. An examination of the stack trace from each node suggests that the most time-consuming cases are those for which a multiplier is used to constrain a $J^P$ sector which was thought to be nonlinearly activated. This is an interesting observation in the context of the strong coupling considerations. In particular, delays occur when new `parallel' or `singular' SiCs of the form
\begin{subequations}
\begin{gather}
  \tensor*[^A]{\chi}{^{\parallel}_{\acu{v}}}\equiv
  \projorthhum[_{\acu{v}}_{nm}^{\ovl{pq}}]{A}\tensor{\mathcal{  R}}{^{nm}_{\ovl{pq}}}\approx 0, \label{canconx}\\
  \tensor*[^E]{\chi}{^{\parallel}_{\acu{v}}}\equiv
  \planck^2\projorthhum[_{\acu{v}}_{n}^{\ovl{pq}}]{E}\tensor{\mathcal{  T}}{^{n}_{\ovl{pq}}}\approx 0,\\
  \tensor*[^A]{\chi}{^{\vDash}_{\acu{v}}}\equiv 
\tensor[^A]{\varphi}{_{\acu{v}}}
 +\frac{8\calpm[A]{\perp\parallel}\alpm[A]{\perp\perp}}{\calpm[A]{\perp\perp}}\projorthhum[_{\acu{v}}_{jk}^{\ovl{lm}}]{A}
  \tensor{\mathcal{  R}}{^{jk}_{\ovl{lm}}}
  \approx 0,
  \label{indcalx}
  \\
  \tensor*[^E]{\chi}{^{\vDash}_{\acu{v}}}\equiv 
\tensor[^E]{\varphi}{_{\acu{v}}}
 +\frac{4\cbetm[E]{\perp\parallel}\betm[E]{\perp\perp}}{\cbetm[E]{\perp\perp}}\planck^2\projorthhum[_{\acu{v}}_{j}^{\ovl{lm}}]{E}
  \tensor{\mathcal{  T}}{^{j}_{\ovl{lm}}}
  \approx 0,
  \label{cancon}
\end{gather}
\end{subequations}
are introduced. It is known from manual calculations in~\cite{mythesis} that these secondaries (specifically their field strength terms) fail to commute with many other constraints, and can produce brackets of surprising complexity.

Physics binaries (\HiGGS{} extension \lstinline!*.thr.mx! files) of all nonlinear brackets identified in this survey, along with plaintext stack traces and clocking times, can be found in the supplemental materials~\cite{supp}.

\subsubsection{Results and prospects for new physics}\label{newphysics}

In this final section, we make some preliminary observations on the new data provided by the `calibration' survey~\cite{supp} -- the bulk of this analysis being reserved for future work.
We focus on the $1^+$ and $1^-$ modes. 

The investigation in~\cite{2002IJMPD..11..747Y} seems to identify the source of strong coupling as follows. In the linear theory without multipliers, rotational PiCs $\tensor[^A]{\varphi}{_{\acu{v}}}$ may be paired off with conjugate PiCs $\tensor[^E]{\varphi}{_{\acu{v}}}$, so as to determine $\tensor[^A]{u}{_{\acu{v}}}$ and $\tensor[^E]{u}{_{\acu{v}}}$ and so terminate both chains. Unpaired PiCs generate SiCs, whose consistency conditions fix the original PiC Hamiltonian multipliers -- all within the same $J^P$ sector. 

As the theory becomes nonlinear, the PiC structure does not, of course, change. However there are generally more PiC-PiC commutators which emerge, and which do not respect the law of conjugate pairs. Consider a theory where the collective $\{\tensor[^A]{\varphi}{_{\acu{v}}}\}$ span $m$ d.o.f, and the $\{\tensor[^E]{\varphi}{_{\acu{v}}}\}$ span $n<m$ d.o.f. Then the emergence of nonlinear commutators can in principle alter the PPM rank so that the rotational consistencies determine up to $n$ rotational Hamiltonian multipliers $\{\tensor[^E]{u}{_{\acu{v}}}\}$, overflowing into $m-n$ SiCs (denoted $\chi_{[m-n]}$ in~\cite{2002IJMPD..11..747Y}). 
The $n$ translational PiCs and $m-n$ SiCs generically fail to commute (again due to nonlinearity) with the $m$ rotational PiCs, whose Hamiltonian multipliers $\{\tensor[^A]{u}{_{\acu{v}}}\}$ they then determine. 

The (reduced) number $m-n$ of SiCs dictates the number of strongly coupled modes in the d.o.f counting. Being unconstrained by the $J^P$ conjugacy, the number $m-n$ may not map on to the $2\ell+1$ integer spin multiplicities, or it may seem to contribute a half-integer d.o.f. These matters should then be clarified by closer study, and identification of FC combinations to restore integer d.o.f.
Critically, the division of translational and rotational PiCs is efficacious because the rotational sector commutes with itself nonlinearly.

Geometric multipliers contribute new primaries by direct analogy to~\eqref{sureprimaries}
\begin{equation}
  \tensor{\phi}{^{ij}_{kl}}\equiv \tensor{\varpi}{^{ij}_{kl}}\approx 0,\quad
\tensor{\phi}{^{i}_{kl}}\equiv \tensor{\varpi}{^{i}_{kl}}\approx 0,
  \label{newp}
\end{equation}
i.e. the multiplier momenta $\tensor{\varpi}{^{ij}_{kl}}$ and $\tensor{\varpi}{^{i}_{kl}}$ (defined as in~\eqref{canonicalmomenta} to be conjugate to $\tensor{\lambda}{^i_{jk}}$ and $\tensor{\lambda}{^{ij}_{kl}}$), which are also resistant to nonlinear commutators. Irreps of the fields $\tensor{\lambda}{^i_{jk}}$ and $\tensor{\lambda}{^{ij}_{kl}}$ appear in other if-constraints, but these produce predictable, linear commutators with the conjugate $J^P$ sectors of the $\tensor{\phi}{^{ij}_{kl}}$ and $\tensor{\phi}{^{i}_{kl}}$. For the case of translational multipliers (as discussed in~\cref{scope}, we suspect these to be safer), an approach might be to use the consistencies of the constrained $\tensor{\phi}{^{i}_{kl}}$ irreps to solve for a large number of $\{\tensor[^E]{u}{_{\acu{v}}}\}$ in the linear theory. This would force the rotational sector into developing the maximum number of SiCs ab initio. Assuming nonlinear commutators between the SiCs and rotational PiCs proliferate as usual, one then expects to solve for all the remaining Hamiltonian multipliers as before, with a generally SC system whose SiCs are not contingent on nonlinear effects. 

The `calibration' survey covers the simple extension of~\eqref{simple_spin_1p} by the condition $\cbet{2}\neq 0$, and we used the resultant brackets when exploring this mechanism in~\cite{smooth}. That theory does not appear to be strongly coupled, but it is also unlikely to be unitary. We close this section by suggesting another option. In~\cite{2002IJMPD..11..747Y} the basic PGT with only $\alp{0}\neq 0$ and $\alp{5}\neq 0$ was briefly considered, in which both $1^+$ and $1^-$ modes were strongly coupled. This theory contains all the translational PiCs\footnote{Recall from~\cref{irreducibledecompositions} that we inherit the $\sothree$ irrep notation from~\cref{traper,trapar,rotper,rotpar}, but just replace the underlying symbol.}: $\pic[]{B0p}$, $\pic[\ovl{ij}]{B1p}$, $\pic[\ovl{i}]{B1m}$ and $\pic[\ovl{ij}]{B2p}$. The rotational PiCs are only conjugate in the $0^+$ and $2^+$ sectors, so the linear theory produces SiCs $\sic[\ovl{ij}]{B1p}$, $\sic[\ovl{i}]{B1m}$, $\sic[]{A0m}$ and $\sic[\ovl{ijk}]{A2m}$. In the nonlinear case, all 12 d.o.f in the rotational and translational sectors are assumed to solve exactly for each others' Hamiltonian multipliers: no SiCs are produced and the total d.o.f rises by two massive vectors $\frac{1}{2}(1+5+3+3)=3+3$.

However by imposing $\cbet{1}\neq 0$, we can obtain all the translational Hamiltonian multipliers except for in the $0^+$ sector -- which remains under traditional conjugacy. In this way, an extra $\sic[\ovl{ij}]{A2p}$ is forced in both regimes, and its natural conjugacy (comparing to the experience of the $0^+$ in~\cite{smooth}) will be $\ncTpic[\ovl{ij}]{B2p}$. A similar structure develops in the $2^-$  sector. Denoting consistency conditions with arrows, we arrive at 
\begin{equation}
\begin{tikzpicture}[baseline={([yshift=-.5ex]current bounding box.center)},vertex/.style={anchor=base,
    inner sep=2pt}]
    \node[vertex,anchor=center] (G1) at (0,0)   {$\ncTpic[\ovl{ij}]{B2p}$};
    \node[vertex,anchor=center] (G1a) at (1.,-1.)   {$\ncTmul[\ovl{ij}]{B2p}$};
    \draw[thick,dotted] (G1.south) to[out=270,in=180] (G1a.west)  ;
    \node[vertex,anchor=center] (G2) at (2,0)   {$\pic[\ovl{ij}]{B2p}$};
    \node[vertex,anchor=center] (G2a) at (3.,-1.)   {$\mul[\ovl{ij}]{B2p}$};
    \draw[thick,dotted] (G2.south) to[out=270,in=180] (G2a.west)  ;
    \node[vertex,anchor=center] (G3) at (2,2)   {$\pic[\ovl{ij}]{A2p}$};
    \node[vertex,anchor=center] (G3a) at (1.,3.)   {$\mul[\ovl{ij}]{A2p}$};
    \draw[thick,dotted] (G3.north) to[out=90,in=0] (G3a.east)  ;
    \node[vertex,anchor=center] (G4) at (0,2)   {$\sic[\ovl{ij}]{A2p}$};
    \draw[->,thick] (G1.east) -- (G2.west)  ;
    \draw[->,thick] (G2.north) -- (G3.south)  ;
    \draw[->,dashed,thick] (G3.west) -- (G4.east)  ;
    \draw[->,thick] (G4.south) -- (G1.north)  ;
    \node[vertex,anchor=center] (E1) at (4,0)   {$\cTpic[\ovl{ijk}]{A2m}$};
    \node[vertex,anchor=center] (E1a) at (5.,-1.)   {$\cTmul[\ovl{ijk}]{A2m}$};
    \draw[thick,dotted] (E1.south) to[out=270,in=180] (E1a.west)  ;
    \node[vertex,anchor=center] (E2) at (6,0)   {$\lorsicpar[\ovl{ijk}]{B2m}$};
    \node[vertex,anchor=center] (E3) at (6,2)   {$\pic[\ovl{ijk}]{A2m}$};
    \node[vertex,anchor=center] (E3a) at (5.,3.)   {$\mul[\ovl{ijk}]{A2m}$};
    \draw[thick,dotted] (E3.north) to[out=90,in=0] (E3a.east)  ;
    \node[vertex,anchor=center] (E4) at (4,2)   {$\sic[\ovl{ijk}]{A2m}$};
    \draw[->,thick,dashed] (E1.east) -- (E2.west)  ;
    \draw[->,thick] (E2.north) -- (E3.south)  ;
    \draw[->,dashed,thick] (E3.west) -- (E4.east)  ;
    \draw[->,thick] (E4.south) -- (E1.north)  ;
  \end{tikzpicture}
  \label{fourstep}
\end{equation}
where solid arrows indicate that a Hamiltonian multiplier is determined, and dashed arrows indicate that a secondary must be constructed. Four-step consistency chains such as in~\eqref{fourstep} are not seen in the original PGT. If the overall SC structure is indeed preserved in the nonlinear theory, it would seem that only two d.o.f overall propagate. This result seems suspicious, and must be very carefully tested -- for example with \HiGGS{}.
If it is true, then the resulting theory might conceivably introduce extra contact interactions to the Einstein--Cartan model, and the overall phenomenological differences with GR would be very interesting to study.

\section{Conclusions}\label{conclusions}

In this paper we have presented the package \HiGGS{}, written for the tensor manipulation suite \xAct{} and the computer algebra software \Mathematica{}. The \HiGGS{} package performs calculations -- such as Poisson brackets -- which frequently arise in the Hamiltonian (canonical) analysis of modified gravity theories with curvature and/or torsion. The current iteration of the package is tailored to the generalised Poincar\'e gauge theory in~\cref{neocon}: given an action of that class, \HiGGS{} can identify primary and secondary constraints, and calculate arbitrary field velocities. For more original torsionful actions, \HiGGS{} may still be used to evaluate brackets, and canonicalise expressions by irreducible decomposition.

Parallelisation is a core feature of \HiGGS{}. We have argued that those aspects of the Dirac--Bergmann Hamiltonian constraint algorithm which can become cumbersome during manual evaluation, are actually very well suited to parallel computing. On a laptop or desktop computer, \HiGGS{} can take advantage of available cores to shorten the analysis of a given theory. Parallelisation is done \emph{within} the \HiGGS{} environment. This capability meshes well with the problem of surveying large numbers of action configurations using high-performance computing (HPC), since it avoids the need to educate general-purpose job scheduling tools (such as \SLURM{}~\cite{10.1007/10968987_3} or \TORQUE{}~\cite{10.1145/1188455.1188464}) about the physical details of the Hamiltonian constraint structure. In our example HPC survey, all job-scheduling decisions were made from within the \HiGGS{} environment, to whole-node granularity.

The HPC survey we have performed here -- whose results will be discussed further in future work -- targets minimal extensions to the Einstein--Cartan theory in which a single extra massive spin-parity $1^+$, $1^-$ or $2^-$ torsion particle is present, i.e. extending the two usual graviton polarisations by three or five extra degrees of freedom (d.o.f). These minimal extensions were believed~\cite{1980PhRvD..21.3269S,1981PhRvD..24.1677S,10.1143/PTP.64.1435} to be ghost free according to the linearised analysis. Subsequent nonlinear Hamiltonian analysis~\cite{2002IJMPD..11..747Y} suggested that the linear regime strongly couples extra modes of spin-parity $1^-$, $1^+$ or $2^+$ respectively, so that these modes spoil the theories' viability. Based on the hypothesis that multiplier fields could selectively suppress these modes -- in the style of teleparallel gravity -- our survey lists all the commutators among all the known primary and secondary `if-constraints' (i.e. constraints which may arise due to a choice of Lagrangian couplings), for the various possible multiplier configurations. Our level of analysis at least matches that of mode activation\footnote{We do not, however, perform the extra step in~\cite{2002IJMPD..11..747Y} of considering the problem of constraint bifurcation using the Poisson matrix pseudodeterminant.} in~\cite{2002IJMPD..11..747Y}, where the primary Poisson matrices of the minimal Einstein--Cartan extensions are obtained. We note that some of these results (which are available in the supplemental materials~\cite{supp}) have already been used in~\cite{smooth}.

While the \HiGGS{} package may be of use to researchers in the ways described above, we must observe some of its many limitations;
\begin{itemize}
  \item The \lstinline!PoissonBracket[]! module automatically expands its operands, rather than first taking advantage of the Leibniz rule wherever those operands are products of covariant factors. This can result in costly attempts by \lstinline!ToNesterForm[]! to covariantise arbitrarily complex brackets.
  \item The \lstinline!Velocity[]! module is directly associated with the canonical Hamiltonian in~\eqref{grandperp} of the generalised Poincar\'e gauge theory in~\eqref{neocon}. This could easily be avoided (i.e. in favour of user-defined Hamiltonia) with Leibniz rule functionality in \lstinline!PoissonBracket[]!.
  \item The \lstinline!PoissonBracket[]! module is incapable of processing the second-order Euler--Lagrange formulation. The lack (to our knowledge) of a general formula in Poincar\'e gauge theory for the second-order bracket was also a limiting factor for our previous analysis in~\cite{chapter4}.
  \item In general, \HiGGS{} relies quite heavily on the gauge-covariant derivatives $\tensor{D}{_\mu}$ or $\tensor{\mathcal{D}}{_{\ovl{k}}}$. The \xAct{} suite already has a very sophisticated functionality to accommodate the definition of such derivatives, which is not exploited by the \HiGGS{} implementation.
  \item In general, \HiGGS{} relies very heavily on the $\soonethree$ and (particularly) the $\othree$ decompositions of tensorial fields. However, these are manually defined in the implementation, for \emph{each decomposed field}. A clear case is made for a general decomposition tool, more closely integrated with the existing functionality in \lstinline!xAct`SymManipulator`!.
  \item Notwithstanding the Lagrangian structure implied by \lstinline!Velocity[]!, inclusion of new dynamical variables constitutes a different problem. Extension to the metric affine gauge theory (MAGT) structure would be a straightforward, if time-consuming, exercise. More work would likely be needed for the inclusion of (e.g. fermionic) matter fields.
  \item The background assumed by \HiGGS{} when the option \lstinline!"ToOrder"->0! or \lstinline!"ToOrder"->1! is passed to modules such as \lstinline!ToNesterForm[]!, is Minkowski spacetime with vanishing background torsion. There are, however, obvious motivations for considering curved backgrounds, such as de Sitter or Schwarzschild. Moreover in~\cite{chapter2,chapter3,chapter4,mythesis} it is argued that non-minimal gravitational gauge theories, which are detatched from the geometrical trinity, only become viable in an attractor background of \emph{constant axial torsion}.
\end{itemize}
Whether these limitations are best addressed by improving the \HiGGS{} implementation, or beginning ab initio with an improved understanding of the challenges posed by canonical computer algebra, remains to be seen. For the moment, we hope at least to have shown that modified gravity is now ready for computer algebra assistance \emph{at scale}. 
In this sense we build on the recent work of Lin, Hobson and Lasenby~\cite{2020arXiv200502228L,Lin2}, who used computer algebra to systematically obtain all ghost and tachyon-free cases of the linearised Poincar\'e gauge theory, and did so with far more limited resources than are brought to bear in this paper.
A future is then suggested in which vague concerns -- viz, a potential for strong coupling demonstrated among a handful of cases -- are no longer valid grounds upon which to dismiss a rich class of theories.

\begin{acknowledgements}
  This work was performed using resources provided by the Cambridge Service for Data Driven Discovery (CSD3) operated by the University of Cambridge Research Computing Service (\href{www.csd3.cam.ac.uk}{www.csd3.cam.ac.uk}), provided by Dell EMC and Intel using Tier-2 funding from the Engineering and Physical Sciences Research Council (capital grant EP/T022159/1), and DiRAC funding from the Science and Technology Facilities Council (\href{www.dirac.ac.uk}{www.dirac.ac.uk}).

  This manuscript was improved by the kind suggestions of Amel Durakovi\'c and Will Handley, and I would like to thank Tom Z\l o\'snik for useful discussions.

	I am grateful for the kind hospitality of Leiden University and the Lorentz Institute, and the support of Girton College, Cambridge.

	The current version of \HiGGS{} incorporates elements of Cyril Pitrou's code from the repository at \href{https://github.com-contrib/examples}{www.github.com/xAct-contrib/examples}.
\end{acknowledgements}

\bibliographystyle{apsrev4-1}
\bibliography{bibliography}

\appendix
\section{Linear velocities of the simple spin-$1^+$ case}\label{appendix}
In this appendix we provide the linearised velocities of the PiCs appearing in the simple $1^+$ extension to Einstein--Cartan theory: 
\begin{lstlisting}[breaklines=true]
In[]:= ViewTheory["simple_spin_1p", "Literature" -> False, "PPM" -> False, "Velocities" -> True]; 
Out[]= |
\vspace{-10pt}
|
 ** DefTheory: Incorporating the binary at svy/simple_spin_1p.thr.mx
|
\vspace{-4pt}
\begin{flushleft}
\includegraphics[width=\linewidth]{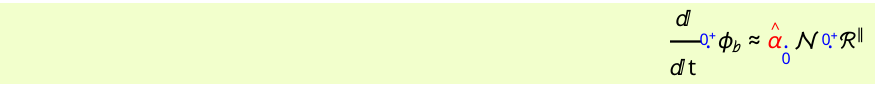}
\end{flushleft}
\vspace{-5pt}
|
|
\vspace{-4pt}
\begin{flushleft}
\includegraphics[width=\linewidth]{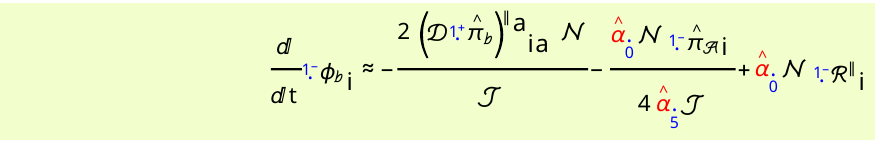}
\end{flushleft}
\vspace{-5pt}
|
|
\vspace{-4pt}
\begin{flushleft}
\includegraphics[width=\linewidth]{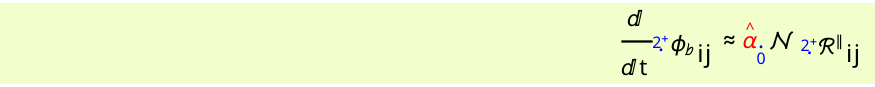}
\end{flushleft}
\vspace{-5pt}
|
|
\vspace{-4pt}
\begin{flushleft}
\includegraphics[width=\linewidth]{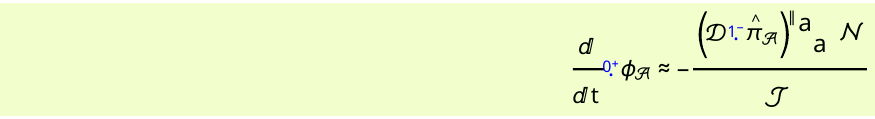}
\end{flushleft}
\vspace{-5pt}
|
|
\vspace{-4pt}
\begin{flushleft}
\includegraphics[width=\linewidth]{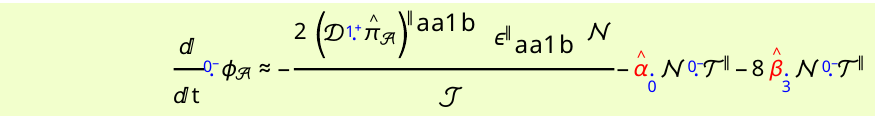}
\end{flushleft}
\vspace{-5pt}
|
|
\vspace{-4pt}
\begin{flushleft}
\includegraphics[width=\linewidth]{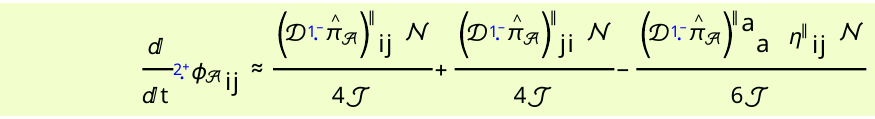}
\end{flushleft}
\vspace{-5pt}
|
|
\vspace{-4pt}
\begin{flushleft}
\includegraphics[width=\linewidth]{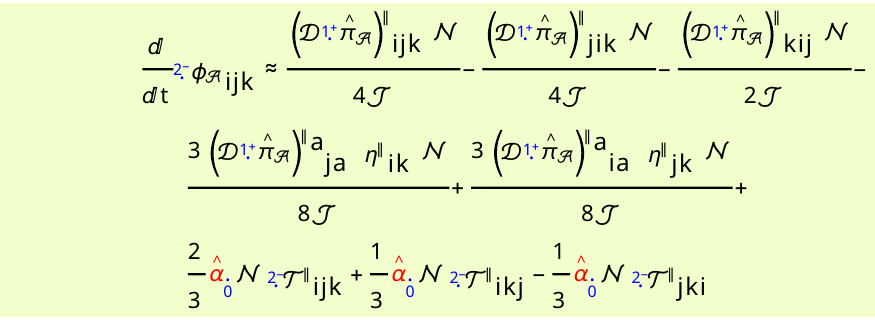}
\end{flushleft}
\vspace{-5pt}
|
\end{lstlisting}
These expressions are expressed on the PiC shell, but not on the sSFC shell. We note in particular the general presence of momentum gradients, which arise at the end of the \HiGGS{} run. In non-minimal theories (i.e. those with PiCs which depend on the field strengths), we can expect gradients of the field strengths to also appear. These are second derivative quantities, even in the first-order formulation of gravity that is PGT: they cannot be further processed by \HiGGS{}, which uses a first-order Euler--Lagrange implementation.
\end{document}